\documentclass[traditabstract]{aa}  

\usepackage{graphicx}
\usepackage{txfonts}
\usepackage{placeins}

\newcommand\ha{${\rm H\alpha}$}
\newcommand\ap{\textsc{autoprof}}
\newcommand\rband{\textit{r}-band}
\newcommand\gband{\textit{g}-band}
\newcommand\grband{\textit{g}, and \textit{r}-band}
\newcommand\grhaband{\textit{g}, \textit{r}, and \ha}
\newcommand\smsd{$\Sigma_*$}
\newcommand\sfrd{$\Sigma_{\rm SFR}$}
\newcommand\angstrom{\mbox{\normalfont\AA}}
\newcommand\hi{H\textsc{i}}
\newcommand\redge{$R_{\rm edge}$} 
\newcommand\rnorm{$R_{\rm norm}$} 

\newcommand\rssfr{$R_{\rm sSFR}$}
\defcitealias{trujillo2020}{T20}
\defcitealias{chamba2022}{C22}
\begin{document}

   \title{A Virgo Environmental Survey Tracing Ionised Gas Emission (VESTIGE)}

   \subtitle{\textbf{XVI. The ubiquity of truncated star-forming disks across the Virgo cluster environment}}

   \author{C.~R.~Morgan\inst{1,2}
          \and
          M.~L. Balogh\inst{1,2}
          \and
          A.~Boselli\inst{3,4}
          \and
          M.~Fossati\inst{5,6}
          \and
          C.~Lawlor-Forsyth\inst{1,2}
          \and
          E.~Sazonova\inst{1,2}
          \and
          P.~Amram\inst{3}
          \and
          M.~Boquien\inst{7}
          \and
          J.~Braine\inst{8}
          \and
          L.~Cortese\inst{9,10}
          \and
          P.~C\^ot\'e\inst{11}
          \and
          J.~C.~Cuillandre\inst{12}
          \and
          L.~Ferrarese\inst{11}
          \and
          S.~Gwyn\inst{11}
          \and
          G.~Hensler\inst{13}
          \and 
          Junais\inst{14}
          \and
          J.~Roediger\inst{11}
          }

   \institute{Waterloo Centre for Astrophysics, University of Waterloo, Waterloo, ON
                N2L 3G1, Canada
              \email{crmorgan@uwaterloo.ca}
         \and
             Department of Physics and Astronomy, University of Waterloo, Waterloo ON N2L 3G1, Canada
         \and
             Aix-Marseille Univ., CNRS, CNES, LAM, Marseille, France
         \and
            INAF - Osservatorio Astronomico di Cagliari, via della Scienza 5, 09047 Selargius , Italy
         \and
            Universit\'a di Milano-Bicocca, Piazza della Scienza 3, 20100 Milano, Italy
         \and
            INAF - Osservatorio Astronomico di Brera, via Brera 28, 21021 Milano, Italy
         \and
            Université Côte d'Azur, Observatoire de la Côte d'Azur, CNRS, Laboratoire Lagrange, 06000, Nice, France
         \and
            Laboratoire d’Astrophysique de Bordeaux, Univ. Bordeaux, CNRS, B18N, allée Geoffroy Saint-Hilaire, 33615 Pessac, France
         \and
            International Centre for Radio Astronomy Research (ICRAR), University of Western Australia, Crawley, WA 6009, Australia
         \and
            Australian Research Council, Centre of Excellence for All Sky Astrophysics in 3 Dimensions (ASTRO 3D), Australia
         \and
            National Research Council of Canada, Herzberg Astronomy and Astrophysics Research Centre, Victoria, BC V9E 2E7, Canada
         \and
            AIM, CEA, CNRS, Universit\'e Paris-Saclay, Universit\'e de Paris, F-91191 Gif-sur-Yvette, France
         \and
            Department of Astrophysics, University of Vienna, Türkenschanzstrasse 17, 1180 Vienna, Austria
         \and
            National Centre for Nuclear Research, Pasteura 7, PL-02-093 Warsaw, Poland
             }

   \date{Received September 15, 1996; accepted March 16, 1997}

% \abstract{}{}{}{}{} 
% 5 {} token are mandatory
 
  \abstract{We examine the prevalence of truncated star-forming disks in the Virgo cluster down to $M_* \simeq 10^7 ~\text{M}_{\odot}$. This work makes use of deep, high-resolution imaging in the \ha{}+[N\textsc{ii}] narrow-band from the Virgo Environmental Survey Tracing Ionised Gas Emission (VESTIGE) and optical imaging from the Next Generation Virgo Survey (NGVS). To aid in understanding the effects of the cluster environment on star formation in Virgo galaxies, we take a physically-motivated approach to define the edge of the star-forming disk via a drop-off in the radial specific star formation rate profile. Comparing with the expected sizes of normal galactic disks provides a measure of how truncated star-forming disks are in the cluster. We find that truncated star-forming disks are nearly ubiquitous across all regions of the Virgo cluster, including beyond the virial radius (0.974 Mpc). The majority of truncated disks at large clustercentric radii are of galaxies likely on first infall. As the intra-cluster medium density is low in this region, it is difficult to explain this population with solely ram-pressure stripping. A plausible explanation is that these galaxies are undergoing starvation of their gas supply before ram-pressure stripping becomes the dominant quenching mechanism. A simple model of starvation shows that this mechanism can produce moderate disk truncations within 1-2~Gyr. This model is consistent with `slow-then-rapid' or `delayed-then-rapid' quenching, where the early starvation mode drives disk truncations without significant change to the integrated star formation rate, and the later ram-pressure stripping mode rapidly quenches the galaxy. The origin of starvation may be in the group structures that exist around the main Virgo cluster, which indicates the importance of understanding pre-processing of galaxies beyond the cluster virial radius.}


   \maketitle
%
%-------------------------------------------------------------------

\section{Introduction}
\label{sec:intro}

Galaxies in the local universe exhibit a bimodality in terms of their colour and star formation rate (SFR). Star-forming (SF) galaxies are blue in colour and are dominated by a disk shape and features such as spiral arms where ongoing star formation occurs. Conversely, `quiescent' galaxies are typically ellipsoidal and red in colour, containing evolved stellar populations with little ongoing star formation. Measurements of the redshift evolution of the stellar mass function (SMF) have shown that the population of quiescent galaxies has built up over time, indicating that galaxies transition or `quench' from actively star-forming to quiescent (e.g. \citealt{faber2007}). Galaxy quenching is known to be a function of mass, as the quiescent fraction is greater at the high end of the SMF in all environments and at all redshifts \citep{peng2010, muzzin2013, lin2014, balogh2016}. 

Galaxies are also known to evolve as a function of their environment \citep{oemler1974, dressler1980, butcher1984}. Galaxies in clusters, particularly those at low redshift, exhibit lower rates of star formation \citep{balogh1997, balogh1998, lewis2002, gomez2003, kauffmann2004, baldry2006, weinmann2006, wetzel2012, gavazzi2013_ha3_3, boselli2023} and are more gas-deficient \citep{giovanelli1985, cayatte1990, gavazzi2005, cortese2011, boselli2014_hrs3, alberts2022} than galaxies in the field. The observed morphology of galaxies also varies by environment, with the fraction of quiescent elliptical galaxies increasing as a function of density \citep{dressler1980, huchra1983, goto2003}. Measurements of this morphology-density relation (e.g. \citealt{dressler1997, sazonova2020}) and the SFR-density relation (e.g. \citealt{quadri2012, cooke2016, kawinwanichakij2017, strazzullo2019, vanderburg2020}) up to $z \sim 2$ indicate that the drivers of environmental quenching in clusters were in place at earlier epochs.
 
A variety of different environmentally-driven mechanisms are known to quench galaxies in clusters (e.g. \citealt{boselli2006_rev, boselli2014_red}). Galaxies can interact with each other through direct mergers which can induce starbursts in the resulting system, alter the morphology of the merged galaxy, and drive AGN feedback \citep{barnes2004, saitoh2009, rich2011, kelkar2020, lotz2021}. Close encounters can strip material to form tidal arms, and can cause gas to lose angular momentum and fall to the centre of the galaxy where a nuclear starburst is induced (`harassment’; \citealt{moore1996, moore1998}). Galaxies falling into clusters will also interact with the hot intra-cluster medium (ICM), during which ram-pressure stripping (RPS) serves to actively strip gas from the disk of the galaxy, removing the fuel necessary for sustained star formation \citep{gunn1972, boselli2022}. Additionally, infalling galaxies can undergo `starvation’ or `strangulation’ as the potential of the galactic dark matter halo becomes subdominant to the cluster potential, preventing the accretion of fresh gas \citep{larson1980}. The existing hot gas contained within the subhalo of the galaxy may also be stripped off via ram-pressure and/or tidal forces \citep{balogh2000, mccarthy2008, kawata2008, bekki2009}, or prevented from cooling due to feedback. Starved galaxies passively quench as they exhaust their remaining cold gas supply.

Each quenching mechanism produces signatures that can be observed and measured to find trends with mass and environment. RPS, for example, can produce elaborate tails of stripped gas that extend beyond the optical disk (e.g. \citealt{gavazzi2001, kenney2004, chung2007} ). In some cases, star formation can occur in the stripped material, producing stars in the tails that can result in a `jellyfish' appearance in the optical (e.g. \citealt{smith2010, ebeling2014, poggianti2016}). Over time, RPS will lead to a truncated SF disk as all the gas is stripped from the outside-in (see \citealt{boselli2022} for a comprehensive review on the effects of RPS). 

In many cases, multiple mechanisms could be quenching galaxies at the same time, or in sequence. Understanding the interplay of various mechanisms in the overall evolutionary picture is an area of ongoing research. For example, while the morphological signatures of RPS may be present in many galaxies, that does not necessarily imply that RPS has been solely responsible for reducing the gas content of the galaxy. A galaxy that has been pre-processed may become even more susceptible to RPS if its gas has been initially displaced by tidal interactions \citep[e.g.][]{serra2023} or its gas content has been reduced by starvation.

\ha{} emission is a powerful tool to observe morphological signatures of environmental quenching mechanisms since it traces star formation on 10~Myr timescales (e.g. \citealt{kennicutt1998_sf}). Early studies such as \citet{moss1993, moss2000} undertook \ha{} surveys of local rich clusters, observing concentrated \ha{} emission that suggested induced star formation due to tidal interactions, particularly for galaxies in denser environments. \citet{gavazzi2006, gavazzi2013_ha3_2} showed that \ha{} emission is correlated with H\textsc{i} gas content in galaxies, and that specific star formation rate (sSFR) is correlated with Hubble type morphology. \citet{koopmann2004_ha_morph} and \citet{fossati2013} compared the sizes of SF disks measured in \ha{} with optical bands to show truncations in the Virgo cluster. More recently, integral field unit (IFU) spectroscopy has become a powerful tool for observing \ha{} emission on a spatially resolved scale across the planes of galaxies. The GASP survey (Gas Stripping Phenomena in galaxies with MUSE; \citealt{poggianti2017}) is a program that has targeted more than 100 nearby jellyfish galaxy candidates to study the kinematics and physical properties of the ionised gas being stripped from the galactic disks. With the advent of IFU spectroscopy and deep, high-resolution imaging with surveys like VESTIGE (the Virgo Environmental Survey Tracing Ionised Gas Emission; \citealt{boselli2018}), detailed \ha{} maps can now be observed for nearby galaxies, allowing star formation and quenching mechanisms to be studied on a spatially resolved scale.

The Virgo cluster is of particular interest to study environmentally-driven quenching mechanisms due in part to its proximity. At a distance of 16.5 Mpc \citep{mei2007}, Virgo is the closest massive galaxy cluster, and has been imaged on sub-kpc scales with many recent surveys including in the UV (GUViCS; \citealt{boselli2011}), optical (NGVS; \citealt{ferrarese2012}), and \ha{} (VESTIGE; \citealt{boselli2018}). Indeed, RPS has been identified as a likely driver of environmental quenching in Virgo through many studies of individual objects and small samples of galaxies (\citealt{koopmann2004_ha_morph, koopmann2006, cortese2012, fossati2013, chung2009, boselli2016, junais2022}; see \citealt{boselli2022} for a review). These studies have identified tails of stripped gas and truncated disks associated with RPS. The majority of SF galaxies in Virgo are also known to be H\textsc{i}-deficient \citep{chamaraux1980, giovanelli1983, gavazzi2005, boselli2023} with truncated gas disks \citep{cayatte1990, cayatte1994, chung2009, boselli2014_hrs3, zabel2022}, indicating the presence of mechanisms to either remove gas or prevent a fresh supply of hot gas from cooling onto galactic disks.

Measuring the size of the SF disk (for example with UV, \ha{} or $24~\rm{\mu m}$ imaging) provides information on the radial extent of star formation, and can be compared to the disk size measured in optical bands to allow a measurement of disk truncation. This is a common method in the literature to determine the presence of an outside-in quenching mechanism (e.g. \citealt{koopmann2004_ha_morph, cortese2012, fossati2013, finn2018}). A popular measurement of the size of the galactic disk is the effective radius, or half-light radius.\footnote{The measurement is often termed as `effective radius' when determined with a S\'ersic fit to a galaxy \citep{devaucouleurs1948, sersic1963}, while `half-light radii' typically is used when the total flux of the galaxy has been determined and used to find the radius which contains half the flux.} The total light of the galaxy can be measured through extrapolation of the flux growth curve, or by using a low-surface brightness isophote. However, the effective radius is sensitive to the concentration of light in the galaxy; this value may be different between the SF and optical disks \citep[e.g.][]{trujillo2020, chamba2020}. Taking the ratio of isophotal radii in SF and optical bands is another method to quantify truncation (e.g. \citealt{cortese2012}), where the isophotal radii provide a more meaningful representation of the extent of the disk. In using a S\'ersic fit to define the radial profile of a galaxy, the effective or isophotal radius are well-defined measures of size, but neither correspond to an actual edge of the disk.

\citealt{chamba2022} (hereafter \citetalias{chamba2022}) identified the edges of galaxies (\redge{}) based on visual identification of turn-off points in their stellar mass surface density and optical colour profiles. They showed that this visually identified edge corresponds to a critical stellar mass surface density that depends weakly on the integrated stellar mass of the galaxy. For massive disk galaxies ($M_* \gtrsim 10^9 ~\text{M}_{\odot}$), this turn-off corresponds with the edge of the star-forming disk, as indicated by a sudden change in slope of the UV surface brightness profiles.  As such, the edge of the stellar disk should correspond with the edge of the SF disk for normal, SF galaxies. A follow-up work, \citealt{chamba2024} found that galaxies in the Fornax cluster have edges that occur at higher stellar mass surface density, giving rise to smaller measured disk sizes.

In addition to physical disk edges, various studies have observed other features that represent deviations from a single exponential fit to a surface brightness profile. Since the work of \citet{freeman1970}, it has been shown that galaxy disks are often best fit by two exponential components, separated by a break, where the outer part of the disk may have a steeper (Type II) or shallower (Type III) slope than the inner part \citep[e.g.][]{pohlen2006, hunter2006, erwin2008, herrmann2013, herrmann2016, watkins2019}. These breaks are often identified in optical bands, though comparisons with star-forming bands (UV and \ha{}) show that in some cases, the breaks are more even more significant in these bands than in the optical \citep[e.g.][]{herrmann2016}, suggesting that these breaks may correspond to the edge of the star-forming disk.

In this work we look to quantify, in the Virgo cluster, the presence of galactic disks with a truncated star-forming component relative to the underlying optical disk. \citet{boselli2020} identified truncated SF disks in a subset of VESTIGE galaxies through the decline in the number density of H\textsc{ii} regions. Other previous studies of truncated SF disks in Virgo were limited to massive galaxies (e.g. \citealt{koopmann2004_ha_morph, boselli2022}) or used size measurements such as the effective radius that do not explicitly trace the edge of the SF disk (e.g. \citealt{fossati2013}). In our approach, we take advantage of the unique NGVS and VESTIGE data to analyse sSFR profiles and identify a physically-meaningful edge to the SF disk. By comparing the edge of the SF disk to the expected disk size based on a comparable field sample, we can quantify how truncated the SF disks are in the Virgo cluster. With our cluster-wide sample that includes all SF galaxies within the cluster virial radius and a sample beyond, we look to provide insights into the different possible quenching pathways that may produce the observed distribution of truncated disks.

A description of the data used in this work is outlined in Section \ref{sec:data}. In Section \ref{sec:methods} we outline the methods used to measure SF disk edges. Section \ref{sec:results} presents the main results of this paper, which are discussed in context in Section \ref{sec:discussion}. We summarise and conclude in Section \ref{sec:conc}. Where necessary, such as in the modelling of Section \ref{sec:discussion}, we adopt a flat $\Lambda$CDM cosmology with $H_0=70~\text{km s}^{-1}~\text{Mpc}^{-1}$ and $\Omega_m=0.3$.

\section{Observations and data reduction}
\label{sec:data}

This study focuses on the Virgo cluster, the nearest massive cluster at a distance of 16.5 Mpc \citep{mei2007}. The main cluster of Virgo (Cluster A) is a virialised system comprised of mostly early-type quiescent galaxies (including dwarfs and massive ellipticals), while several smaller substructures merging into the cluster contain more SF systems \citep{binggeli1987, gavazzi1999, solanes2002, boselli2014_guvics}. Early studies of the dynamics of Virgo derived a mass profile for the cluster with $R_{200}=1.55~\text{Mpc}$ \citep[e.g.][]{mclaughlin1999}.  However, recent work deriving the mass profile from X-ray observations have measured a Navarro-Frenk-White (NFW; \citealt{navarro1996}) profile with $R_{200}=0.974~\text{Mpc}$ and $M_{200}=1.05\times 10^{14}~\text{M}_{\odot}$ \citep{simionescu2017}. These latter values will be adopted throughout this work.

\subsection{NGVS imaging}

The Next Generation Virgo Survey (NGVS) is an optical survey of the entire Virgo cluster out to 1.55 Mpc \citep{ferrarese2012}. Using MegaCam on the Canada-France-Hawaii Telescope (CFHT), NGVS imaged a 104~deg$^2$ footprint of the Virgo cluster in the $u^*, \, g, \, i$ and $z$ bands (and partially in the $r$ band). In this work, we make use of the \gband{} images from NGVS to determine radial $g-r$ colour profiles for galaxies observed with the VESTIGE survey (with \rband{} imaging coming from VESTIGE). Total integration times of $3170~\text{s}$ were used to reach point source depths of $25.9~\text{mag}$ ($10\sigma$) and surface brightness (SB) depths of $29~\text{mag arcsec}^{-2}$ in the \gband{}. 

CFHT MegaCam consists of an array of 40 CCDs with pixel scale of $0.187''~\text{pixel}^{-1}$, and observations were carried out under seeing conditions of $<1''$ (FWHM). 
The NGVS images used in this work were processed using Elixir-LSB, a pipeline designed for NGVS that employs specific methods for detection of low surface brightness features \citep{ferrarese2012}. Photometric calibration and astrometric correction of the data follows the MegaCam procedures detailed in \citet{gwyn2008}.

\begin{table*} 

\caption{Properties of Virgo cluster substructures (taken from \citealt{boselli2023_lfha})}
\label{tab:sub}
\centering
\begin{tabular}{cccccccccc}
\hline
\hline
Substructure & RA & DEC  & Radius  & Vel range  & Distance  & $\langle v \rangle$  &  $\sigma$   & Central galaxy & \# of  \\ 
 & deg & deg  & deg & km/s & Mpc & km s$^{-1}$   &  km s$^{-1}$  &  & galaxies\\ \hline

Cluster A  &  187.71  &  12.39  & 5.383 & $<3000$ &  16.5 &  955 &  799 & M87 & 182 \\
Cluster B  &  187.44  &  8.00  & 3.334 & $<3000$ &  23 &  1134 &  464 & M49 & 115 \\
Cluster C  &  190.85  &  11.45  & 0.7 & $<3000$ &  16.5 &  1073 &  545 & M60 & 5 \\
W Cloud  &  185.00  &  5.80  & 1.2 & $ 1000 < v <3000$ &  32 &  2176 &  416 & NGC~4261  & 18 \\
W$'$ Cloud &  186.00  &  7.20  & 0.8 & $<2000$ &  23 &  1019 &  416 & NGC~4365 & 17 \\
M Cloud  &  183.00  &  13.40  & 1.5 & $1500 < v <3000$ &  32 &  2109 &  280 & NGC~4168 & 25 \\
LV Cloud  &  184.00  &  13.40  & 1.5 & $<400$ &  16.5 &  85 &  208 & NGC~4216  & 22 \\

\hline\hline
    \end{tabular}
\label{tab:regions}

\end{table*}

\subsection{VESTIGE imaging}

In addition to \gband{} images from NGVS, this work makes use of images obtained for the VESTIGE survey. VESTIGE is a blind, narrow-band (NB) \ha{} survey covering 104~deg$^2$ of the Virgo cluster following closely the footprint defined by NGVS. Imaging was carried out with MegaCam on CFHT using the NB MP9603 filter ($\lambda_c=6591~\angstrom $; $\Delta\lambda = 106~\angstrom$) as well as the broadband $r$ filter. At the low redshift of Virgo ($-300 \leq v_{\rm hel} \leq 3000~\text{km s}^{-1}$), the NB filter contains the \ha{} Balmer line at $\lambda=6563~\angstrom$ as well as the two [N\textsc{ii}] lines at $\lambda=6548,6583~\angstrom$. Integration times of 2~hours in the NB filter have been used to achieve depths of $f(\text{H}\alpha)\simeq 4 \times 10^{-17}~\text{erg s}^{-1}~ \text{cm}^{-2}$ at $5\sigma$ for point sources and $\Sigma(\text{H}\alpha)\simeq 2 \times 10^{-18}~\text{erg s}^{-1}~ \text{cm}^{-2}~ \text{arcsec}^{-2}$ ($1\sigma$ after smoothing the data to $\sim \! 3''$ resolution) for extended sources. \rband{} exposures are 12~minutes, achieving a point source depth of $24.5~\text{mag}$ ($5\sigma)$ and a SB limit of $25.8~\text{mag arcsec}^{-2}$ ($1\sigma$ for scales comparable to the size of structures in target galaxies, $\sim30~\text{arcsec}$). At the time of writing, the survey is 76\% complete, with observations having been taken in excellent seeing conditions (FWHM = $0.76^{\prime\prime} \pm 0.07^{\prime\prime}$). At the distance of Virgo, $1^{\prime\prime}$ is equal to $80~\text{pc}$. The full observing strategy is discussed in detail in \citet{boselli2018}.

Stellar continuum subtraction of the NB images is described in \citet{boselli2019} and is done using a combination of the \rband{} images and $g-r$ colours.

\subsection{Galaxy identification}
\ha{} emitting sources in VESTIGE images were identified as described in \citet{boselli2023}. We summarise the method here:

\begin{enumerate}

    \item[1.] Counterparts to galaxies in the Virgo Cluster Catalogue (VCC; \citealt{binggeli1985}) with redshifts in the range of Virgo (307 objects). 

    \item[2.] Counterparts of the H\textsc{i} detections from the ALFALFA survey \citep{giovanelli2005} with redshifts in the range of Virgo (37 more objects). 

    \item[3.] Counterparts of galaxies identified as Virgo members based on the scaling relations determined using NGVS data in \citet{ferrarese2012, ferrarese2020} and \citet{lim2020} (31 more objects). 

    \item[4.] Four bright galaxies outside the VCC and NGVS footprints with extended \ha{} emission. 

    \item[5.] Five additional line emitter objects not previously identified, but whose extended emission suggests they are local bright compact dwarfs.

\end{enumerate}

\noindent Thus, the total number of Virgo members with \ha{} emission in the VESTIGE survey is 384. spanning a stellar mass range $10^6 \text{M}_{\odot} < M_* < 10^{12}\text{M}_{\odot}$. The morphology distribution of the sample is $61\%$ late-type spirals, $5\%$ early-type/lenticular, $20\%$ late-type dwarf (including blue-compact dwarfs), $9\%$ early-type dwarf, and $5\%$ with unclassified morphologies. %%we will likely exclude certain objects based on axis ratio, stellar mass/flux, early type galaxies, etc

\subsubsection{Galaxy membership in subcluster regions}

\label{sec:subclusters}

The Virgo cluster is made up of several subclusters and clouds currently in the process of merging into Cluster A which features M87 at its centre. We make use of the most recent information of the Virgo substructures as in \citet{boselli2023_lfha}, which we summarise in Table \ref{tab:sub}. The distances to the substructures are based on measurements from \citet{gavazzi1999} who employed Fundamental Plane and Tully-Fisher relations for redshift-independent distance determination. However, for consistency with \citet{boselli2023_lfha} and other works from the VESTIGE collaboration, we use a distance of 16.5~Mpc \citep{mei2007} for Clusters A and C and the Low-Velocity cloud instead of 17~Mpc which is used in previous studies \citep{gavazzi1999}. Additionally, a sole galaxy (VCC~357) has a measured velocity of $3008 ~\text{km s}^{-1}$ which places it outside the Virgo range of $-300 ~\text{km s}^{-1} \leq v_{\rm hel} \leq 3000 ~\text{km s}^{-1}$. However, we consider VCC~357 to be a Virgo member, following \citet{boselli2023_lfha}. 

The radii in Table \ref{tab:sub} for Clusters A and B are $R_{200}$ values derived in \citet{mclaughlin1999} and \citet{ferrarese2012}; we consider these values upper limits on the virial radii and use them only to assign galaxies to their likely subclusters. Galaxies are assigned potential membership to substructures based on 2D position and radial velocity information following the argument of \citet{boselli2014_guvics} and in consistency with \citet{boselli2023_lfha}. In overlapping regions, galaxies are assigned to the smaller substructure, except in the overlap between Clusters A and B where galaxies are assigned based on proximity to the centres of the subclusters.

\subsection{Supplementary data}

\subsubsection{Stellar masses}

Stellar masses were also derived in \citet{boselli2023}, and we briefly highlight the process here. Spectral energy distributions (SEDs) were fit using the Code Investigating GALaxy Emission (CIGALE; \citealt{boquien2019}). A delayed star formation history (SFH) with a recent burst or quench was used, along with \citet{bruzual2003} stellar population synthesis (SPS) models and a \citet{chabrier2003} initial mass function (IMF). SEDs were fit to multi-frequency observations including (where available) GALEX far-UV and near-UV bands, NGVS $u$, $g$, $r$, $i$ and $z$ bands (or SDSS magnitudes where needed; \citealt{kim2014}), WISE 22 $\rm \mu m$ and \textit{Herschel} at 100, 160, 250, 350 and 500 $\rm \mu m$. The resulting stellar masses have uncertainties $<0.15 ~\rm dex$, with a mean uncertainty of $0.07 ~\rm dex$. Results were compared with SED fitting done on NGVS data using PROSPECTOR \citep{johnson2021}, with excellent agreement found.

\subsubsection{H\textsc{i}-deficiency}

\hi{}-deficiency is defined as the logarithmic difference between the observed mass of \hi{} gas in a galaxy compared to the expected \hi{} mass for isolated galaxies of similar size and morphology. The process for determining the \hi{}-deficiency parameter is detailed in \citet{boselli2023}. The calibration used is from a local sample of $\sim \! 8000$ galaxies from the work of \citet{cattorini2023}. Where available, deep observations from the GOLDMine database \citep{gavazzi2003} were used to determine \hi{} mass, and otherwise ALFALFA \citep{haynes2018} data was used. The scatter in the calibration corresponds to $\sim 0.26~\text{dex}$, and galaxies with \hi{}-deficiency $\geq 0.4$ ($\sim 1.5\sigma$) are considered with high probability to be gas deficient, perturbed systems.

\begin{figure*}
\centering
\includegraphics[width=17cm]{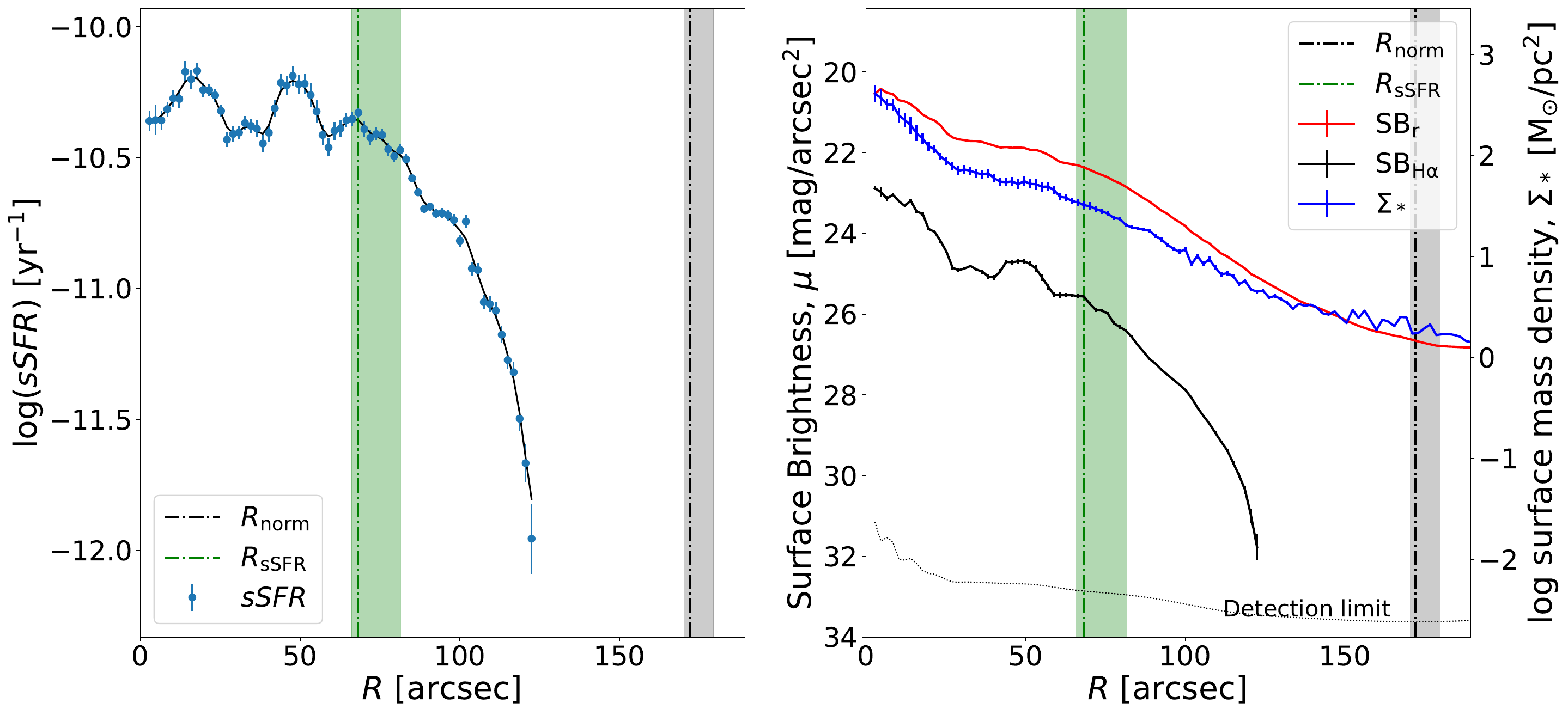}
\caption{\textit{Left}: sSFR profile for VCC~865 (NGC~4396; blue data points with a smoothed profile shown as a black curve). \textit{Right}: Surface brightness profiles for \ha{} NB (black) and r-band emission (red), along with the \smsd{} profile (blue). The dashed black profile shows the detection limit for the \ha{} profile. In both panels, the green vertical line shows the median SF disk edge, \rssfr{}, and the shaded green band the uncertainty range based on the $16^{\rm th}$ and $84^{\rm th}$ percentiles computed using a Monte Carlo simulation with 1000 iterations. The vertical black line shows the expected edge, \rnorm{}, of the galaxy and the shaded grey band uncertainty range based on the $16^{\rm th}$ and $84^{\rm th}$ percentiles of the Monte Carlo simulation. }
\label{fig:VCC865_prof}
\end{figure*}

\section{Methods}
\label{sec:methods}

The pipeline used for measuring SF disks in VESTIGE galaxies is outlined in the following. Throughout this section, we use VCC~865 (NGC~4396) as a case study and refer back to Fig.~\ref{fig:VCC865_prof} to exemplify each step of the pipeline. VCC~865 has a stellar mass of $10^{9.4}\text{M}_{\odot}$. It has a SFR that places it on the SFMS, yet appears to be somewhat gas deficient, and is likely on first infall into the cluster based on its location in phase space. For these reasons, we use VCC~865 as an illustrative example of a galaxy that is most relevant to our analysis, as further discussed in Sects. \ref{sec:results} and \ref{sec:discussion}.

\subsection{Extraction of radial profiles}

Our analysis relies on using radial sSFR profiles to determine the edge of the star-forming disk. sSFR is determined using SFR and stellar mass surface density (\sfrd{} and \smsd{}) profiles, which we obtain using \grhaband{} surface brightness profiles. 

\subsubsection{Creating object masks}

\label{sec:masks}

In order to properly extract SB profiles of our target galaxies, we need to mask any contaminating stars or background galaxies. For each galaxy in VESTIGE, we cut a postage stamp around the object with a size of $10 \, r_{\rm e, i}$ on each axis, where $r_{\rm e, i}$ are \textit{i}-band effective radii coming from S\'ersic profile and growth curve fitting performed on NGVS images (in the few case where $r_{\rm e, i}$ is unavailable from NGVS, we use $r_{\rm e, i}=60^{\prime\prime}$). We use the Python package \textsc{photutils} \citep{photutils} to create object masks for our postage stamps. Since our galaxies span a wide range of sizes and luminosities and our data quality and depth pick up small structures and low-SB features, we find a single masking run does not suffice. 

We choose to employ a `hot, cool, cold' masking routine based on \citet{galametz2013} and \citet{sazonova2021}. First, the images are passed through a 2D top-hat smoothing kernel with a size equal to $0.2R_{\rm e,i}$. The `hot' run uses a very high detection threshold, small minimum area and a small amount of deblending to identify the brightest peaks in the image. The `cool' run follows, using a low detection threshold, large area and strong deblending. Using these first two runs, we mask `cool' regions that are far from the galaxy centre, and mask nearby `cool' regions where a `hot' peak was identified (this attempts to mask areas around saturated stars, for example), except for the central hot peak associated with the galaxy. Finally, we perform the `cold' run which uses the mask created in the previous step and performs one more detection run using a low threshold, no deblending and a minimum area of 20 pixels. 

We found that this routine worked well based on a visual inspection of all masks. In certain cases where over- or under-deblending occurred, we re-ran the masking routine, altering certain parameters as necessary until a good mask was produced. We show a table of the \textsc{photutils} parameters used in each run in Appendix \ref{app:mask}. For a more detailed explanation of the `hot, cool, cold' masking technique, we refer the reader to \citet{sazonova2021}.

\subsubsection{Radial surface brightness extraction}

The next step in our analysis is to extract surface brightness profiles of VESTIGE galaxies in the \textit{g, r} and \ha{} bands ($\mu_{\rm r}, \mu_{\rm g}$ and $ \mu_{\rm H{\rm \alpha}}$). We use the non-parametric surface brightness extraction pipeline \ap{} \citep{stone2021}. \ap{} is based on the work of \citet{jedrzejewski1987} but with addition of machine learning regularisation techniques. The software provides a full pipeline for background fitting, centroid finding, masking and surface brightness extraction, and is proven to be proficient is extraction of low-surface brightness features. We use the deep \textit{g}-band images in our initial \ap{} run, and set the object centre to be the same as the optical positions used in \citet{boselli2023}, determined with NGVS \citep{ferrarese2012}. We run \ap{} with a fixed isoband width of 10 pixels ($1.87^{\prime\prime}$); full width of band) and sample linearly in increments of 10 pixels, applying the masks created as described in Sect. \ref{sec:masks}. \ap{} varies the ellipticity and position angles of the isophotes with each radial step as the shape of the galaxy changes. Since the structure of SF galaxies is complex, especially with high-resolution data, varying the ellipticity and position angle in this way provides a more accurate measure of the galaxy shape in each radial bin. We then use the \ap{} forced photometry method to force the extracted \textit{g}-band profile onto the \rband{} and \ha{} images in order to obtain the same radial sampling, ellipticities and position angles in all three bands. 

In the right panel of Fig.~\ref{fig:VCC865_prof} we show radial SB profiles for the \rband{} in red and the NB \ha{} in black. The black dotted line shows the \ha{} detection limit based on the noise in the image and the size of our sampled annuli. Uncertainties on surface brightness values are computed by \ap{} dividing the inter-percentile range (between the $16^{th}$ and $84^{th}$ percentiles) by $2\sqrt{N}$ where $N$ is the number of pixels contributing to the SB measurement.

\begin{figure}
\centering
\includegraphics[width=\columnwidth]{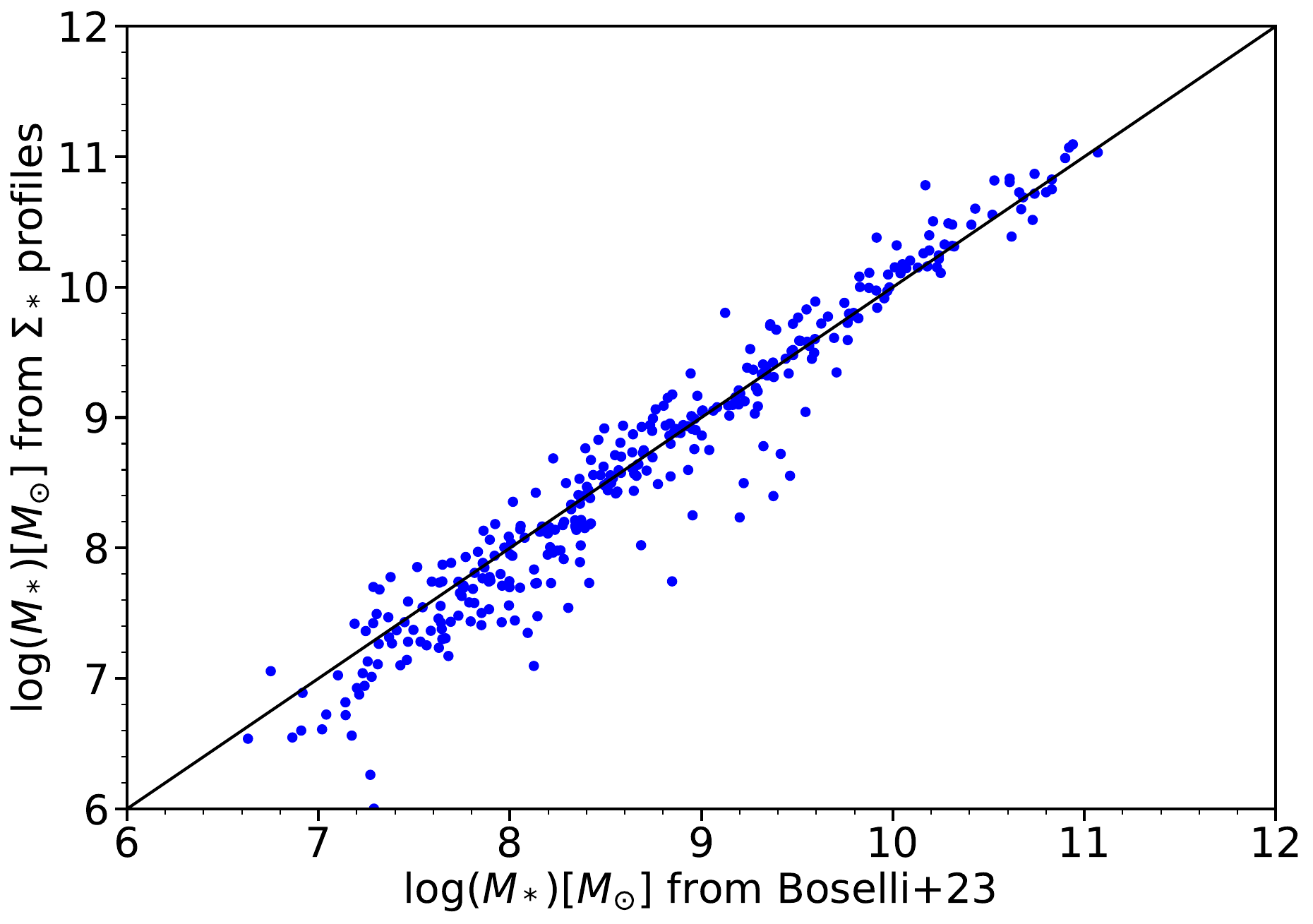}
\caption{Recovered total stellar masses obtained by numerically integrating \smsd{} profiles up to \rnorm{} compared to stellar masses obtained by \citet{boselli2023} through SED fitting}
\label{fig:mass_check}
\end{figure}

\begin{figure*}
\centering
\includegraphics[width=16.cm]{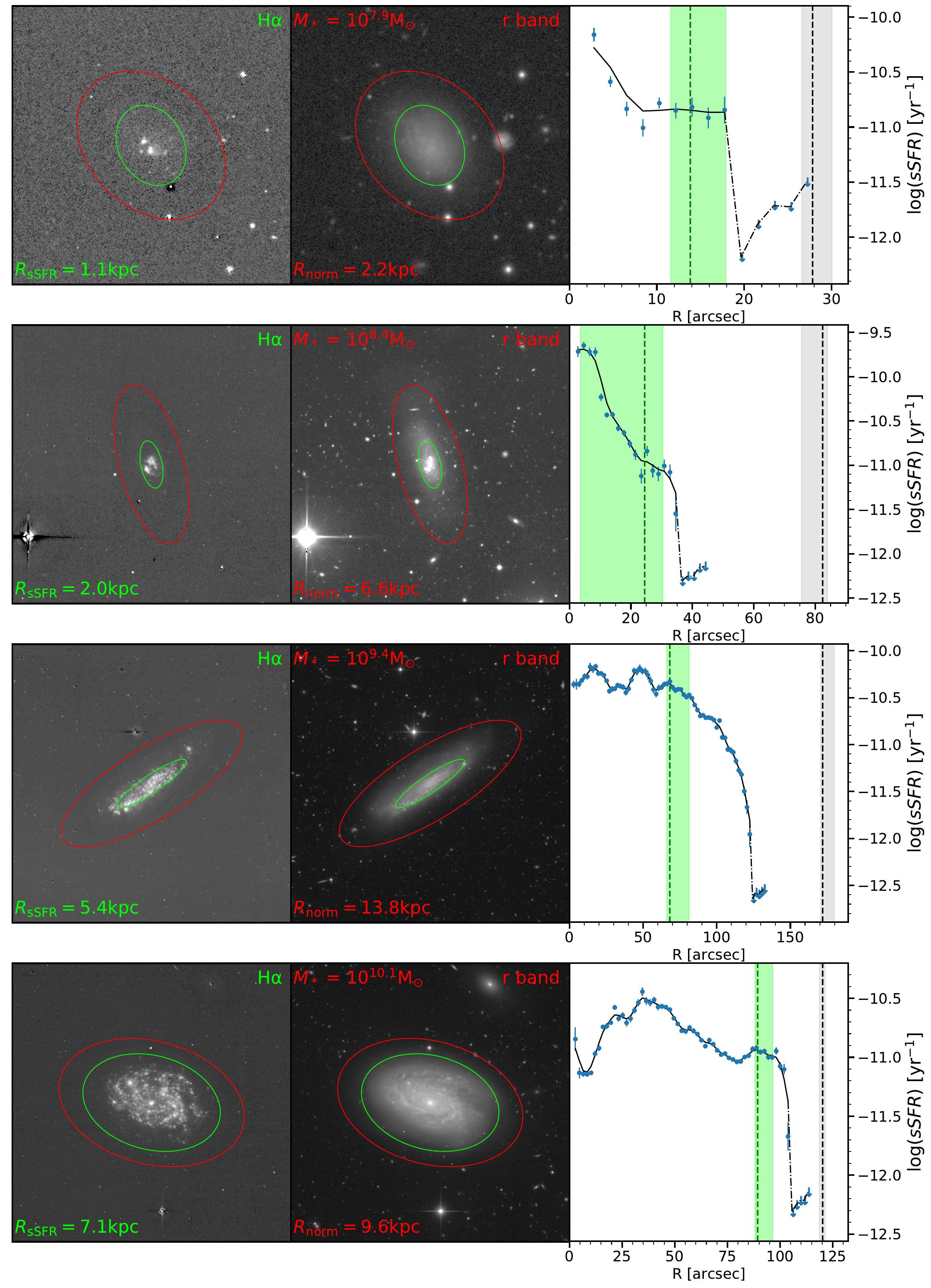}
\caption{Continuum-subtracted \ha{} emission (\textit{left}), \rband{} optical emission (\textit{centre}), and sSFR profile (\textit{right}) for VCC~304, VCC~2037, VCC~865 (NGC~4396), VCC~157 (NGC~4208). In the left and centre postage stamps, the green ellipse outlines \rssfr{} while the red ellipse outlines \rnorm{}. On the sSFR plots (\textit{right}), we show \rssfr{} and its associated uncertainty range as a green vertical line and band, and \rnorm{} and its uncertainty range as a grey vertical line and band. Also included are the upper limits (down arrows) in the next five radial bins in which \ha{} is not detected (but stellar mass is); this represents the upper limit to the sSFR based on the \ha{} detection limits.}
\label{fig:stamps}
\end{figure*}

\renewcommand{\thefigure}{\arabic{figure} (Cont.)}
\addtocounter{figure}{-1}
\begin{figure*}
\centering
\includegraphics[width=16.cm]{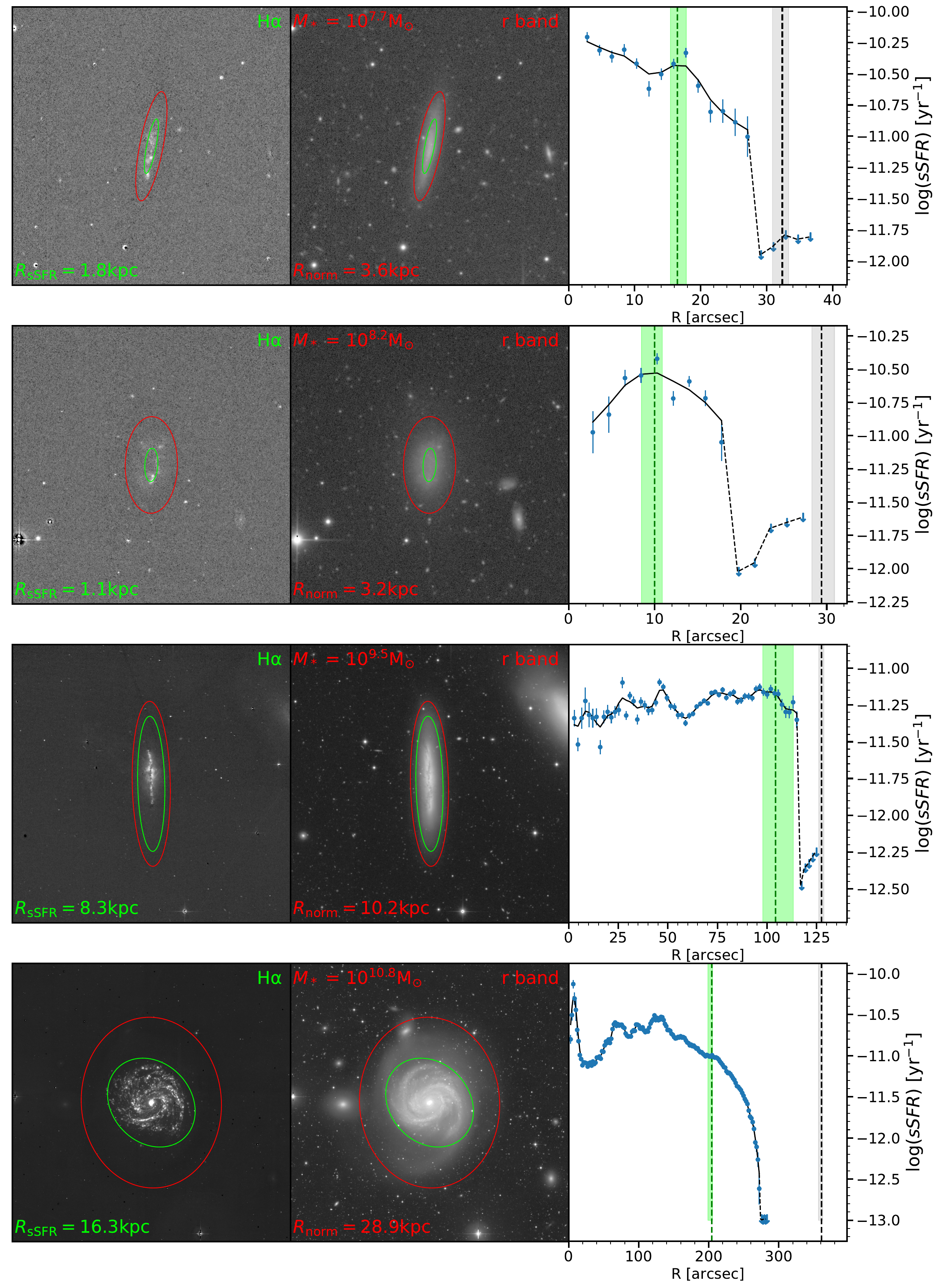}
\caption{Continuum-subtracted \ha{} emission (\textit{left}), \rband{} optical emission (\textit{centre}), and sSFR profile (\textit{right}) for VCC~1605, VCC~565, VCC~1868 (NGC~4607), VCC~596 (NGC~4321). In the left and centre postage stamps, the green ellipse outlines \rssfr{} while the red ellipse outlines \rnorm{}. On the sSFR plots (\textit{right}), we show \rssfr{} and its associated uncertainty range as a green vertical line and band, and \rnorm{} and its uncertainty range as a grey vertical line and band. Also included are the upper limits (down arrows) in the next five radial bins in which \ha{} is not detected (but stellar mass is); this represents the upper limit to the sSFR based on the \ha{} detection limits.}
\label{fig:stamps2}
\end{figure*}

\renewcommand{\thefigure}{\arabic{figure}}

\subsubsection{Stellar-mass surface density profiles}

To obtain \smsd{} profiles, we turn to the mass-to-light versus colour relations (MLCRs) calibrated in \citet{roediger2015}. These calibrations provide linear relations between colour and mass-to-light ratio $(M_*/L)_{\rm \lambda}$ for a given waveband, $\lambda$, based on either \citet{bruzual2003} or \citet{conroy2009} (FSPS) stellar population synthesis (SPS) models. It follows that:
\begin{equation}
    \log{(M_*/L)_{\rm \lambda}} = m_{\rm \lambda} \times ({\rm colour}) + b_{\rm \lambda}.
	\label{eq:ML}
\end{equation}

\noindent We choose to use  $\lambda = r$, $g-r$ colour and the \citet{bruzual2003} SPS models with a \citet{chabrier2003} IMF, so $m_{\rm \lambda}=1.629$ and $b_{\rm \lambda}=-0.792$ (see Table A1 of \citealt{roediger2015}). After calculating $\log{(M_*/L)_{\rm r}}$ at each radial step from the \grband{} SB profiles, we transform our \rband{} SB profile into \smsd{} using:
\begin{equation}
    \log{\Sigma_*} = \log{(M_*/L)_{\rm r}} + 0.4\! \left(m_{\rm abs, \odot , r} - \mu_{\rm r} \right) + 8.63,
	\label{eq:smsd}
\end{equation}

\noindent where $m_{\rm abs, \odot , r}=4.64$ \citep{willmer2018} and \smsd{} has units of $\rm M_{\odot}/{\rm pc^2}$. The right panel of Fig.~\ref{fig:VCC865_prof} shows the \smsd{} profile for VCC~865 in blue, using the secondary y-axis on the right.

While Eq. \ref{eq:smsd} provides a convenient measure of stellar mass, it does not account for variations in dust content amongst individual galaxies. Larger-than-average dust leads to an underestimated luminosity, biasing the mass low.  This is partially mitigated by the fact that the larger dust content also reddens the light, therefore increasing the mass-to-light ratio inferred from observed colours. The net effect is that there is a small, systematic uncertainty in our stellar mass profiles. Based on radial dust attenuation profiles from the Calar Alto Legacy Integral Field Area (CALIFA: \citealt{sanchez2012, gonzalez-delgado2015}) and Mapping Nearby Galaxies at Apache Point (MaNGA: e.g. \citealt{bundy2015, goddard2017, greener2020}), we expect that, in the most extreme cases, the radial variation in a dust profile is $\pm 0.5~\text{mag}$ from the weighted average. Propagating this effect through Eqs. \ref{eq:ML} and \ref{eq:smsd} shows that the attributed uncertainty on $\log{(\Sigma_*)}$ is $\lesssim 0.3~\text{dex}$. 

As a consistency check of our \smsd{} profiles, we numerically integrate the \smsd{} profiles out to \rnorm{} (as defined in \ref{sec:rnorm}) to obtain the total stellar mass for the galaxy, and compare the results to the stellar masses obtained from SED fitting by \citep{boselli2023}, finding good agreement as shown in Fig.~\ref{fig:mass_check}. The mean difference in log-log space between our recovered stellar mass value and that measured in \citet{boselli2023} is 0.02 with a scatter of 0.21 after removing $3\sigma$ outliers. Our \smsd{} profiles are corrected for inclination such that each profile describes a face-on galaxy, allowing consistent determination of galaxy edges based on critical \smsd{} value. We describe the inclination correction in Appendix \ref{app:inc}.

\subsubsection{SFR profiles}

To transform \ha{} SB profiles into SFR surface density (\sfrd{}) profiles, we use the calibration of \citet{calzetti2010} (which assumes solar metallicity and constant SFR) modified to include a Chabrier IMF \citep{chabrier2003}, as in \citet{boselli2023}. This model produces the relation
\begin{equation}
    {\rm SFR} \, [{\rm M_{\odot} \, yr^{-1}}] = 5.01 \times 10^{-42} L \! \left( {\rm H\alpha} \right) [\rm erg \, s^{-1}].
\end{equation}

\noindent \ha{} luminosities were determined based on the distance to the subcluster regions to which each galaxy belongs, based on the classifications in Section \ref{sec:subclusters}. We do not apply corrections for internal dust attenuation or [N\textsc{ii}] contamination. Global corrections would not alter the shape of the radial profiles being studied, and deriving spatially-resolved dust attenuation curves is beyond the scope of this work. As a check that dust corrections are not necessary for our results, we applied spatially resolved attenuation corrections ($A_{\rm H\alpha}$ as a function of radius) from CALIFA \citep{sanchez2012, gonzalez-delgado2015} and MaNGA \citep[e.g.][]{bundy2015, goddard2017, greener2020} to a subset of our galaxies and re-ran our algorithm for finding \rssfr{}. We found no changes in our measured \rssfr{}, showing that typical dust gradients would have no effect on our measured values.

\subsection{Quantifying the degree of truncation}

\label{sec:ratio}

We measure the edges of the star-forming disks of VESTIGE galaxies by identifying a turn-off point in their sSFR profiles, as described in detail below (Sect. \ref{sec:rssfr}). In order to determine how truncated VESTIGE star-forming disks are relative to the field, we would ideally measure the edges of star-forming disks from an isolated field sample matched on colour and stellar mass. However, the VESTIGE sample does not extend beyond the Virgo cluster environment, and existing large \ha{} field samples have comparatively shallow depths. Instead, we compare our star-forming disk sizes to the expected sizes of disks based on the field sample of \citetalias{chamba2022}. At least for massive galaxies on the SFMS, \citetalias{chamba2022} show that the optical edge corresponds to the edge of the SF disk, evidenced by breaks in the UV profile. We describe this process in more detail in Sect. \ref{sec:rnorm}.

\subsubsection{Measuring the edge of the star-forming disk}

\label{sec:rssfr}

We take inspiration from the methods of \citetalias{chamba2022} who determined the edges of galactic disks by looking for breaks in surface brightness and optical colour profiles. Combining our constructed SFR and stellar mass density profiles, we calculate specific star formation rate (sSFR) profiles determined by:

\begin{equation}
    \text{sSFR}(r)=\frac{\text{SFR}(r)}{\Sigma_*(r)}.
    % sSFR(r)=\frac{SFR(r)}{\Sigma_*(r)}
\end{equation}
Uncertainties on the sSFR profiles are determined by compounding the uncertainties associate with the SFR and \smsd{} profiles. Because the \smsd{} profiles are so precisely defined, the uncertainties on the sSFR values are typically small relative to even local variations in the sSFR profile.

For disk galaxies on the star-forming main sequence (SFMS), sSFR profiles are shown to be flat for galaxies with $M_*<10^{10.5}~\rm{M_{\odot}}$ (the majority of our sample), with central suppressions in massive SF galaxies (e.g. \citealt{belfiore2018}). This makes it straightforward to identify a drastic change in the shape of the profile, which signifies a change in the SFR distribution relative to the underlying stellar mass distribution. This means that features such as inner disk breaks will not be captured unless they correspond to a SF disk edge, as desired.

 We then define \rssfr{} as the last local maximum in the sSFR profile before it drops off, ignoring small peaks that may occur while the sSFR profile is dropping off. Because VESTIGE imaging has high-resolution, our profiles are sensitive to small structures across the galaxy. We choose to smooth our profiles with a 1D Gaussian filter with $\sigma=1$ to reduce the risk that a peak due to small-scale structure is determined to be the turn-off point. In certain extreme cases, galaxies are found to have centralised star formation when visually inspected. This observation results in a very truncated disk according to our criteria, though the sSFR profile may not have any detectable local maxima. Thus, we add a condition in our algorithm to check if the sSFR profile falls by at least 1~dex over the range of radii for which sSFR values are measured. If this is the case, we identify the smallest radial value as the edge of the SF disk (\rssfr{}$=2.805^{\prime\prime}$). 
 
 In Fig.~\ref{fig:VCC865_prof}, \rssfr{} is shown with its associated uncertainty range (Sect. \ref{sec:uncert}) as a green band and green dashed line. In this example, \rssfr{} clearly corresponds to a the location where a sharp decline in the sSFR profile begins. While there is still non-negligible star formation beyond this radius, the large change in slope represents a well-defined edge.  In other galaxies (e.g. top row in Fig.~\ref{fig:stamps}) the drop is even more severe, with no detectable star formation beyond \rssfr{}. 
 
 There exists in the literature many different methods for detecting features such as disk breaks which in some cases may correspond with the SF disk edges that we measure \citep[e.g.][]{hunter2006, herrmann2013, herrmann2016}. While beyond the scope of this paper, a study comparing the features identified by these different methods and the physical interpretations of each would be an interesting future work.

\begin{figure*}
\centering
\includegraphics[width=16cm]{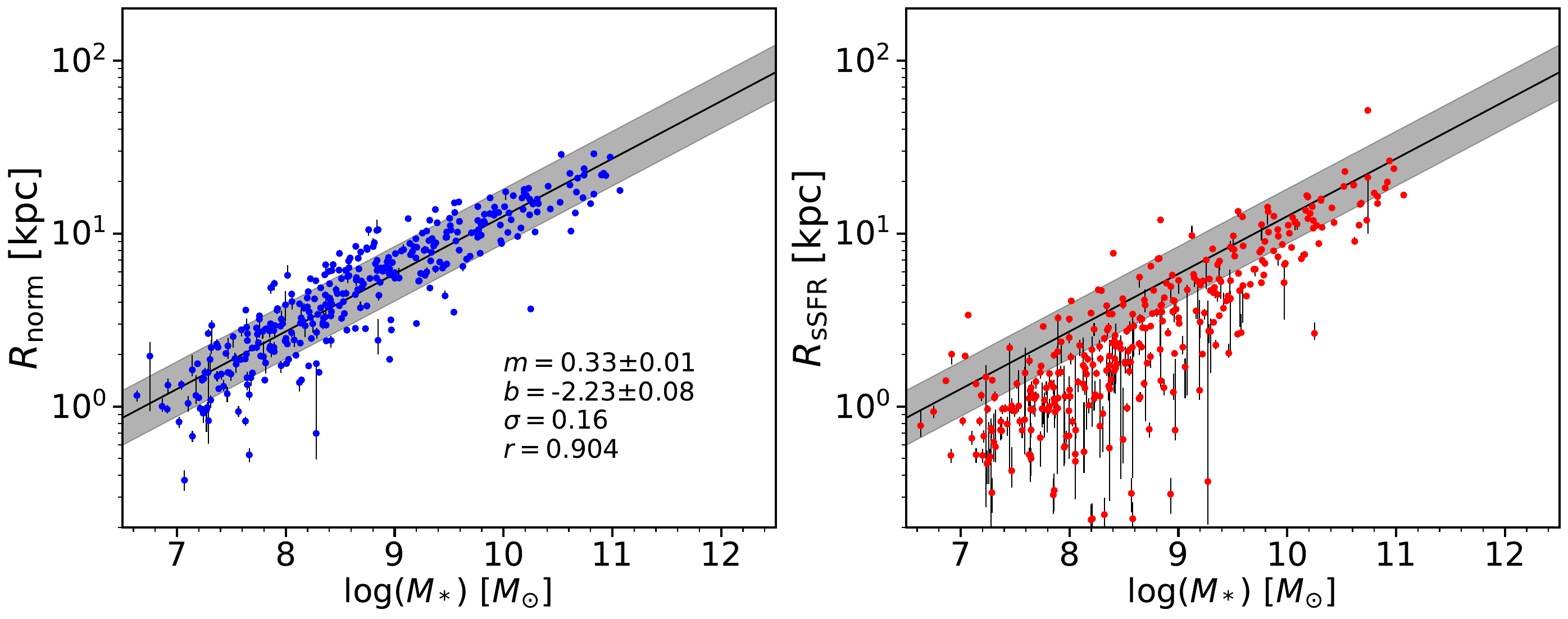}
\caption{\textit{Left}: Size-stellar mass relation for galaxies with identified $R_{\rm norm}$. We fit a linear relation to the data and show that a slope of $m=0.34 \pm 0.01$ is measured, consistent with \citetalias{chamba2022}. The grey shaded region shows the $1\sigma$ scatter in the linear relation. \textit{Right}: Size-stellar mass relation using \rssfr{}, showing the linear fit to the \rnorm{}-mass relation}
\label{fig:size_mass}
\end{figure*}

\subsubsection{Determining the expected edges of galaxies}

\label{sec:rnorm}

We wish to define an `expected' edge for isolated counterparts to our cluster galaxies. We want to define this edge in a way that is insensitive to properties of the Virgo galaxies, such as colour profiles, that may be affected by environment. Therefore, we define \rnorm{}, the edge of the optical disk for an unperturbed galaxy of a given stellar mass, by first fitting a piecewise function to the $\Sigma_* \! (R_{\rm edge} )$ versus $M_*$ relation from \citetalias{chamba2022} to find the critical \smsd{} value for a galaxy of a given stellar mass. This function is:
\begin{equation}
    \Sigma_*(R_{\rm edge}) = \begin{cases}
    0.6, & \log{(M_*)} < 9.6 ~\text{M}_{\odot}\\
    10^{\left( 0.3\log{(M_*)} - 3.1 \right)}, & \log{(M_*)} \geq 9.6 ~\text{M}_{\odot}
    \end{cases},
\label{eq:edge_density}
\end{equation}

\noindent where \smsd{} has units of $\text{M}_{\odot}~\text{pc}^{-2}$. Using inclination-corrected \smsd{} profiles, we can determine \rnorm{} by finding the point at which the profile drops below the threshold \smsd{} value. In Fig.~\ref{fig:VCC865_prof} we show \rnorm{} as a black line with a grey band for its uncertainty. Figure 9 of \citealt{chamba2022} shows the ratio between \redge{} to the effective radius as a function of stellar mass. The ratio is $\sim 4$ for all stellar masses but contains significant scatter, stemming from the tighter size-mass relation that is produced when using \redge{}.

\subsubsection{Quantifying uncertainties}

\label{sec:uncert}

The measurement of \rssfr{} is sensitive to the noise in the sSFR profile, the radial bin size, and the presence of small local maxima in the sSFR profile. Similarly, \rnorm{} is sensitive to the noise in the \smsd{} profile and the radial bin size. To quantify the uncertainties in these measurements, we perform a Monte Carlo simulation with 1000~iterations for each galaxy. During each iteration, we draw sSFR values for each radial bin from a Gaussian centred on the calculated values with a scale equal to the measured noise. We then run the algorithm for finding the turn-off point as described in Sect. \ref{sec:rssfr}, and draw the final value for \rssfr{} from a uniform distribution spanning the width of the radial bin. For each iteration, we also draw the \smsd{} values in each radial bin from a Gaussian centred on the measured value and with a width equal to the noise. The \smsd{} threshold value has an associated uncertainty from our piecewise fit to the \citetalias{chamba2022} relation, so we also draw that value from a Gaussian distribution centred at the measured value with a width equal to the uncertainty. We then can determine \rnorm{}, which we again draw from a uniform distribution equal to the width of the radial bin. Now with 1000~values for both \rssfr{} and \rnorm{}, we have 1000~values for the ratio between the two and choose the accepted value to be the median and use the $16^{\rm th}$ and $84^{\rm th}$ percentile values as the upper and lower limits on the error.

Fig.~\ref{fig:stamps} shows \ha{} and \rband{} imaging for eight galaxies spanning a range of stellar masses, with ellipses outlining \rssfr{} and \rnorm{}. The ellipticities and position angles used for these ellipses are the local values determined by \ap{} at those radial points. Further details of these example radial profiles are shown in Appendix \ref{app:moreplots}, where we show the equivalent of Fig. \ref{fig:VCC865_prof} for the seven additional galaxies in Fig. \ref{fig:stamps}. On the right-hand side of the postage stamps are the sSFR profiles for these galaxies, again highlighting \rssfr{} and \rnorm{}. On the sSFR plots, we have included the upper limits of sSFR (based on \ha{} detection limits) for the next five radial bins in which \ha{} is not detected (but stellar mass is). This helps to highlight cases where, even in the absence of an obvious local peak to define \rssfr{}, an edge is still clearly present. 

These figures display the variety of shapes that sSFR profiles exhibit due to local peaks of star formation or off-centre star formation, and show the success of our algorithm at detecting a physical \rssfr{} in each of these cases. This success is validated by the green ellipses on the continuum-subtracted \ha{} images which capture the \ha{} emission of the galaxies very well.

\subsection{Final sample of galaxies}

The full VESTIGE sample consists of 384 Virgo galaxies with \ha{} emission. We have removed a subset of galaxies from our sample based on the following criteria:

\begin{itemize}
    \item Three massive elliptical galaxies with \ha{} emission (M84, M86, M87)
    \item Six additional faint galaxies for which the \ap{} routine did not produce good surface brightness profiles
    \item 22 additional galaxies with less than three data points in their measured sSFR profile
    \item 18 additional galaxies for which \rssfr{} is clearly unphysical based on inspection of radial profiles (Appendix \ref{app:peculiar})

\end{itemize}

\noindent In total, we have removed 49 galaxies, leaving us with a sample size of 335 down to $M_* \simeq 10^7 \rm{M}_{\odot}$.

\section{Results}
\label{sec:results}

\subsection{Galaxy size as a function of stellar mass}

We first consider the size-stellar mass relation of our sample using \rnorm{} and \rssfr{} as our measurements of size, shown in Fig.~\ref{fig:size_mass}. This puts our \rnorm{} measurements in context with the results of \citetalias{chamba2022}. Fitting a linear relation to the log-log size-stellar mass relation, we find a slope of $0.33 \pm 0.01$ with a scatter of $0.16$. This is similar to the fits of \citetalias{chamba2022} who found slopes of $0.31 \pm 0.01$, $0.27 \pm 0.02$ and $0.32 \pm 0.03$ (with scatter of $ \sim 0.1$) when fitting their whole sample, giant spirals and dwarf SF galaxies, respectively. We find that our \rnorm{} measurements are systematically smaller than the \citetalias{chamba2022} \redge{} values by $\sim 10 \%$. In contrast, \citet{chamba2024} found that Fornax cluster galaxies are $\sim 50\%$ smaller than isolated galaxies, as \redge{} (defined in the same way as \citetalias{chamba2022}) occurs at a higher stellar mass surface density. This is expected in the case where cluster galaxies have had their SF disks truncated. This justifies our use of the field-derived critical mass density for determining \rnorm{}. 

The relationship between \rssfr{} and stellar mass does not follow a clear linear relation and has much more scatter than the size-mass relation with \rnorm{}. This is expected in the case where galaxies with \rssfr{} $<$ \rnorm{} are undergoing quenching mechanisms which may be mass dependent. 

\begin{figure}
\includegraphics[width=\columnwidth]{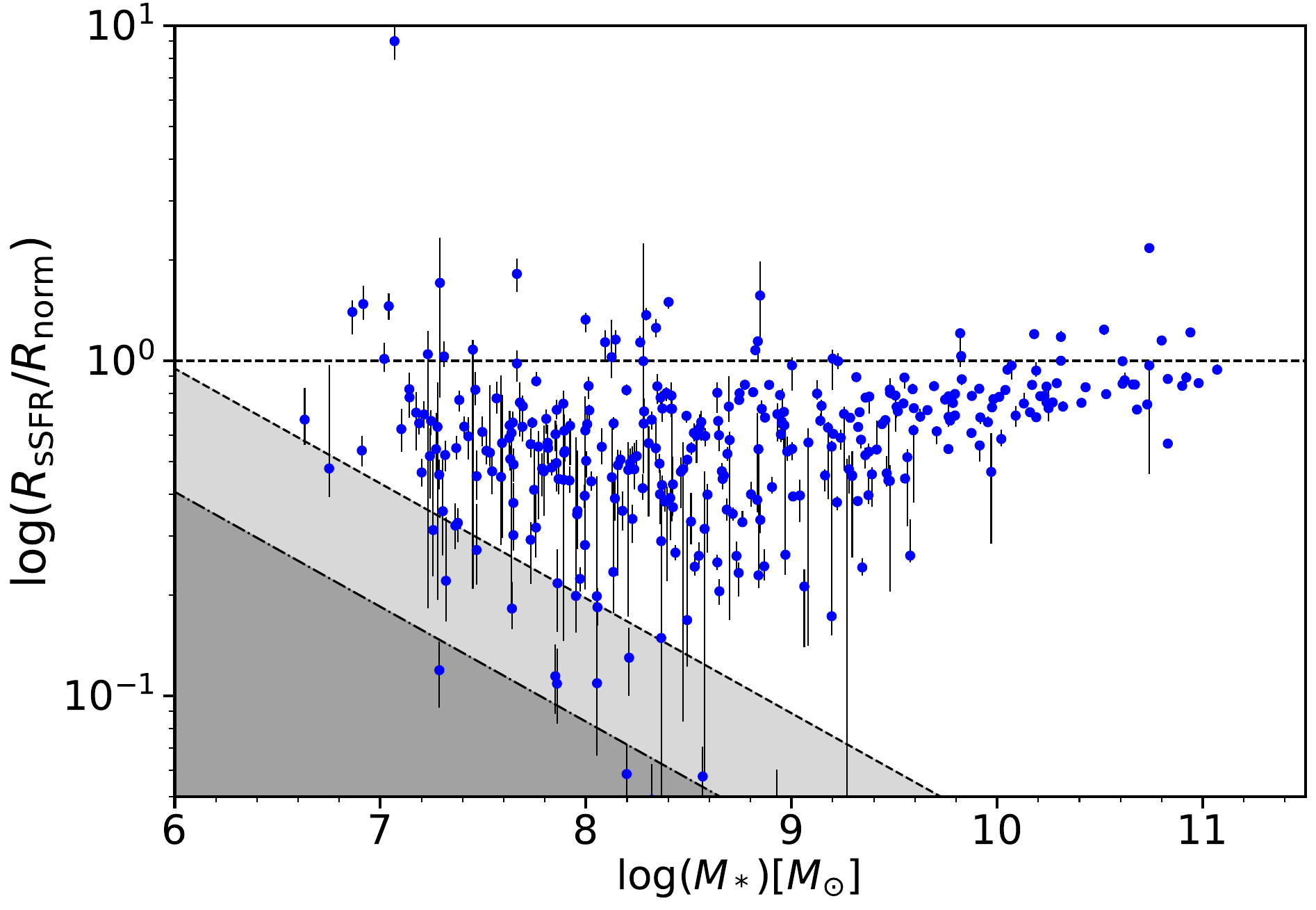}
\caption{\rssfr{}/\rnorm{} as a function of stellar mass. The two grey shaded regions show limits on the measurements of \rssfr{}/\rnorm{}. We have excluded any galaxies where the sSFR profile spans less than three data points, so the dashed black line has $R_{\rm sSFR} = 6.545''$. In principle, any sSFR profile with more than three data points could still have a truncation at the first data point, so the dot-dashed line has $R_{\rm sSFR} = 2.805''$. In both cases, $R_{\rm norm}$ is determined from the size mass relation in Fig.~\ref{fig:size_mass} at a distance of 16.5~Mpc.}
\label{fig:trunc_mass}
\end{figure}

We plot \rssfr{}/\rnorm{} against stellar mass in Fig.~\ref{fig:trunc_mass}. Galaxies of all masses tend to have \rssfr{}/\rnorm{} $< 1$. There is a trend whereby galaxies with lower stellar masses have smaller values of \rssfr{}/\rnorm{} on average, though there is much scatter. Shown in grey bands are the detection limits based on our methods. The first results from our removal of any galaxies whose sSFR profiles contained less than three data points (\rssfr{}$=6.545^{\prime\prime}$). The second detection limit is at the radius for the first data point in the profiles, \rssfr{}$=2.805^{\prime\prime}$, since any galaxy could have a measured edge at the first data point.

\subsection{\texorpdfstring{$R_{\rm sSFR}/R_{\rm norm}$ as a function of environmental indicators}{Rssfr/Rnorm as a function of environmental indicators}  }

\begin{figure*}
\centering
\resizebox{\hsize}{!}{\includegraphics{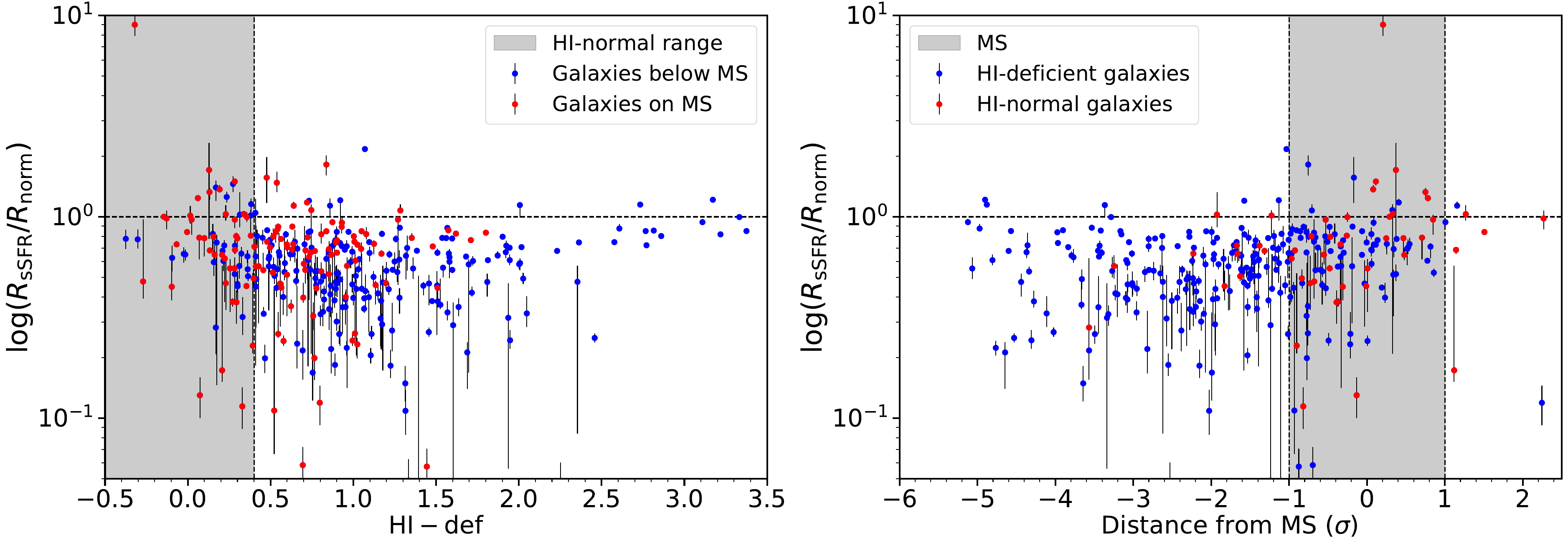}}
\caption{\textit{Left}: \rssfr{}/\rnorm{} as a function of H\textsc{i}-deficiency, with galaxies located on or above the SFMS shown in red and galaxies below the SFMS shown in blue. The range of \hi{}-normal galaxies is shaded. \textit{Right}: \rssfr{}/\rnorm{} as a function of distance from the SFMS, with \hi{}-normal galaxies shown in red and \hi{}-deficient galaxies shown in blue. The SFMS is shown as a shaded region.}
\label{fig:trunc_hidef_MS}
\end{figure*}

In Fig.~\ref{fig:trunc_hidef_MS} we plot \rssfr{}/\rnorm{} as a function of both H\textsc{i}-deficiency and distance from the SFMS \citep[determined in][]{boselli2023}. On the \hi{}-deficiency plot we indicate which galaxies are on/above the SFMS and which are below. Similarly on the SFMS plot we indicate which galaxies are \hi{}-normal and which are \hi{}-deficient. While most \hi{}-normal galaxies are on the SFMS, the converse is not true: many galaxies on the SFMS are \hi{}-deficient. We observe that most galaxies have \rssfr{}/\rnorm{} $<1$ regardless of \hi{} content or SFR. Galaxies that are more H\textsc{i}-deficient or that are below the SFMS have more substantially truncated disks.  Even for many galaxies on the SFMS, however, we find that \rssfr{}/\rnorm{} $<1$. We interpret the ubiquity of truncated disks across the range of \hi{}-deficiency and distance from the SFMS as evidence that nearly all Virgo galaxies are truncated relative to the field. For galaxies on the SFMS with normal gas content, the data points scatter around \rssfr{}/\rnorm{} $\sim 0.8$. While we expect that even the most normal galaxies in our sample may be somewhat affected by the Virgo environment, without a direct field comparison we cannot say for certain that galaxies with $0.8 < $ \rssfr{}/\rnorm{} $ < 1$ are truncated with respect to the field. Being mindful of this caveat, in further analysis we focus on galaxies with \rssfr{}/\rnorm{} $<0.8$ We will further discuss these results in context in Sect. \ref{sec:discussion}. 

A peculiar feature in Fig.~\ref{fig:trunc_hidef_MS} is the handful of galaxies that are extremely H\textsc{i}-deficient (H\textsc{i}-def  $> 2.5$) and well below the SFMS with only mildly truncated disks. Upon inspection, some of these galaxies appear to be starbursts and feature low-SB ionised gas emission stemming from a central burst of star formation. Our measure of disk truncation is not able to pick these galaxies out, however they do not really have SF disks anymore at all. We show an example in Appendix \ref{app:peculiar}.

\subsection{\texorpdfstring{\rssfr{}/\rnorm{} across the Virgo environment}{Rssfr/Rnorm across the Virgo environment}}

The Virgo cluster is made up of a main cluster (Cluster A), along with another large structure (Cluster B) and several smaller substructures outlined in Table \ref{tab:regions}. Throughout our discussion of the main cluster structure, we adopt the NFW model of \citet{simionescu2017} to describe the mass profile of the cluster. This fit infers $R_{200}=0.974~\text{Mpc}$.

\begin{figure}
\centering
\includegraphics[width=\columnwidth]{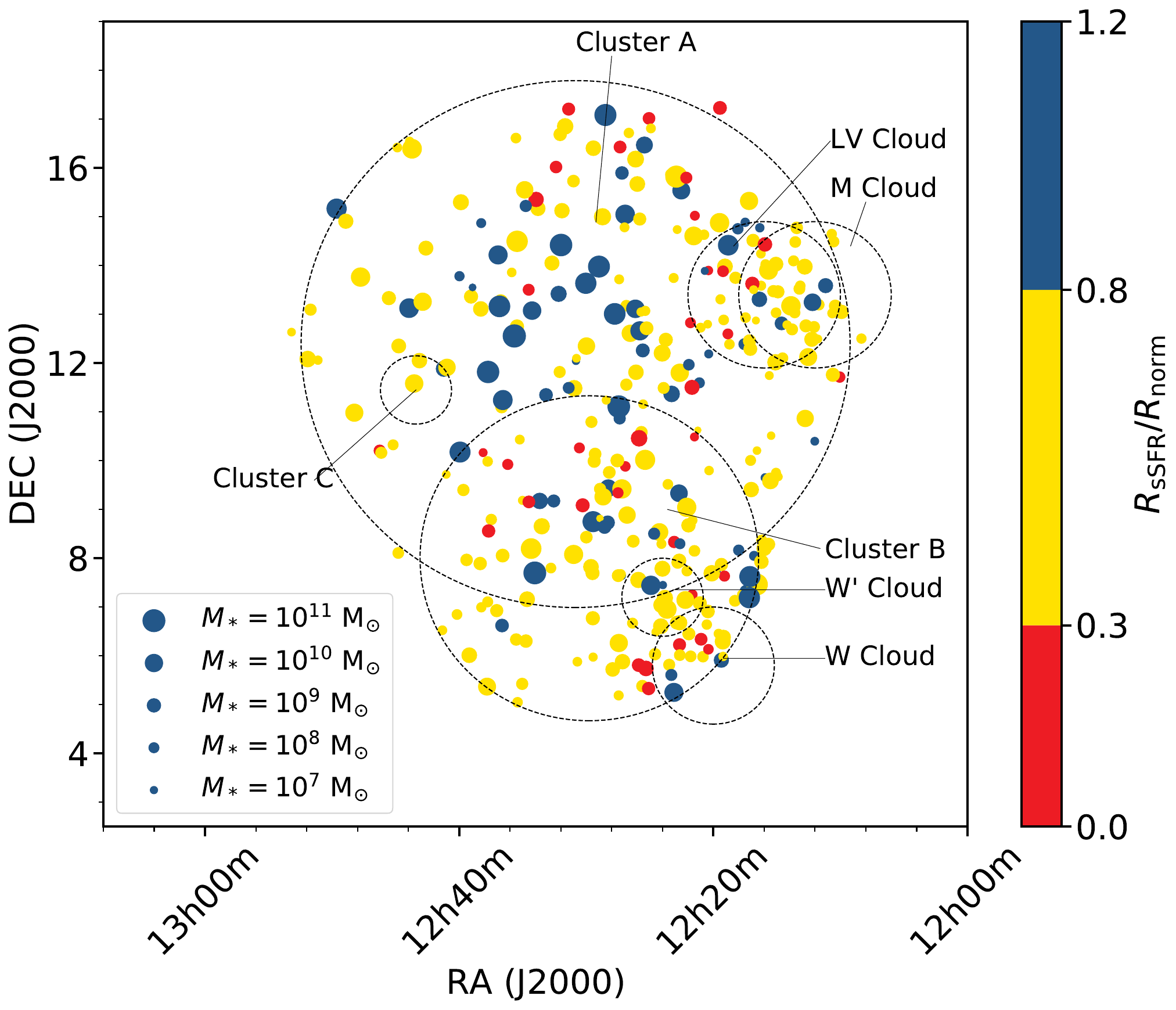}
\caption{Right ascension and declination footprint of VESTIGE galaxies with measured values of \rssfr{}/\rnorm{}. Overplotted are the seven identified regions associated with the Virgo cluster: Clusters A, B and C, the W and W$'$ clouds, the M cloud and the Low-Velocity cloud, with radii as noted in Table \ref{tab:regions}. Data points are sized proportionally to the stellar mass of the galaxy, and coloured based on the value of \rssfr{}/\rnorm{}.}
\label{fig:footprint}
\end{figure}

\begin{figure}
\centering
\includegraphics[width=0.8\columnwidth]{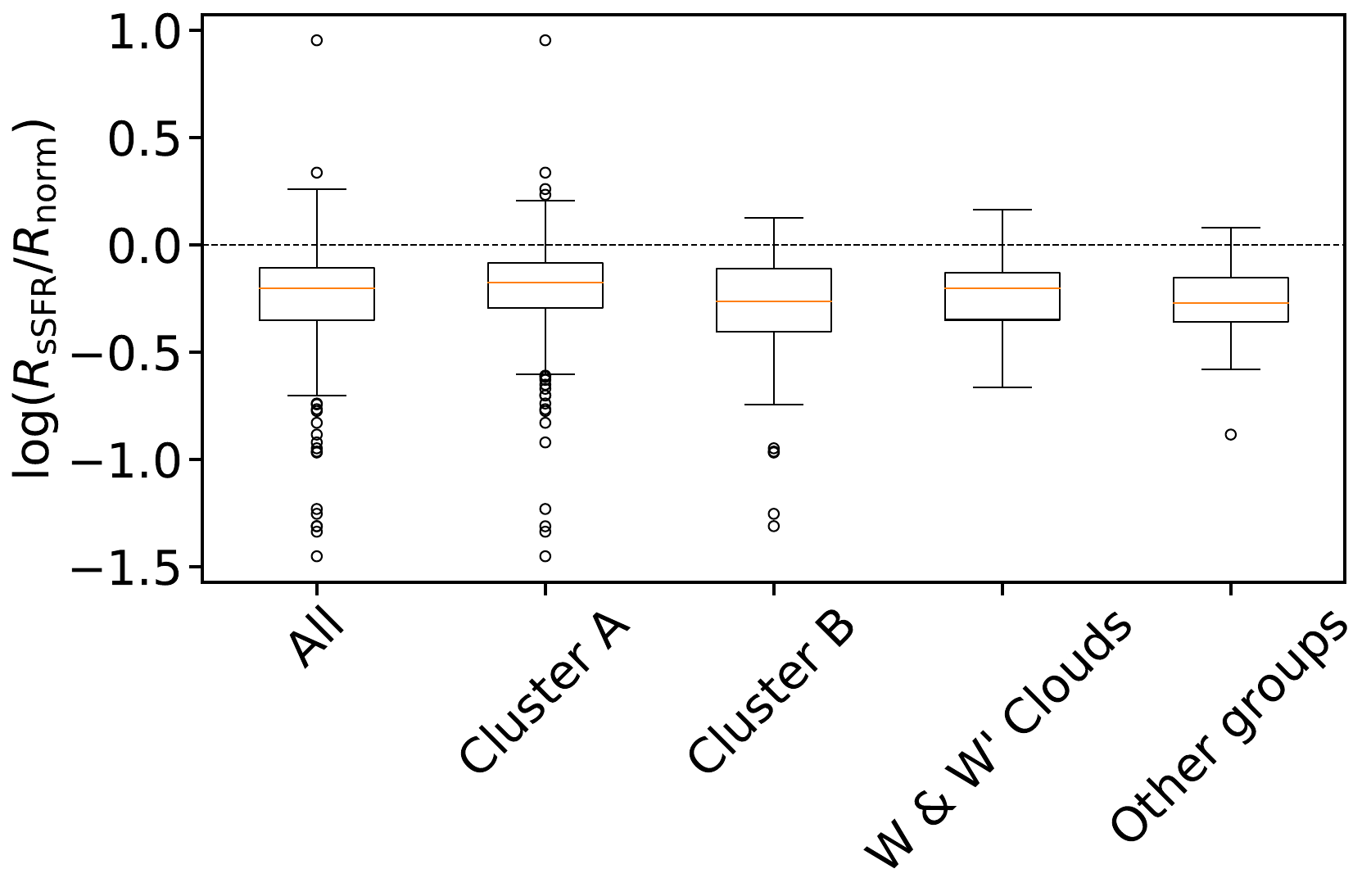}
\caption{Box-and-whisker plot showing the median and first and third quartiles of the distributions of \rssfr{}/\rnorm{} in the entire sample and four groupings of Virgo cluster substructures. `Other groups' covers the M and low-velocity clouds as well as Cluster C.}
\label{fig:box}
\end{figure}

The footprint of the Virgo cluster with its substructures is shown in Fig.~\ref{fig:footprint}. Each galaxy is shown at its 2D position in the cluster region, with the data points sized according to the stellar mass of the galaxy and coloured according to the value of \rssfr{}/\rnorm{} The radii of the circles depicting different substructures are those listed in Table \ref{tab:regions}. We use a discrete colour mapping on Fig.~\ref{fig:footprint} to separate galaxies into three categories based on their values of \rssfr{}/\rnorm{}. Blue data points have $R_{\rm sSFR}/R_{\rm norm} > 0.8$ and are considered at most mildly truncated. Since we have not used a control sample to directly determine the scatter in \rssfr{}/\rnorm{} for normal SF galaxies, we will not focus on these mildly truncated galaxies in our analysis.  Yellow data points ($0.3 < R_{\rm sSFR}/R_{\rm norm} \leq 0.8$) are considered moderately truncated and red points ($R_{\rm sSFR}/R_{\rm norm} \leq 0.3$) are considered extremely truncated. We can see from Fig.~\ref{fig:footprint} that there are groupings of galaxies with at least moderately truncated disks in the M and Low-Velocity clouds, as well as in and around the W and W$'$ cloud regions. If these galaxies are truly a part of these substructures, this result indicates the possible importance of pre-processing in group structures beginning to perturb galaxies before they fall into the main cluster. We assign galaxies to their most likely substructure as outlined in Sect. \ref{sec:subclusters}. In Fig.~\ref{fig:box} we show a box-and-whisker plot displaying the median and interquartile range of $R_{\rm sSFR}/R_{\rm norm}$ for five groupings of Virgo substructures: all regions, Cluster A alone, Cluster B alone, the W and W$'$ clouds together, and the M cloud, low-velocity cloud and Cluster C together. In each grouping, the median value of $R_{\rm sSFR}/R_{\rm norm}$ is less than unity, with the scatter varying across the groupings.

We also look at the distribution of galaxies in projected phase-space (PPS) in Fig.~\ref{fig:PPS_fracs}. To do so, we plot the radial velocities of galaxies (as collected in \citealt{boselli2023}) relative to the cluster against the 2D distance from the cluster centre (M87). While PPS diagrams often normalise the x-axis by $R_{200}$ and the y-axis by the velocity dispersion ($\sigma$), we use physical (unnormalised) coordinates to ensure our interpretation of the results is not sensitive to the assumed value of $R_{200}$ and $\sigma$. The data points are colour-coded according to $R_{\rm sSFR}/R_{\rm norm}$, as in Fig.~\ref{fig:footprint}. Evidence of truncations can be found throughout the cluster, including in the area beyond the virial radius of 0.974 Mpc. Interestingly, we do not find a strong trend between degree of truncation (based on the colours of the data points) and location in PPS. This contrasts sharply with the distribution of quenched galaxies relative to star forming galaxies, illustrated as the background greyscale shading.  This is created by comparing the 2D kernel density estimation (KDE) of the VESTIGE H$\alpha$ sample distribution to that of the full NGVS sample, for the mass range $M_* > 10^8~\text{M}_{\odot}$ where NGVS is mostly redshift complete.  

As expected, the VESTIGE detection fractions are lower in regions of phase space near the centre of the cluster and with lower velocities, corresponding to ancient infall regions where galaxies have interacted with the cluster environment likely on multiple passes through the centre. If these ancient infallers have quenched, we would not expect detectable \ha{} emission and thus they are absent from the VESTIGE sample. On the contrary, the VESTIGE detection fractions are highest in the infall regions of the cluster, at higher relative velocities and far from the centre.

\section{Discussion}

\label{sec:discussion}

\subsection{Modelling the radial profiles of galaxies}
\label{sec:model}

In this section we model the radial stellar and gaseous components of a normal SF galaxy (prior to any quenching) and explore how they change under two different simple quenching models. Starting with a chosen stellar mass, we use the scaling relations of \citet{huang2012} to obtain an H\textsc{i} mass and integrated SFR for our galaxy. With the relations from Sect. \ref{sec:methods}, we obtain the critical \smsd{} value and edge radius (\rnorm{}; \rssfr{} at $t=0$) of our galaxy. This allows us to determine the scale radius for an exponential \smsd{} profile. For simplicity, we assume that the SFR profile will have the same scale radius, though in practice the scale radii for the two components could differ. Following the sub-kpc Kennicutt-Schmidt law of \citet{bigiel2008}, the SFR profile of the model galaxy is proportional to the H\textsubscript{2} profile. The H\textsc{i} profile is flat at $8 ~\text{M}_{\odot}~\text{pc}^{-2}$ which is a good approximation based on the observations of \citet{bigiel2008}. 

Observations comparing gas density profiles and \ha{} profiles have identified that breaks in \ha{} profiles correspond to a critical gas density for star formation \citep[e.g.][]{kennicutt1989}, which can signify the edge of the SF disk. Additionally, through our analysis we have used \rnorm{} to define the expected edge of the SF disk for a normal, isolated galaxy. By incorporating the critical gas threshold for star formation, our SFR (and therefore H\textsubscript{2}) profiles drop off abruptly at \rnorm{}, where \rssfr{}$=$\rnorm{} at $t=0$ and the threshold gas density is the value of the total gas density at \rnorm{} at $t=0$. The stellar mass surface density profile also has a break at \rnorm{}. However, unlike \rssfr{}, the location of this break will not change as we evolve the model.

In our simple model, the break at the gas density threshold is abrupt, with no star formation occurring beyond this radius; in reality, galaxy disk edges are less drastic due (for example) to localised peaks of dense gas fuelling small amounts of star formation in the outer disk. This simplification does not change our results as our goal is to explore how \rssfr{} changes with time in different quenching scenarios. As gas is consumed and/or removed from the galaxy, the total gas density will decrease, causing the position of \rssfr{} to change.

\subsection{Ram-pressure stripping}

\label{sec:rps}

The effects of ram-pressure stripping (RPS) are known to be present in the Virgo cluster based on studies of individual or small samples of galaxies that display evidence of stripped H\textsc{i} gas through direct observations of the H\textsc{i} (e.g. \citealt{gavazzi2001, kenney2004, chung2007, sorgho2017, minchin2019, boselli2023_vic}) or observations of radio emission (e.g. \citealt{vollmer2003, vollmer2007, vollmer2010, vollmer2013, crowl2005, chyzy2007, kantharia2008}) or ionised gas (e.g. \citealt{yoshida2002, boselli2016, boselli2018_4424, fossati2018, boselli2021}) in the stripped material. A multitude of simulation studies have shown that RPS can deplete the gas reservoir of a galaxy on timescales of $<1~\rm{Gyr}$ (e.g. \citep{vollmer2001, boselli2006, steinhauser2016, boselli2016, fossati2018, junais2022}. However, substantial removal of star-forming gas may not occur until the galaxy approaches first pericentre as the combination of a dense ICM and high relative speeds are necessary to remove gas from a deep galactic potential well. Other high-resolution simulations have shown that while removal of the star-forming gas is rapid, removal of the entire gas reservoir may require timescales of $>1~\rm{Gyr}$ \citep[e.g.][]{tonnesen2007}. 

Additionally, the multi-phase nature of gas plays a role in the efficiency of stripping, with denser gas being harder to strip \citep[e.g.][]{tonnesen2009}. Our simple model below neglects any direct stripping of the molecular gas and assumes that H$_{2}$ depletion occurs only through star formation. Once the H\textsc{i} gas is entirely removed after a stripping event, the H$_{2}$ gas cannot be replenished. In fact, stripping of molecular gas has been shown to occur in Virgo \citep{watts2023}. Factoring this into our model would only serve to speed up the RPS timescales, which will not affect the interpretation of this model since we will note that the RPS timescales are much faster than starvation-only models.

To explore the range of effectiveness of RPS throughout the cluster, we invoke a simple model to determine where the combined ICM density and galaxy velocity becomes large enough to overcome the gravitational restoring force holding the gas in the galactic disk. Using a $M_*=10^{9.5}~\rm{M_{\odot}}$ model galaxy, we can calculate the restoring force from each component. While previous works have considered only the stellar disk component of the restoring force acting to hold the \hi{} gas (e.g. \citealt{yoon2017, boselli2022}), we loosely follow \citet{roberts2019} and consider the force from the stellar disk, dark matter halo, and the self-gravity of the gas itself.\footnote{\citet{roberts2019} also factored in a bulge component of the stellar mass based on bulge-disk decomposition, but we model the stellar component of our galaxies as a disk only.} Therefore, the restoring force is given as:
\begin{equation}
    f = \left[g_*(\rm r) + g_{\rm H\textsc{i}}(r) + g_{\rm H\textsubscript{2}}(r) + g_{\rm DM}(r)\right]\,\Sigma_{\rm H\textsc{i}}(r),
\label{eq:restore}
\end{equation}

\noindent where $g_*(\rm r)$, $g_{\rm H\textsc{i}}(r)$, $g_{\rm H\textsubscript{2}}(r)$ and $g_{\rm DM}(r)$ are the contributions to the gravitational acceleration from the different galactic components (stars, H\textsc{i}, H\textsubscript{2} and dark matter) and $\Sigma_{\rm H\textsc{i}}(r)$ is the surface density of H\textsc{i} gas. We leave further details of the model to Appendix \ref{app:model}.

\begin{figure}
\includegraphics[width=\columnwidth]{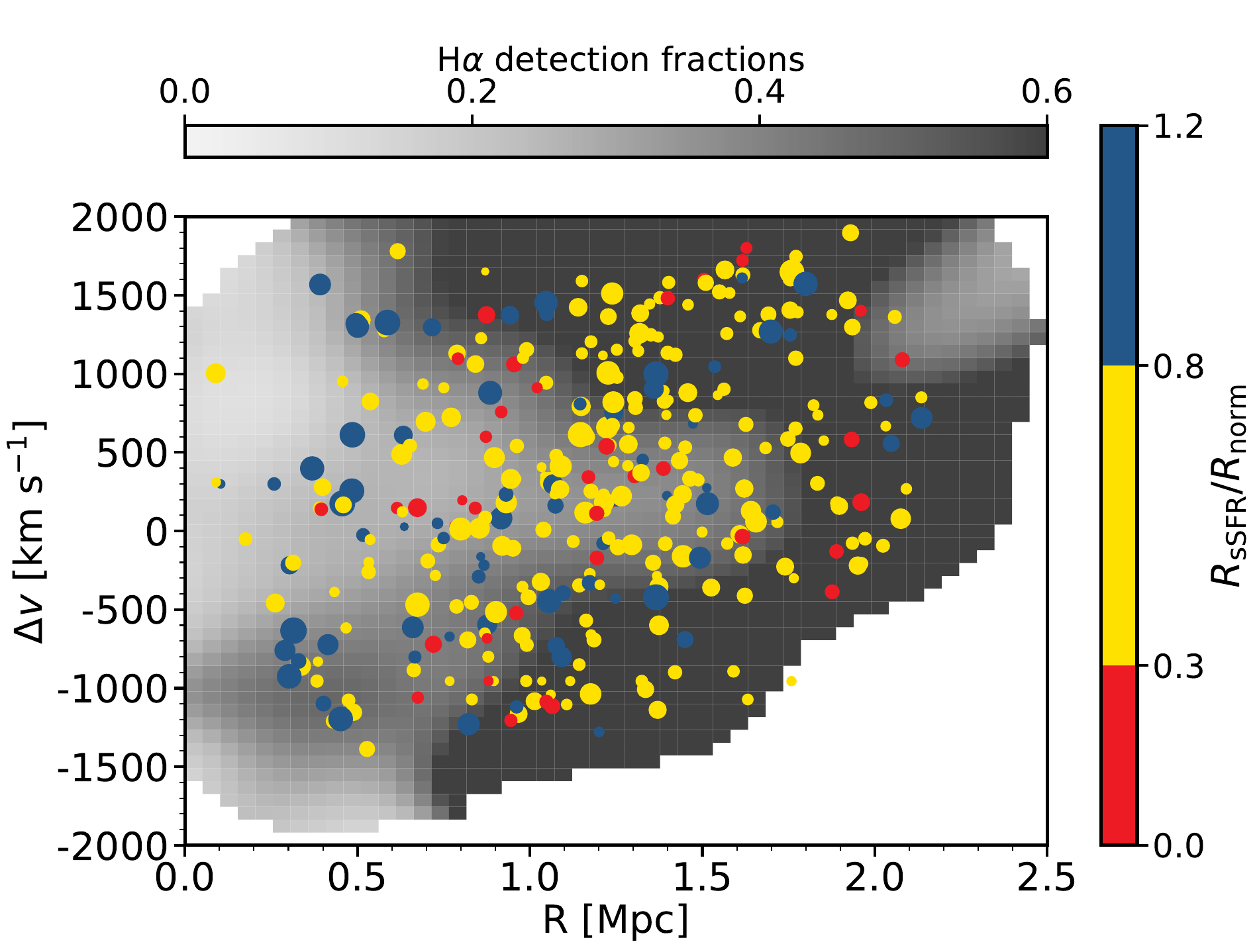}
\caption{PPS diagram of VESTIGE galaxies with VESTIGE \ha{} detection fractions computed using the ratio of 2D Gaussian KDEs of the VESTIGE and NGVS distributions. Data points are sized proportionally to the stellar mass of the galaxy, and coloured based on the value of \rssfr{}/\rnorm{}. While the detection fractions show the expected structure, truncated disks exist across the cluster environment.}
\label{fig:PPS_fracs}
\end{figure}

\begin{figure}
\includegraphics[width=\columnwidth]{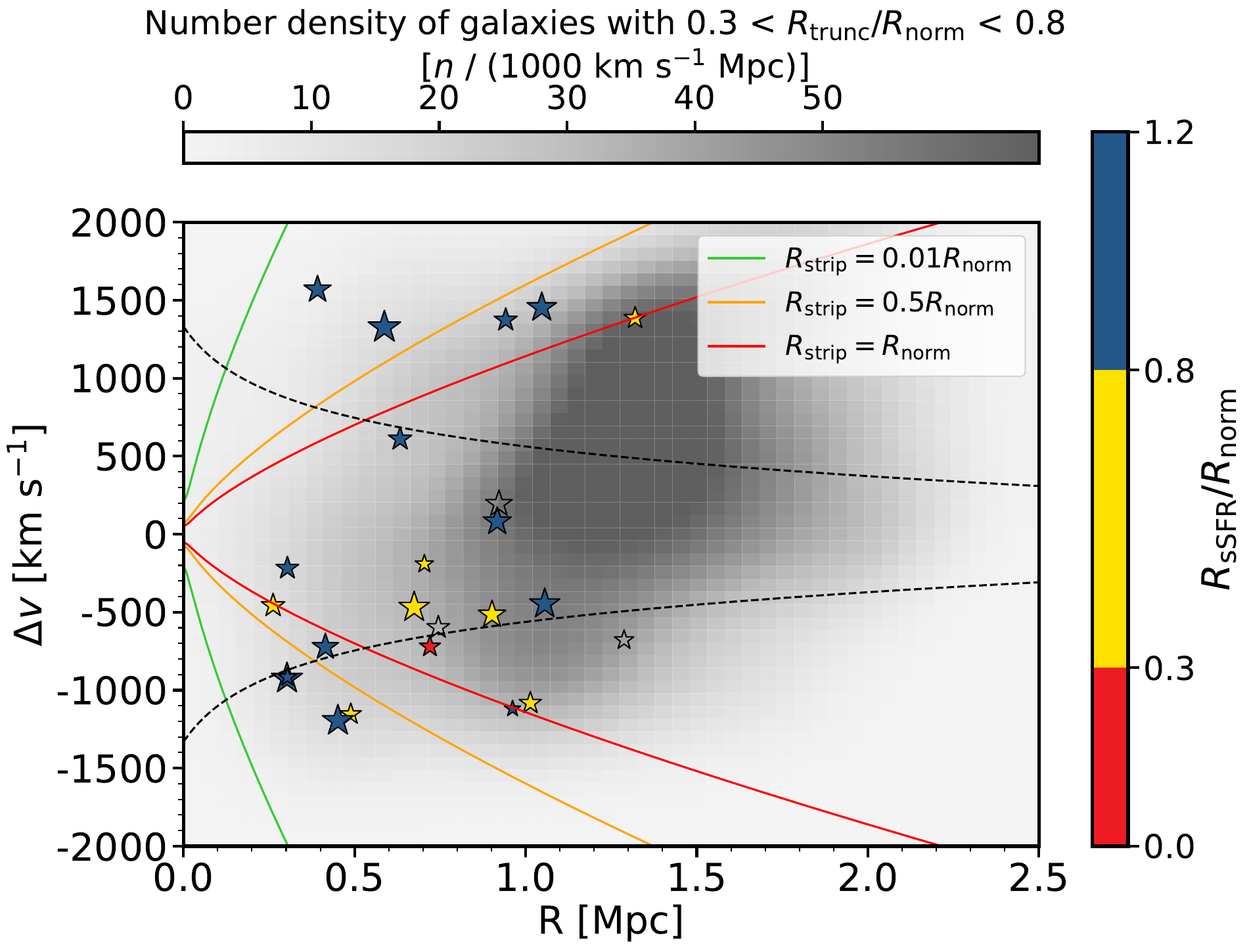}
\caption{PPS diagram showing the region within the escape velocity of the cluster as the region between the two dashed black lines. The escape velocity is determined based on the NFW profile of \citet{simionescu2017}. The green, yellow, and red lines show the contours where a $M_*=10^{9.5}~\rm{M_{\odot}}$ galaxy could have its gas removed by RPS from 0.01, 0.5, and 1\rnorm{}, respectively, assuming isotropic orbits. Data points are sized proportionally to the stellar mass of the galaxy, and coloured based on the value of \rssfr{}/\rnorm{}. The background greyscale shows the phase-space number density of galaxies with $0.3 < $\rssfr{}/\rnorm{} $<0.7$ determined using a 2D Gaussian KDE. }
\label{fig:PPS_RPS_cands}
\end{figure}

Ram-pressure is determined by $\rho_{\rm ICM} v^2 $, where $\rho_{\rm ICM}$ is the density of the ICM and $v$ is the relative velocity of the galaxy within the cluster. We can then calculate the strength of the ram-pressure at any point in phase space. We assume that a galaxy with some observed radial velocity, $v_{\rm r}$, has a three-dimensional velocity $v = \sqrt{3}v_{\rm r}$, since the ram-pressure depends on this three-dimensional velocity through the ICM. While this factor is statistically sound for a population of galaxies on isotropic orbits, the orbits of individual galaxies may vary.  A more realistic analysis of the ram-pressure experienced by infalling Virgo galaxies was reviewed by \citet{boselli2022}, using a simulation to characterise accurate orbital parameters.

With this model, we can calculate whether the ram-pressure at a given 2D point in projected phase space is sufficient to overcome the restoring force of the galaxy at a given galactocentric radius. We show a second PPS diagram in Fig.~\ref{fig:PPS_RPS_cands}, where the x- and y-axes are the same as in Fig.~\ref{fig:PPS_fracs}. The two dashed black curves on the plot indicate the escape velocity of the cluster based on the potential from the \citet{simionescu2017} NFW profile. Also plotted are, for a $M_*=10^{9.5}~\rm{M_{\odot}}$ galaxy modelled as in section \ref{sec:model}, the lines in phase space where the ram-pressure from the ICM can strip gas from the optical edge (\rnorm{}) of the galaxy (red), from 0.5\rnorm{} (amber) and from the centre (0.01\rnorm{}; green). The data points on Fig.~\ref{fig:PPS_RPS_cands} are galaxies previously identified to be undergoing RPS (\citealt{boselli2022} and references therein). These galaxies have a wide range of \hi{}-deficiencies and are located across the cluster environment; however, the most perturbed galaxies are preferentially located closer to the cluster centre, and at high relative velocities, near the contours where we expect a significant fraction of gas to be stripped from galaxies on typical orbits. There are also several 
galaxies undergoing RPS that are located beyond the red contours, where our model predicts ram pressure is typically not effective. This is unsurprising, as the strength of RPS on individual objects depends on the shapes of orbits and the exact radial profiles of the galactic components, which can differ significantly from our statistical model assumptions. 

The background greyscale of Fig.~\ref{fig:PPS_RPS_cands} shows the phase space number density of galaxies with moderate truncations ($0.3<$\rssfr{}/\rnorm{} $<0.8$), determined using a 2D Gaussian KDE. A dense region of truncated disks is located just outside the virial radius (between 1 and 1.5 Mpc) and with relative velocities between 0 and 1500 km s$^{-1}$. Statistically, many of these galaxies are likely to be on their first infall based on their location in phase space \citep{rhee2017}. This region of phase space can also contain backsplash galaxies. However, if the galaxies in our sample had already passed through the densest region of the ICM in the cluster centre they likely would have been completely stripped due to RPS. In fact, we find the quenched fractions to be lower in these regions (Fig.~\ref{fig:PPS_fracs}), again supporting that many of these galaxies are likely on their first infall into the cluster. While RPS may still be effective for some galaxies in this region of PPS, depending on their precise orbital geometries and restoring forces, the ubiquity of truncated disks in these regions suggests other effects may also be at play.

\subsection{Starvation}
\label{sec:starve}

In a simplified steady-state scenario, a galaxy forms stars from molecular gas which is replenished by atomic gas, which is in turn replenished by hot gas that has accreted from the cosmic web onto the galactic halo. If accretion of gas ceases \citep[e.g.][]{larson1980} or halo gas is removed via stripping (`harassment': \citealt{moore1996, moore1998}) when the galaxy becomes a satellite of a more dominant system, the galaxy will eventually exhaust its reservoir of both atomic and molecular gas through star formation and become quenched. As such, starvation is typically thought of as a process which uniformly lowers the gas density and suppresses star formation across the galactic disk, thus producing an anaemic, but not necessarily truncated, disk (e.g. \citealt{boselli2006}). However, starvation models typically do not account for observations that suggest that there is a critical total gas density threshold required to fuel local star formation. We show below that a simple starvation model that factors in a gas density threshold for star formation will, in time, produce a galaxy with a truncated disk. 

We demonstrate this with a simple toy model: At $t<0$ (where $t=0$ is the time at which starvation begins) a galaxy is in a `steady state' of gas accretion where the radial gas density profiles of atomic and molecular gas remain constant over time due to accretion of hot halo gas; this in turn fuels a constant \sfrd{} profile. The radial profiles for the galactic components are determined as outlined in Sect. \ref{sec:model}. 
We then allow our model to evolve from $t=0$. At this point, we assume the galaxy is completely unable to accrete fresh gas, and thus the H\textsc{i} reservoir cannot be replenished as it converts into H\textsubscript{2}. The model assumes that as long as there is H\textsc{i} in the galaxy then the H\textsubscript{2} profile (and thus the SFR profile) will remain constant\footnote{This prescription is likely an oversimplification as H\textsc{i}-deficient galaxies have been shown to have reduced star formation efficiency; e.g. \citealt{villanueva2022, brown2023}.}. At each timestep, stars form from the H\textsubscript{2} gas and the H\textsubscript{2} is replenished by the H\textsc{i}. Once the H\textsc{i} has been used up, subsequent star formation will deplete the remaining molecular gas, which will in turn reduce the SFR, moving the galaxy off of the main sequence. 

\begin{figure}
\centering
\includegraphics[width=\columnwidth]{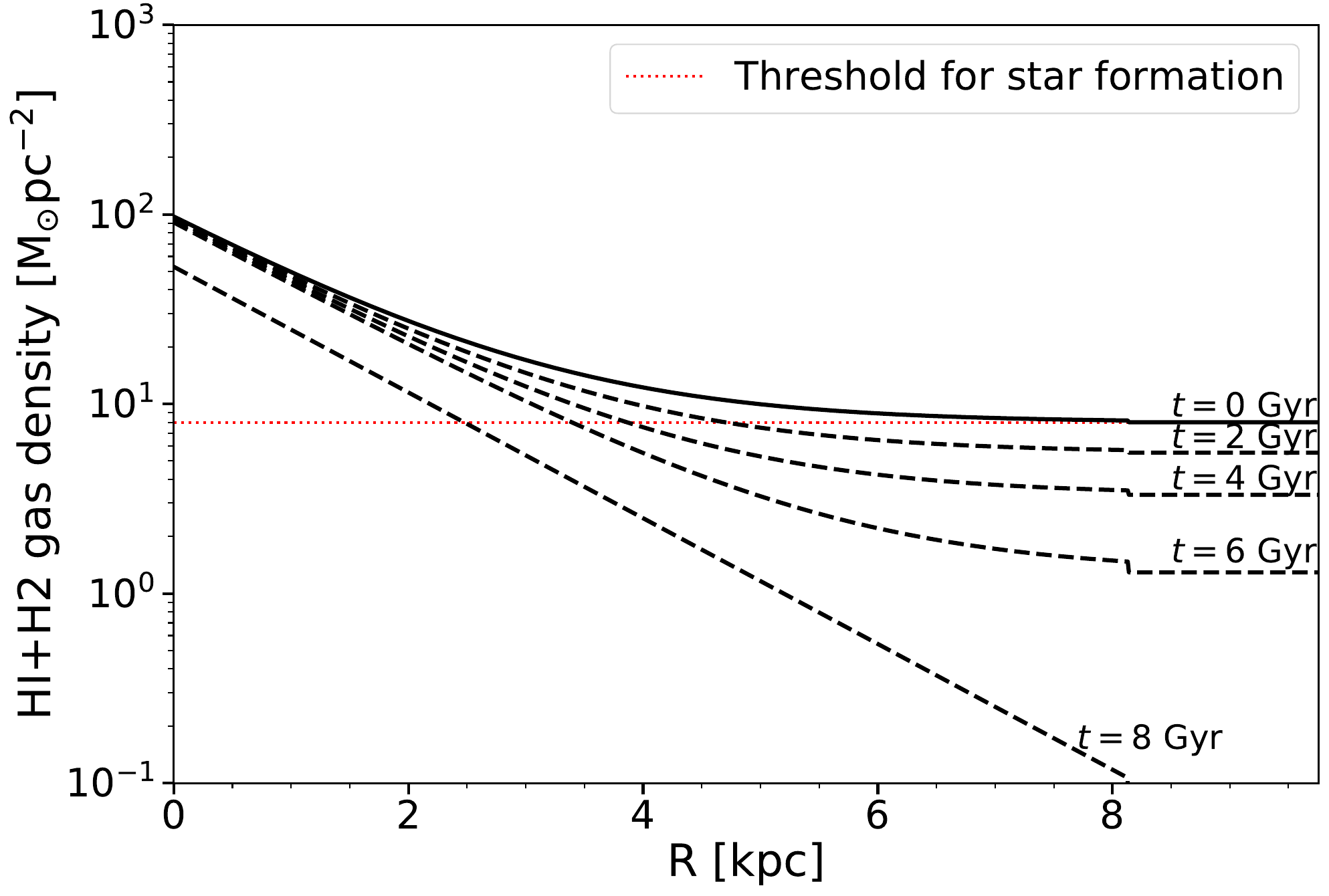}
\caption{Evolution of total gas density ($\Sigma_{\rm HI} + \Sigma_{\rm H2}$) over 8 Gyr for a $M_*=10^{9.5}~\rm{M_{\odot}}$ galaxy undergoing simple starvation. The threshold gas density for star formation (explained in Appendix \ref{app:model})} is shown as a red dashed line. By the $t=8~\text{Gyr}$ curve, all \hi{} gas has been consumed and as such the H$_2$ is not being replenished, causing the total gas density to decrease at all radii.
\label{fig:tot_gas}
\end{figure}

In Fig.~\ref{fig:tot_gas} we show the evolution of the total gas density profile in this model, for a $M_*=10^{9.5}~\rm{M_{\odot}}$ galaxy over 8~Gyr of starvation. Also plotted is the threshold gas density for star formation. Therefore, the points where the red line crosses each black curve are the points where the galaxy would be truncated at each timestep. In this model, while starvation takes a long time to move a galaxy off the SFMS and completely deplete its gas reservoir (as shown in Appendix \ref{app:model}), moderate truncation can still happen after just 1-2~Gyr. While this model is based on oversimplified assumptions about the gas distribution, consumption, and redistribution, as a toy model it effectively illustrates that any model of starvation that globally reduces the total gas density profile will also truncate the star-forming disk, if a critical gas threshold for star formation is included. 

Quantitatively, these results will depend on the treatment of starvation in the model, and also on the stellar mass of the galaxy, but qualitatively the behaviour would remain the same. In future work a gas threshold for star formation should be incorporated into a more sophisticated model, like the chemo-spectophotometric model of \citet{boselli2006}. Such a model should then be tested in order to see if it can also re-produce observed colour and metallicity gradients in galaxies.

\begin{figure}
\includegraphics[width=\columnwidth]{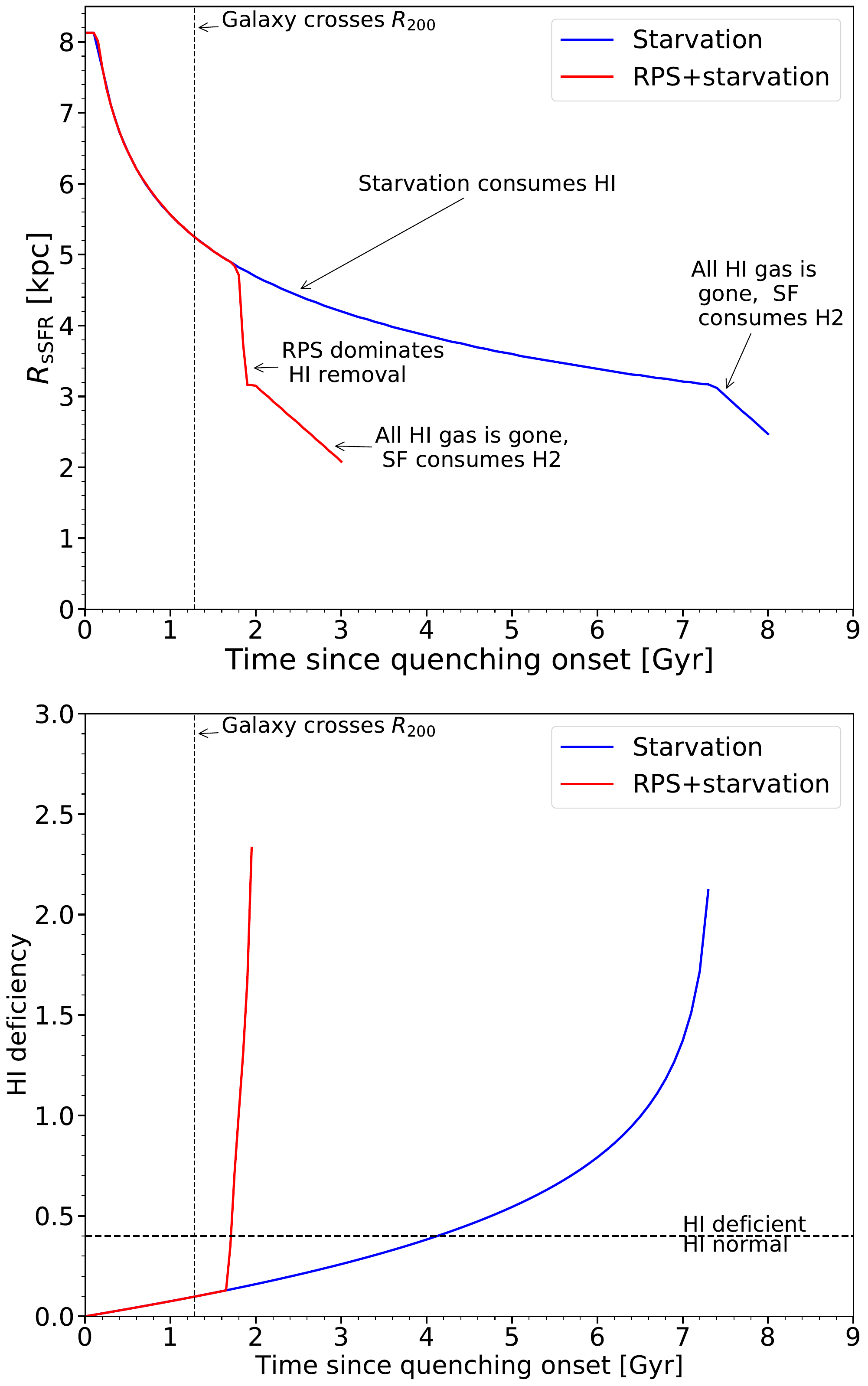}
\caption{Evolution through time of \rssfr{} (\textit{top}) and H\textsc{i}-deficiency (\textit{bottom}), for a $M_*=10^{9.5}~\rm{M_{\odot}}$ galaxy undergoing a simple starvation model (blue curves) and a galaxy falling into Virgo radially and experiencing RPS along with starvation (red curves). The dashed black horizontal line at H\textsc{i}-def=0.4 represents the point above which a galaxy is considered H\textsc{i}-deficient.}
\label{fig:RPS_starve}
\end{figure}

\subsection{Ram-pressure stripping and starvation in context}

In Fig.~\ref{fig:RPS_starve} we show a representation of the effects on \rssfr{} and H\textsc{i} content of our starvation model compared to our RPS+starvation model for a galaxy falling radially into the cluster. Previous studies have indicated that starvation and removal of hot halo gas can be efficient as far out as $5R_{200}$ in clusters (e.g. \citealt{bahe2013}). While we have not directly modelled the halo gas removal here, we assume for simplicity that starvation begins at $2R_{200}$, at which point the galaxy can no longer accrete fresh gas from the cosmic web. This choice of starting point is not meant to predict exactly where or how starvation begins for individual galaxies, but rather to show how starvation onset beyond $R_{200}$ produces galaxies with truncated disks before ram-pressure is sufficient to strip gas directly from the galactic disk. 

For a $M_*=10^{9.5}~\text{M}_{\odot}$ galaxy in our model, it takes $\sim 1.3~\text{Gyr}$ for the galaxy to fall radially from $2R_{200}$ to $R_{200}$, during which time the size of the star-forming disk shrinks significantly, by a factor of $\sim 2$. In the absence of any ram pressure stripping or other effects, the disk would continue to shrink gradually for several Gyr, as shown by the blue line.  However, for this galaxy on a radial orbit our model predicts ram-pressure stripping begins at $t \sim 1.8~\text{Gyr}$, when the galaxy is at $\sim 0.5R_{200}$, and is the dominant quenching force until $t \sim 1.95~\text{Gyr}$. This rapid timescale ($\sim 150~\text{Myr}$) for RPS is broadly consistent with the findings of other recent studies (e.g. \citealt{boselli2006, boselli2021, fossati2018}). At this point, when $R_{\rm trunc}=3.1~\text{kpc}$, the H\textsubscript{2} gas becomes dominant in the gas profile, and shortly after the remaining H\textsc{i} gas from the galactic centre is removed by RPS. Once the galaxy is completely H\textsc{i}-deficient, the consumption of H\textsubscript{2} takes over and slowly truncates the disk further, moving it off the main sequence as the SFR decreases with decreasing H$_{2}$ content\footnote{We have not considered the effects of RPS on the H$_{2}$ in our model; if RPS also removed H$_{2}$ then the quenching of the galaxy in this final stage would be even more rapid.}. 

This shows that there is the potential for a galaxy to become completely stripped of \hi{} by its first pericentre approach in the cluster. The removal of \hi{} by RPS is most pronounced near the cluster centre, however, which makes our starvation model a natural choice to explain galaxies near and beyond $R_{200}$ that appear to be on first infall with already-truncated disks.

Our toy model indicates that, while starvation has a long timescale to completely quench a galaxy, it may contribute to considerable disk truncation in only 1-2~Gyr. This means a galaxy can already have a truncated disk before RPS becomes dominant, especially if starvation begins early, when the galaxy is well beyond the virial radius. This implies a `slow-then-rapid' quenching sequence, whereby galaxies begin slow quenching via starvation and, at a critical point during first infall, RPS becomes dominant and rapidly quenches the galaxy. These findings are consistent with those of \citet{roberts2019}, who found that the quenched fraction of galaxies in nearby clusters increases sharply at a threshold ICM density corresponding to the point where galaxies are susceptible to having the majority of their gas removed by RPS. 

We argue that this `slow-then-rapid' quenching scenario is also consistent with the `delayed-then-rapid' scenario of \citet{wetzel2012}. In our case, the early starvation quenching mode is the `delayed' phase because it does not rapidly quench star formation in the galaxy. The global SFR and \hi{} content of the galaxy do decrease during the time that starvation is dominant ($\sim 1.8~\text{Gyr}$), but slowly enough that the integrated SFR drops by only $\sim 10\%$ ($\sim 0.07 ~\text{M}_{\odot}~\text{yr}^{-1}$ for the modelled galaxy in Fig.~\ref{fig:RPS_starve}). This means the galaxy would remain on the SFMS during the starvation phase, according to the main sequence fit defined in \citet{boselli2023}. 

We find this to be an attractive model to explain the dense region of moderately truncated disks beyond the virial radius of the main cluster as seen in Fig.~\ref{fig:PPS_RPS_cands}. The galaxies in this region have a range of \hi{}-deficiencies and SFRs. The precise effects of starvation on the global gas content and SFR will depend on the details of the starvation mechanism and the time since starvation onset. Galaxies of different stellar masses will experience qualitatively the same effects, though the timescales may vary due to differences in gas content and SFR. While reproducing the parameters of individual galaxies is beyond the scope of this work, the peculiar trends seen in Fig. \ref{fig:trunc_hidef_MS} can be qualitatively explained with this model. In Fig. \ref{fig:trunc_hidef_MS}, galaxies with moderate truncations can be seen with only mild \hi{}-deficiencies and on the SFMS. Additionally, while SFMS galaxies are scattered across the range of \hi{}-deficiencies, most \hi{}-normal galaxies are on the SFMS. A slower process such as starvation truncating a disk, removing \hi{} gas and then moving a galaxy of the main sequence successively over long timescales explains these trends since there is a high chance of observing a given galaxy in one of these three stages. Starvation before $R_{200}$ could occur when a galaxy is a part of a smaller substructure, providing an explanation for the large number of truncated disks seen in the cloud structures of Virgo (Fig.~\ref{fig:footprint}). Other pre-processing mechanisms such as gravitational effects and even RPS (shown to be present in some groups and smaller clusters: \citealt{roberts2021_group, serra2023}) could occur in these regions as well, and should be explored further with dedicated studies of pre-processing in Virgo substructures.

\section{Summary and Conclusions}
\label{sec:conc}

We have used spatially resolved \ha{} imaging from VESTIGE, coupled with optical data from NGVS to measure the edge of star-forming disks of galaxies across the entire footprint of the Virgo cluster, for $M_*>10^{7}~\rm{M_{\odot}}$. To quantify the expected sizes of disks for isolated counterparts (\rnorm{}), we have developed methodology inspired by the physically motivated definition for the edge of a galactic disk outlined by \citet{chamba2022}. We have used \textsc{autoprof} to measure surface brightness profiles in \ha{}, as well as the $r$ and \gband{}s, allowing us to construct \smsd{} profiles and sSFR profiles. From our sSFR profiles, we have identified the turn-off point (\rssfr{}) based on the derivative of the profiles. We summarise our key results below:

\begin{itemize}

\item Disk truncations, where \rssfr{}/\rnorm{} $<1$, appear ubiquitous across the Virgo cluster and as a function of stellar mass, \hi{}-deficiency and distance from the SFMS. However, we are cautious to assume that galaxies with minor disk truncations (\rssfr{}/\rnorm{} $>0.8$) are truly truncated due to a lack of control sample and the presence of star-forming, gas-normal galaxies with minor truncations (Sect. \ref{sec:ratio}). \\

\item We find that moderate disk truncations are more common at lower stellar mass, greater H\textsc{i}-deficiency, and for galaxies that have fallen off the SFMS. However, these trends are not strong and contain considerable scatter (Figs. \ref{fig:trunc_mass}, \ref{fig:trunc_hidef_MS}).  \citealt{chamba2024} measured the edges of galaxies using a method analogous to \citealt{chamba2022} and found that galaxies in the Fornax cluster are systematically smaller than field counterparts, with a similar trend with stellar mass to our findings.  \\

\item Disk truncations occur to varying degrees across the entire cluster footprint, including in areas outside of the main cluster where galaxies may be infalling in groups (Fig.~\ref{fig:footprint}). The ubiquity of truncated disks across the cluster environment was also noted for the Fornax cluster by \citealt{chamba2024}. We note as well that our sample consists only of galaxies with detectable \ha{} emission, most of which are therefore star-forming systems. This means that galaxies that are entirely quenched and would therefore have no measurable \rssfr{} are not shown in our results. A comparison with galaxy densities from NGVS shows that indeed, the quenched fraction increases close to the cluster centre, so the truncated disks we measure are not representative of all galaxies in these regions (Fig.~\ref{fig:PPS_fracs}). This is consistent with results from \citet{boselli2014_guvics}. \\

\item The high number of galaxies with moderately truncated disks in regions with low quenched fractions and low ICM density implies additional drivers giving a head-start to quenching before RPS becomes the dominant gas removal mechanism. \\

\item Galaxies on first infall may experience starvation before RPS, whether only in the main cluster or in smaller structures prior to cluster infall. We have shown that a starvation model can produce disk truncations when including a threshold gas density for star formation. Starvation beginning early, followed by a dominant RPS phase suggests a `slow-then-rapid' quenching sequence, consistent with the findings of \citet{roberts2019}. The nature of this quenching sequence is also broadly consistent with the `delayed-then-rapid' mechanism described in \citet{wetzel2012}, since the early starvation phase only mildly affects the integrated SFR of the galaxy before the RPS phase rapidly quenches it. Our results show that galaxies in the W and W$'$ Cloud regions have truncated disks, and a further study of group pre-processing prior to infall into the main Virgo cluster is critical to understanding the full evolutionary sequence of galaxies in these dense environments.\\

\end{itemize}

\begin{acknowledgements}
We thank the anonymous referee for their constructive feedback that has helped us to greatly improve this manuscript. \\

We are grateful to the whole CFHT team who assisted us in the preparation and in the execution of the observations and in the calibration and data reduction: 
Todd Burdullis, Daniel Devost, Bill Mahoney, Nadine Manset, Andreea Petric, Simon Prunet, Kanoa Withington. \\ 

This research was supported by an NSERC  Discovery grant to MLB. MB acknowledges support by the ANID BASAL project FB210003. J is grateful to the Polish National Science Centre grant UMO-2018/30/E/ST9/00082 and the Polish National Agency for Academic Exchange POLONIUM grant BPN/BFR/2022/1/00005.
\\ 

The preparation of this manuscript and much of the research behind it took place at the University of Waterloo, on the traditional territory of the Neutral, Anishinaabeg and Haudenosaunee peoples. The campus is located on the Haldimand Tract, the land granted to the Six Nations that includes six miles on each side of the Grand River. The authors from the University of Waterloo community acknowledge the privilege to work on these traditional lands. We also acknowledge that the data for this research was collected at CFHT on the summit of Mauna Kea, a location with great cultural significance to the Native Hawaiian community.  \\

\indent \textit{Software:} \textsc{astropy}  \citep{astropy, astropy2, astropy3}, \textsc{autoprof}  \citep{stone2021}, \textsc{ipython} \citep{ipython}, \textsc{matplotlib} \citep{matplotlib}, \textsc{numpy} \citep{numpy}, \textsc{pandas} \citep{pandas, pandasv1.0.1}, \textsc{photutils} \citep{photutils}, \textsc{scipy} \citep{scipy}, \textsc{topcat} \citep{topcat}.
\end{acknowledgements}

\bibliographystyle{aa} % style aa.bst
\bibliography{mybib} % your references Yourfile.bib

\begin{thebibliography}{174}
\expandafter\ifx\csname natexlab\endcsname\relax\def\natexlab#1{#1}\fi

\bibitem[{{Alberts} {et~al.}(2022){Alberts}, {Adams}, {Gregg}, {Pope},
  {Williams}, \& {Eisenhardt}}]{alberts2022}
{Alberts}, S., {Adams}, J., {Gregg}, B., {et~al.} 2022, \apj, 927, 235

\bibitem[{{Astropy Collaboration} {et~al.}(2022){Astropy Collaboration},
  {Price-Whelan}, {Lim}, {Earl}, {Starkman}, {Bradley}, {Shupe}, {Patil},
  {Corrales}, {Brasseur}, {N{\"o}the}, {Donath}, {Tollerud}, {Morris},
  {Ginsburg}, {Vaher}, {Weaver}, {Tocknell}, {Jamieson}, {van Kerkwijk},
  {Robitaille}, {Merry}, {Bachetti}, {G{\"u}nther}, {Aldcroft},
  {Alvarado-Montes}, {Archibald}, {B{\'o}di}, {Bapat}, {Barentsen},
  {Baz{\'a}n}, {Biswas}, {Boquien}, {Burke}, {Cara}, {Cara}, {Conroy},
  {Conseil}, {Craig}, {Cross}, {Cruz}, {D'Eugenio}, {Dencheva}, {Devillepoix},
  {Dietrich}, {Eigenbrot}, {Erben}, {Ferreira}, {Foreman-Mackey}, {Fox},
  {Freij}, {Garg}, {Geda}, {Glattly}, {Gondhalekar}, {Gordon}, {Grant},
  {Greenfield}, {Groener}, {Guest}, {Gurovich}, {Handberg}, {Hart},
  {Hatfield-Dodds}, {Homeier}, {Hosseinzadeh}, {Jenness}, {Jones}, {Joseph},
  {Kalmbach}, {Karamehmetoglu}, {Ka{\l}uszy{\'n}ski}, {Kelley}, {Kern},
  {Kerzendorf}, {Koch}, {Kulumani}, {Lee}, {Ly}, {Ma}, {MacBride}, {Maljaars},
  {Muna}, {Murphy}, {Norman}, {O'Steen}, {Oman}, {Pacifici}, {Pascual},
  {Pascual-Granado}, {Patil}, {Perren}, {Pickering}, {Rastogi}, {Roulston},
  {Ryan}, {Rykoff}, {Sabater}, {Sakurikar}, {Salgado}, {Sanghi}, {Saunders},
  {Savchenko}, {Schwardt}, {Seifert-Eckert}, {Shih}, {Jain}, {Shukla}, {Sick},
  {Simpson}, {Singanamalla}, {Singer}, {Singhal}, {Sinha}, {Sip{\H{o}}cz},
  {Spitler}, {Stansby}, {Streicher}, {{\v{S}}umak}, {Swinbank}, {Taranu},
  {Tewary}, {Tremblay}, {de Val-Borro}, {Van Kooten}, {Vasovi{\'c}}, {Verma},
  {de Miranda Cardoso}, {Williams}, {Wilson}, {Winkel}, {Wood-Vasey}, {Xue},
  {Yoachim}, {Zhang}, {Zonca}, \& {Astropy Project Contributors}}]{astropy3}
{Astropy Collaboration}, {Price-Whelan}, A.~M., {Lim}, P.~L., {et~al.} 2022,
  \apj, 935, 167

\bibitem[{{Astropy Collaboration} {et~al.}(2018){Astropy Collaboration},
  {Price-Whelan}, {Sip{\H{o}}cz}, {G{\"u}nther}, {Lim}, {Crawford}, {Conseil},
  {Shupe}, {Craig}, {Dencheva}, {Ginsburg}, {VanderPlas}, {Bradley},
  {P{\'e}rez-Su{\'a}rez}, {de Val-Borro}, {Aldcroft}, {Cruz}, {Robitaille},
  {Tollerud}, {Ardelean}, {Babej}, {Bach}, {Bachetti}, {Bakanov}, {Bamford},
  {Barentsen}, {Barmby}, {Baumbach}, {Berry}, {Biscani}, {Boquien}, {Bostroem},
  {Bouma}, {Brammer}, {Bray}, {Breytenbach}, {Buddelmeijer}, {Burke},
  {Calderone}, {Cano Rodr{\'\i}guez}, {Cara}, {Cardoso}, {Cheedella}, {Copin},
  {Corrales}, {Crichton}, {D'Avella}, {Deil}, {Depagne}, {Dietrich}, {Donath},
  {Droettboom}, {Earl}, {Erben}, {Fabbro}, {Ferreira}, {Finethy}, {Fox},
  {Garrison}, {Gibbons}, {Goldstein}, {Gommers}, {Greco}, {Greenfield},
  {Groener}, {Grollier}, {Hagen}, {Hirst}, {Homeier}, {Horton}, {Hosseinzadeh},
  {Hu}, {Hunkeler}, {Ivezi{\'c}}, {Jain}, {Jenness}, {Kanarek}, {Kendrew},
  {Kern}, {Kerzendorf}, {Khvalko}, {King}, {Kirkby}, {Kulkarni}, {Kumar},
  {Lee}, {Lenz}, {Littlefair}, {Ma}, {Macleod}, {Mastropietro}, {McCully},
  {Montagnac}, {Morris}, {Mueller}, {Mumford}, {Muna}, {Murphy}, {Nelson},
  {Nguyen}, {Ninan}, {N{\"o}the}, {Ogaz}, {Oh}, {Parejko}, {Parley}, {Pascual},
  {Patil}, {Patil}, {Plunkett}, {Prochaska}, {Rastogi}, {Reddy Janga},
  {Sabater}, {Sakurikar}, {Seifert}, {Sherbert}, {Sherwood-Taylor}, {Shih},
  {Sick}, {Silbiger}, {Singanamalla}, {Singer}, {Sladen}, {Sooley},
  {Sornarajah}, {Streicher}, {Teuben}, {Thomas}, {Tremblay}, {Turner},
  {Terr{\'o}n}, {van Kerkwijk}, {de la Vega}, {Watkins}, {Weaver}, {Whitmore},
  {Woillez}, {Zabalza}, \& {Astropy Contributors}}]{astropy2}
{Astropy Collaboration}, {Price-Whelan}, A.~M., {Sip{\H{o}}cz}, B.~M., {et~al.}
  2018, \aj, 156, 123

\bibitem[{{Astropy Collaboration} {et~al.}(2013){Astropy Collaboration},
  {Robitaille}, {Tollerud}, {Greenfield}, {Droettboom}, {Bray}, {Aldcroft},
  {Davis}, {Ginsburg}, {Price-Whelan}, {Kerzendorf}, {Conley}, {Crighton},
  {Barbary}, {Muna}, {Ferguson}, {Grollier}, {Parikh}, {Nair}, {Unther},
  {Deil}, {Woillez}, {Conseil}, {Kramer}, {Turner}, {Singer}, {Fox}, {Weaver},
  {Zabalza}, {Edwards}, {Azalee Bostroem}, {Burke}, {Casey}, {Crawford},
  {Dencheva}, {Ely}, {Jenness}, {Labrie}, {Lim}, {Pierfederici}, {Pontzen},
  {Ptak}, {Refsdal}, {Servillat}, \& {Streicher}}]{astropy}
{Astropy Collaboration}, {Robitaille}, T.~P., {Tollerud}, E.~J., {et~al.} 2013,
  \aap, 558, A33

\bibitem[{{Bah{\'e}} {et~al.}(2013){Bah{\'e}}, {McCarthy}, {Balogh}, \&
  {Font}}]{bahe2013}
{Bah{\'e}}, Y.~M., {McCarthy}, I.~G., {Balogh}, M.~L., \& {Font}, A.~S. 2013,
  \mnras, 430, 3017

\bibitem[{{Baldry} {et~al.}(2006){Baldry}, {Balogh}, {Bower}, {Glazebrook},
  {Nichol}, {Bamford}, \& {Budavari}}]{baldry2006}
{Baldry}, I.~K., {Balogh}, M.~L., {Bower}, R.~G., {et~al.} 2006, \mnras, 373,
  469

\bibitem[{{Balogh} {et~al.}(2016){Balogh}, {McGee}, {Mok}, {Muzzin}, {van der
  Burg}, {Bower}, {Finoguenov}, {Hoekstra}, {Lidman}, {Mulchaey}, {Noble},
  {Parker}, {Tanaka}, {Wilman}, {Webb}, {Wilson}, \& {Yee}}]{balogh2016}
{Balogh}, M.~L., {McGee}, S.~L., {Mok}, A., {et~al.} 2016, \mnras, 456, 4364

\bibitem[{{Balogh} {et~al.}(1997){Balogh}, {Morris}, {Yee}, {Carlberg}, \&
  {Ellingson}}]{balogh1997}
{Balogh}, M.~L., {Morris}, S.~L., {Yee}, H.~K.~C., {Carlberg}, R.~G., \&
  {Ellingson}, E. 1997, \apjl, 488, L75

\bibitem[{{Balogh} {et~al.}(2000){Balogh}, {Navarro}, \& {Morris}}]{balogh2000}
{Balogh}, M.~L., {Navarro}, J.~F., \& {Morris}, S.~L. 2000, \apj, 540, 113

\bibitem[{{Balogh} {et~al.}(1998){Balogh}, {Schade}, {Morris}, {Yee},
  {Carlberg}, \& {Ellingson}}]{balogh1998}
{Balogh}, M.~L., {Schade}, D., {Morris}, S.~L., {et~al.} 1998, \apjl, 504, L75

\bibitem[{{Barnes}(2004)}]{barnes2004}
{Barnes}, J.~E. 2004, \mnras, 350, 798

\bibitem[{{Bekki}(2009)}]{bekki2009}
{Bekki}, K. 2009, \mnras, 399, 2221

\bibitem[{{Belfiore} {et~al.}(2018){Belfiore}, {Maiolino}, {Bundy}, {Masters},
  {Bershady}, {Oyarz{\'u}n}, {Lin}, {Cano-Diaz}, {Wake}, {Spindler}, {Thomas},
  {Brownstein}, {Drory}, \& {Yan}}]{belfiore2018}
{Belfiore}, F., {Maiolino}, R., {Bundy}, K., {et~al.} 2018, \mnras, 477, 3014

\bibitem[{{Bigiel} {et~al.}(2008){Bigiel}, {Leroy}, {Walter}, {Brinks}, {de
  Blok}, {Madore}, \& {Thornley}}]{bigiel2008}
{Bigiel}, F., {Leroy}, A., {Walter}, F., {et~al.} 2008, \aj, 136, 2846

\bibitem[{{Binggeli} {et~al.}(1985){Binggeli}, {Sandage}, \&
  {Tammann}}]{binggeli1985}
{Binggeli}, B., {Sandage}, A., \& {Tammann}, G.~A. 1985, \aj, 90, 1681

\bibitem[{{Binggeli} {et~al.}(1987){Binggeli}, {Tammann}, \&
  {Sandage}}]{binggeli1987}
{Binggeli}, B., {Tammann}, G.~A., \& {Sandage}, A. 1987, \aj, 94, 251

\bibitem[{{Boquien} {et~al.}(2019){Boquien}, {Burgarella}, {Roehlly}, {Buat},
  {Ciesla}, {Corre}, {Inoue}, \& {Salas}}]{boquien2019}
{Boquien}, M., {Burgarella}, D., {Roehlly}, Y., {et~al.} 2019, \aap, 622, A103

\bibitem[{{Boselli} {et~al.}(2006){Boselli}, {Boissier}, {Cortese}, {Gil de
  Paz}, {Seibert}, {Madore}, {Buat}, \& {Martin}}]{boselli2006}
{Boselli}, A., {Boissier}, S., {Cortese}, L., {et~al.} 2006, \apj, 651, 811

\bibitem[{{Boselli} {et~al.}(2011){Boselli}, {Boissier}, {Heinis}, {Cortese},
  {Ilbert}, {Hughes}, {Cucciati}, {Davies}, {Ferrarese}, {Giovanelli},
  {Haynes}, {Baes}, {Balkowski}, {Brosch}, {Chapman}, {Charmandaris},
  {Clemens}, {Dariush}, {De Looze}, {di Serego Alighieri}, {Duc}, {Durrell},
  {Emsellem}, {Erben}, {Fritz}, {Garcia-Appadoo}, {Gavazzi}, {Grossi},
  {Jord{\'a}n}, {Hess}, {Huertas-Company}, {Hunt}, {Kent}, {Lambas},
  {Lan{\c{c}}on}, {MacArthur}, {Madden}, {Magrini}, {Mei}, {Momjian}, {Olowin},
  {Papastergis}, {Smith}, {Solanes}, {Spector}, {Spekkens}, {Taylor},
  {Valotto}, {van Driel}, {Verstappen}, {Vlahakis}, {Vollmer}, \&
  {Xilouris}}]{boselli2011}
{Boselli}, A., {Boissier}, S., {Heinis}, S., {et~al.} 2011, \aap, 528, A107

\bibitem[{{Boselli} {et~al.}(2014{\natexlab{a}}){Boselli}, {Cortese},
  {Boquien}, {Boissier}, {Catinella}, {Gavazzi}, {Lagos}, \&
  {Saintonge}}]{boselli2014_hrs3}
{Boselli}, A., {Cortese}, L., {Boquien}, M., {et~al.} 2014{\natexlab{a}}, \aap,
  564, A67

\bibitem[{{Boselli} {et~al.}(2016){Boselli}, {Cuillandre}, {Fossati},
  {Boissier}, {Bomans}, {Consolandi}, {Anselmi}, {Cortese}, {C{\^o}t{\'e}},
  {Durrell}, {Ferrarese}, {Fumagalli}, {Gavazzi}, {Gwyn}, {Hensler}, {Sun}, \&
  {Toloba}}]{boselli2016}
{Boselli}, A., {Cuillandre}, J.~C., {Fossati}, M., {et~al.} 2016, \aap, 587,
  A68

\bibitem[{{Boselli} {et~al.}(2018{\natexlab{a}}){Boselli}, {Fossati},
  {Consolandi}, {Amram}, {Ge}, {Sun}, {Anderson}, {Boissier}, {Boquien},
  {Buat}, {Burgarella}, {Cortese}, {C{\^o}t{\'e}}, {Cuillandre}, {Durrell},
  {Epinat}, {Ferrarese}, {Fumagalli}, {Galbany}, {Gavazzi},
  {G{\'o}mez-L{\'o}pez}, {Gwyn}, {Hensler}, {Kuncarayakti}, {Marcelin}, {Mendes
  de Oliveira}, {Quint}, {Roediger}, {Roehlly}, {Sanchez}, {Sanchez-Janssen},
  {Toloba}, {Trinchieri}, \& {Vollmer}}]{boselli2018_4424}
{Boselli}, A., {Fossati}, M., {Consolandi}, G., {et~al.} 2018{\natexlab{a}},
  \aap, 620, A164

\bibitem[{{Boselli} {et~al.}(2023{\natexlab{a}}){Boselli}, {Fossati},
  {C{\^o}t{\'e}}, {Cuillandre}, {Ferrarese}, {Gwyn}, {Amram}, {Ayromlou},
  {Balogh}, {Bellusci}, {Boquien}, {Gavazzi}, {Hensler}, {Longobardi},
  {Nelson}, {Pillepich}, {Roediger}, {Sanchez-Janssen}, {Sun}, \&
  {Trinchieri}}]{boselli2023_lfha}
{Boselli}, A., {Fossati}, M., {C{\^o}t{\'e}}, P., {et~al.} 2023{\natexlab{a}},
  \aap, 675, A123

\bibitem[{{Boselli} {et~al.}(2018{\natexlab{b}}){Boselli}, {Fossati},
  {Ferrarese}, {Boissier}, {Consolandi}, {Longobardi}, {Amram}, {Balogh},
  {Barmby}, {Boquien}, {Boulanger}, {Braine}, {Buat}, {Burgarella}, {Combes},
  {Contini}, {Cortese}, {C{\^o}t{\'e}}, {C{\^o}t{\'e}}, {Cuillandre},
  {Drissen}, {Epinat}, {Fumagalli}, {Gallagher}, {Gavazzi}, {Gomez-Lopez},
  {Gwyn}, {Harris}, {Hensler}, {Koribalski}, {Marcelin}, {McConnachie},
  {Miville-Deschenes}, {Navarro}, {Patton}, {Peng}, {Plana}, {Prantzos},
  {Robert}, {Roediger}, {Roehlly}, {Russeil}, {Salome}, {Sanchez-Janssen},
  {Serra}, {Spekkens}, {Sun}, {Taylor}, {Tonnesen}, {Vollmer}, {Willis},
  {Wozniak}, {Burdullis}, {Devost}, {Mahoney}, {Manset}, {Petric}, {Prunet}, \&
  {Withington}}]{boselli2018}
{Boselli}, A., {Fossati}, M., {Ferrarese}, L., {et~al.} 2018{\natexlab{b}},
  \aap, 614, A56

\bibitem[{{Boselli} {et~al.}(2020){Boselli}, {Fossati}, {Longobardi},
  {Boissier}, {Boquien}, {Braine}, {C{\^o}t{\'e}}, {Cuillandre}, {Epinat},
  {Ferrarese}, {Gavazzi}, {Gwyn}, {Hensler}, {Plana}, {Roehlly}, {Schimd},
  {Sun}, \& {Trinchieri}}]{boselli2020}
{Boselli}, A., {Fossati}, M., {Longobardi}, A., {et~al.} 2020, \aap, 634, L1

\bibitem[{{Boselli} {et~al.}(2019){Boselli}, {Fossati}, {Longobardi},
  {Consolandi}, {Amram}, {Sun}, {Andreani}, {Boquien}, {Braine}, {Combes},
  {C{\^o}t{\'e}}, {Cuillandre}, {Duc}, {Emsellem}, {Ferrarese}, {Gavazzi},
  {Gwyn}, {Hensler}, {Peng}, {Plana}, {Roediger}, {Sanchez-Janssen}, {Sarzi},
  {Serra}, \& {Trinchieri}}]{boselli2019}
{Boselli}, A., {Fossati}, M., {Longobardi}, A., {et~al.} 2019, \aap, 623, A52

\bibitem[{{Boselli} {et~al.}(2023{\natexlab{b}}){Boselli}, {Fossati},
  {Roediger}, {Boquien}, {Fumagalli}, {Balogh}, {Boissier}, {Braine}, {Ciesla},
  {C{\^o}t{\'e}}, {Cuillandre}, {Ferrarese}, {Gavazzi}, {Gwyn}, {Junais},
  {Hensler}, {Longobardi}, \& {Sun}}]{boselli2023}
{Boselli}, A., {Fossati}, M., {Roediger}, J., {et~al.} 2023{\natexlab{b}},
  \aap, 669, A73

\bibitem[{{Boselli} {et~al.}(2022){Boselli}, {Fossati}, \& {Sun}}]{boselli2022}
{Boselli}, A., {Fossati}, M., \& {Sun}, M. 2022, \aapr, 30, 3

\bibitem[{{Boselli} \& {Gavazzi}(2006)}]{boselli2006_rev}
{Boselli}, A. \& {Gavazzi}, G. 2006, \pasp, 118, 517

\bibitem[{{Boselli} \& {Gavazzi}(2014)}]{boselli2014_red}
{Boselli}, A. \& {Gavazzi}, G. 2014, \aapr, 22, 74

\bibitem[{{Boselli} {et~al.}(2021){Boselli}, {Lupi}, {Epinat}, {Amram},
  {Fossati}, {Anderson}, {Boissier}, {Boquien}, {Consolandi}, {C{\^o}t{\'e}},
  {Cuillandre}, {Ferrarese}, {Galbany}, {Gavazzi}, {G{\'o}mez-L{\'o}pez},
  {Gwyn}, {Hensler}, {Hutchings}, {Kuncarayakti}, {Longobardi}, {Peng},
  {Plana}, {Postma}, {Roediger}, {Roehlly}, {Schimd}, {Trinchieri}, \&
  {Vollmer}}]{boselli2021}
{Boselli}, A., {Lupi}, A., {Epinat}, B., {et~al.} 2021, \aap, 646, A139

\bibitem[{{Boselli} {et~al.}(2023{\natexlab{c}}){Boselli}, {Serra}, {de
  Gasperin}, {Vollmer}, {Amram}, {Edler}, {Fossati}, {Consolandi},
  {C{\^o}t{\'e}}, {Cuillandre}, {Ferrarese}, {Gwyn}, {Postma}, {Boquien},
  {Braine}, {Combes}, {Gavazzi}, {Hensler}, {Miville-Deschenes}, {Murgia},
  {Roediger}, {Roehlly}, {Smith}, {Zhang}, \& {Zabel}}]{boselli2023_vic}
{Boselli}, A., {Serra}, P., {de Gasperin}, F., {et~al.} 2023{\natexlab{c}},
  \aap, 676, A92

\bibitem[{{Boselli} {et~al.}(2014{\natexlab{b}}){Boselli}, {Voyer}, {Boissier},
  {Cucciati}, {Consolandi}, {Cortese}, {Fumagalli}, {Gavazzi}, {Heinis},
  {Roehlly}, \& {Toloba}}]{boselli2014_guvics}
{Boselli}, A., {Voyer}, E., {Boissier}, S., {et~al.} 2014{\natexlab{b}}, \aap,
  570, A69

\bibitem[{Bradley {et~al.}(2021)Bradley, Sipőcz, Robitaille, Tollerud,
  Vinícius, Deil, Barbary, Wilson, Busko, Günther, Cara, Conseil, Bostroem,
  Droettboom, Bray, Bratholm, Lim, Barentsen, Craig, Rathi, Pascual, Perren,
  Donath, Georgiev, de~Val-Borro, Kerzendorf, Bach, Quint, Souchereau, \&
  Weaver}]{photutils}
Bradley, L., Sipőcz, B., Robitaille, T., {et~al.} 2021, astropy/photutils:
  1.0.2, Zenodo, doi:10.5281/zenodo.4453725

\bibitem[{{Brown} {et~al.}(2023){Brown}, {Roberts}, {Thorp}, {Ellison},
  {Zabel}, {Wilson}, {Bah{\'e}}, {Bisaria}, {Bolatto}, {Boselli}, {Chung},
  {Cortese}, {Catinella}, {Davis}, {Jim{\'e}nez-Donaire}, {Lagos}, {Lee},
  {Parker}, {Smith}, {Spekkens}, {Stevens}, {Villanueva}, \&
  {Watts}}]{brown2023}
{Brown}, T., {Roberts}, I.~D., {Thorp}, M., {et~al.} 2023, arXiv e-prints,
  arXiv:2308.10943

\bibitem[{{Bruzual} \& {Charlot}(2003)}]{bruzual2003}
{Bruzual}, G. \& {Charlot}, S. 2003, \mnras, 344, 1000

\bibitem[{{Bundy} {et~al.}(2015){Bundy}, {Bershady}, {Law}, {Yan}, {Drory},
  {MacDonald}, {Wake}, {Cherinka}, {S{\'a}nchez-Gallego}, {Weijmans}, {Thomas},
  {Tremonti}, {Masters}, {Coccato}, {Diamond-Stanic}, {Arag{\'o}n-Salamanca},
  {Avila-Reese}, {Badenes}, {Falc{\'o}n-Barroso}, {Belfiore}, {Bizyaev},
  {Blanc}, {Bland-Hawthorn}, {Blanton}, {Brownstein}, {Byler}, {Cappellari},
  {Conroy}, {Dutton}, {Emsellem}, {Etherington}, {Frinchaboy}, {Fu}, {Gunn},
  {Harding}, {Johnston}, {Kauffmann}, {Kinemuchi}, {Klaene}, {Knapen},
  {Leauthaud}, {Li}, {Lin}, {Maiolino}, {Malanushenko}, {Malanushenko}, {Mao},
  {Maraston}, {McDermid}, {Merrifield}, {Nichol}, {Oravetz}, {Pan}, {Parejko},
  {Sanchez}, {Schlegel}, {Simmons}, {Steele}, {Steinmetz}, {Thanjavur},
  {Thompson}, {Tinker}, {van den Bosch}, {Westfall}, {Wilkinson}, {Wright},
  {Xiao}, \& {Zhang}}]{bundy2015}
{Bundy}, K., {Bershady}, M.~A., {Law}, D.~R., {et~al.} 2015, \apj, 798, 7

\bibitem[{{Butcher} \& {Oemler}(1984)}]{butcher1984}
{Butcher}, H. \& {Oemler}, A., J. 1984, \apj, 285, 426

\bibitem[{{Calzetti} {et~al.}(2010){Calzetti}, {Wu}, {Hong}, {Kennicutt},
  {Lee}, {Dale}, {Engelbracht}, {van Zee}, {Draine}, {Hao}, {Gordon},
  {Moustakas}, {Murphy}, {Regan}, {Begum}, {Block}, {Dalcanton}, {Funes}, {Gil
  de Paz}, {Johnson}, {Sakai}, {Skillman}, {Walter}, {Weisz}, {Williams}, \&
  {Wu}}]{calzetti2010}
{Calzetti}, D., {Wu}, S.~Y., {Hong}, S., {et~al.} 2010, \apj, 714, 1256

\bibitem[{{Cattorini} {et~al.}(2023){Cattorini}, {Gavazzi}, {Boselli}, \&
  {Fossati}}]{cattorini2023}
{Cattorini}, F., {Gavazzi}, G., {Boselli}, A., \& {Fossati}, M. 2023, \aap,
  671, A118

\bibitem[{{Cayatte} {et~al.}(1994){Cayatte}, {Kotanyi}, {Balkowski}, \& {van
  Gorkom}}]{cayatte1994}
{Cayatte}, V., {Kotanyi}, C., {Balkowski}, C., \& {van Gorkom}, J.~H. 1994,
  \aj, 107, 1003

\bibitem[{{Cayatte} {et~al.}(1990){Cayatte}, {van Gorkom}, {Balkowski}, \&
  {Kotanyi}}]{cayatte1990}
{Cayatte}, V., {van Gorkom}, J.~H., {Balkowski}, C., \& {Kotanyi}, C. 1990,
  \aj, 100, 604

\bibitem[{{Chabrier}(2003)}]{chabrier2003}
{Chabrier}, G. 2003, \pasp, 115, 763

\bibitem[{{Chamaraux} {et~al.}(1980){Chamaraux}, {Balkowski}, \&
  {Gerard}}]{chamaraux1980}
{Chamaraux}, P., {Balkowski}, C., \& {Gerard}, E. 1980, \aap, 83, 38

\bibitem[{{Chamba} {et~al.}(2024){Chamba}, {Hayes}, \& {LSST Dark Energy
  Science Collaboration}}]{chamba2024}
{Chamba}, N., {Hayes}, M.~J., \& {LSST Dark Energy Science Collaboration}.
  2024, \aap, 689, A28

\bibitem[{{Chamba} {et~al.}(2020){Chamba}, {Trujillo}, \&
  {Knapen}}]{chamba2020}
{Chamba}, N., {Trujillo}, I., \& {Knapen}, J.~H. 2020, \aap, 633, L3

\bibitem[{{Chamba} {et~al.}(2022){Chamba}, {Trujillo}, \&
  {Knapen}}]{chamba2022}
{Chamba}, N., {Trujillo}, I., \& {Knapen}, J.~H. 2022, \aap, 667, A87

\bibitem[{{Chung} {et~al.}(2009){Chung}, {van Gorkom}, {Kenney}, {Crowl}, \&
  {Vollmer}}]{chung2009}
{Chung}, A., {van Gorkom}, J.~H., {Kenney}, J. D.~P., {Crowl}, H., \&
  {Vollmer}, B. 2009, \aj, 138, 1741

\bibitem[{{Chung} {et~al.}(2007){Chung}, {van Gorkom}, {Kenney}, \&
  {Vollmer}}]{chung2007}
{Chung}, A., {van Gorkom}, J.~H., {Kenney}, J. D.~P., \& {Vollmer}, B. 2007,
  \apjl, 659, L115

\bibitem[{{Chy{\.z}y} {et~al.}(2007){Chy{\.z}y}, {Ehle}, \& {Beck}}]{chyzy2007}
{Chy{\.z}y}, K.~T., {Ehle}, M., \& {Beck}, R. 2007, \aap, 474, 415

\bibitem[{{Conroy} {et~al.}(2009){Conroy}, {Gunn}, \& {White}}]{conroy2009}
{Conroy}, C., {Gunn}, J.~E., \& {White}, M. 2009, \apj, 699, 486

\bibitem[{{Cooke} {et~al.}(2016){Cooke}, {Hatch}, {Stern}, {Rettura},
  {Brodwin}, {Galametz}, {Wylezalek}, {Bridge}, {Conselice}, {De Breuck},
  {Gonzalez}, \& {Jarvis}}]{cooke2016}
{Cooke}, E.~A., {Hatch}, N.~A., {Stern}, D., {et~al.} 2016, \apj, 816, 83

\bibitem[{{Cortese} {et~al.}(2012){Cortese}, {Boissier}, {Boselli}, {Bendo},
  {Buat}, {Davies}, {Eales}, {Heinis}, {Isaak}, \& {Madden}}]{cortese2012}
{Cortese}, L., {Boissier}, S., {Boselli}, A., {et~al.} 2012, \aap, 544, A101

\bibitem[{{Cortese} {et~al.}(2011){Cortese}, {Catinella}, {Boissier},
  {Boselli}, \& {Heinis}}]{cortese2011}
{Cortese}, L., {Catinella}, B., {Boissier}, S., {Boselli}, A., \& {Heinis}, S.
  2011, \mnras, 415, 1797

\bibitem[{{Crowl} {et~al.}(2005){Crowl}, {Kenney}, {van Gorkom}, \&
  {Vollmer}}]{crowl2005}
{Crowl}, H.~H., {Kenney}, J. D.~P., {van Gorkom}, J.~H., \& {Vollmer}, B. 2005,
  \aj, 130, 65

\bibitem[{{de Vaucouleurs}(1948)}]{devaucouleurs1948}
{de Vaucouleurs}, G. 1948, Annales d'Astrophysique, 11, 247

\bibitem[{{Diemer} \& {Kravtsov}(2015)}]{diemer2015}
{Diemer}, B. \& {Kravtsov}, A.~V. 2015, \apj, 799, 108

\bibitem[{{Dressler}(1980)}]{dressler1980}
{Dressler}, A. 1980, \apj, 236, 351

\bibitem[{{Dressler} {et~al.}(1997){Dressler}, {Oemler}, {Couch}, {Smail},
  {Ellis}, {Barger}, {Butcher}, {Poggianti}, \& {Sharples}}]{dressler1997}
{Dressler}, A., {Oemler}, Augustus, J., {Couch}, W.~J., {et~al.} 1997, \apj,
  490, 577

\bibitem[{{Ebeling} {et~al.}(2014){Ebeling}, {Stephenson}, \&
  {Edge}}]{ebeling2014}
{Ebeling}, H., {Stephenson}, L.~N., \& {Edge}, A.~C. 2014, \apjl, 781, L40

\bibitem[{{Erwin} {et~al.}(2008){Erwin}, {Pohlen}, \& {Beckman}}]{erwin2008}
{Erwin}, P., {Pohlen}, M., \& {Beckman}, J.~E. 2008, \aj, 135, 20

\bibitem[{{Faber} {et~al.}(2007){Faber}, {Willmer}, {Wolf}, {Koo}, {Weiner},
  {Newman}, {Im}, {Coil}, {Conroy}, {Cooper}, {Davis}, {Finkbeiner}, {Gerke},
  {Gebhardt}, {Groth}, {Guhathakurta}, {Harker}, {Kaiser}, {Kassin},
  {Kleinheinrich}, {Konidaris}, {Kron}, {Lin}, {Luppino}, {Madgwick},
  {Meisenheimer}, {Noeske}, {Phillips}, {Sarajedini}, {Schiavon}, {Simard},
  {Szalay}, {Vogt}, \& {Yan}}]{faber2007}
{Faber}, S.~M., {Willmer}, C.~N.~A., {Wolf}, C., {et~al.} 2007, \apj, 665, 265

\bibitem[{{Ferrarese} {et~al.}(2012){Ferrarese}, {C{\^o}t{\'e}}, {Cuillandre},
  {Gwyn}, {Peng}, {MacArthur}, {Duc}, {Boselli}, {Mei}, {Erben}, {McConnachie},
  {Durrell}, {Mihos}, {Jord{\'a}n}, {Lan{\c{c}}on}, {Puzia}, {Emsellem},
  {Balogh}, {Blakeslee}, {van Waerbeke}, {Gavazzi}, {Vollmer}, {Kavelaars},
  {Woods}, {Ball}, {Boissier}, {Courteau}, {Ferriere}, {Gavazzi},
  {Hildebrandt}, {Hudelot}, {Huertas-Company}, {Liu}, {McLaughlin}, {Mellier},
  {Milkeraitis}, {Schade}, {Balkowski}, {Bournaud}, {Carlberg}, {Chapman},
  {Hoekstra}, {Peng}, {Sawicki}, {Simard}, {Taylor}, {Tully}, {van Driel},
  {Wilson}, {Burdullis}, {Mahoney}, \& {Manset}}]{ferrarese2012}
{Ferrarese}, L., {C{\^o}t{\'e}}, P., {Cuillandre}, J.-C., {et~al.} 2012, \apjs,
  200, 4

\bibitem[{{Ferrarese} {et~al.}(2020){Ferrarese}, {C{\^o}t{\'e}}, {MacArthur},
  {Durrell}, {Gwyn}, {Duc}, {S{\'a}nchez-Janssen}, {Santos}, {Blakeslee},
  {Boselli}, {Boyer}, {Cantiello}, {Courteau}, {Cuillandre}, {Emsellem},
  {Erben}, {Gavazzi}, {Guhathakurta}, {Huertas-Company}, {Jord{\'a}n},
  {Lan{\c{c}}on}, {Liu}, {Mei}, {Mihos}, {Peng}, {Puzia}, {Roediger}, {Schade},
  {Taylor}, {Toloba}, \& {Zhang}}]{ferrarese2020}
{Ferrarese}, L., {C{\^o}t{\'e}}, P., {MacArthur}, L.~A., {et~al.} 2020, \apj,
  890, 128

\bibitem[{{Finn} {et~al.}(2018){Finn}, {Desai}, {Rudnick}, {Balogh}, {Haynes},
  {Jablonka}, {Koopmann}, {Moustakas}, {Peng}, {Poggianti}, {Rines}, \&
  {Zaritsky}}]{finn2018}
{Finn}, R.~A., {Desai}, V., {Rudnick}, G., {et~al.} 2018, \apj, 862, 149

\bibitem[{{Fossati} {et~al.}(2013){Fossati}, {Gavazzi}, {Savorgnan},
  {Fumagalli}, {Boselli}, {Guti{\'e}rrez}, {Hern{\'a}ndez Toledo},
  {Giovanelli}, \& {Haynes}}]{fossati2013}
{Fossati}, M., {Gavazzi}, G., {Savorgnan}, G., {et~al.} 2013, \aap, 553, A91

\bibitem[{{Fossati} {et~al.}(2018){Fossati}, {Mendel}, {Boselli}, {Cuillandre},
  {Vollmer}, {Boissier}, {Consolandi}, {Ferrarese}, {Gwyn}, {Amram}, {Boquien},
  {Buat}, {Burgarella}, {Cortese}, {C{\^o}t{\'e}}, {C{\^o}t{\'e}}, {Durrell},
  {Fumagalli}, {Gavazzi}, {Gomez-Lopez}, {Hensler}, {Koribalski}, {Longobardi},
  {Peng}, {Roediger}, {Sun}, \& {Toloba}}]{fossati2018}
{Fossati}, M., {Mendel}, J.~T., {Boselli}, A., {et~al.} 2018, \aap, 614, A57

\bibitem[{{Freeman}(1970)}]{freeman1970}
{Freeman}, K.~C. 1970, \apj, 160, 811

\bibitem[{{Galametz} {et~al.}(2013){Galametz}, {Grazian}, {Fontana},
  {Ferguson}, {Ashby}, {Barro}, {Castellano}, {Dahlen}, {Donley}, {Faber},
  {Grogin}, {Guo}, {Huang}, {Kocevski}, {Koekemoer}, {Lee}, {McGrath}, {Peth},
  {Willner}, {Almaini}, {Cooper}, {Cooray}, {Conselice}, {Dickinson}, {Dunlop},
  {Fazio}, {Foucaud}, {Gardner}, {Giavalisco}, {Hathi}, {Hartley}, {Koo},
  {Lai}, {de Mello}, {McLure}, {Lucas}, {Paris}, {Pentericci}, {Santini},
  {Simpson}, {Sommariva}, {Targett}, {Weiner}, {Wuyts}, \& {CANDELS
  Team}}]{galametz2013}
{Galametz}, A., {Grazian}, A., {Fontana}, A., {et~al.} 2013, \apjs, 206, 10

\bibitem[{{Gavazzi} {et~al.}(2006){Gavazzi}, {Boselli}, {Cortese}, {Arosio},
  {Gallazzi}, {Pedotti}, \& {Carrasco}}]{gavazzi2006}
{Gavazzi}, G., {Boselli}, A., {Cortese}, L., {et~al.} 2006, \aap, 446, 839

\bibitem[{{Gavazzi} {et~al.}(2003){Gavazzi}, {Boselli}, {Donati}, {Franzetti},
  \& {Scodeggio}}]{gavazzi2003}
{Gavazzi}, G., {Boselli}, A., {Donati}, A., {Franzetti}, P., \& {Scodeggio}, M.
  2003, \aap, 400, 451

\bibitem[{{Gavazzi} {et~al.}(2001){Gavazzi}, {Boselli}, {Mayer},
  {Iglesias-Paramo}, {V{\'\i}lchez}, \& {Carrasco}}]{gavazzi2001}
{Gavazzi}, G., {Boselli}, A., {Mayer}, L., {et~al.} 2001, \apjl, 563, L23

\bibitem[{{Gavazzi} {et~al.}(1999){Gavazzi}, {Boselli}, {Scodeggio}, {Pierini},
  \& {Belsole}}]{gavazzi1999}
{Gavazzi}, G., {Boselli}, A., {Scodeggio}, M., {Pierini}, D., \& {Belsole}, E.
  1999, \mnras, 304, 595

\bibitem[{{Gavazzi} {et~al.}(2005){Gavazzi}, {Boselli}, {van Driel}, \&
  {O'Neil}}]{gavazzi2005}
{Gavazzi}, G., {Boselli}, A., {van Driel}, W., \& {O'Neil}, K. 2005, \aap, 429,
  439

\bibitem[{{Gavazzi} {et~al.}(2013{\natexlab{a}}){Gavazzi}, {Fumagalli},
  {Fossati}, {Galardo}, {Grossetti}, {Boselli}, {Giovanelli}, \&
  {Haynes}}]{gavazzi2013_ha3_2}
{Gavazzi}, G., {Fumagalli}, M., {Fossati}, M., {et~al.} 2013{\natexlab{a}},
  \aap, 553, A89

\bibitem[{{Gavazzi} {et~al.}(2013{\natexlab{b}}){Gavazzi}, {Savorgnan},
  {Fossati}, {Dotti}, {Fumagalli}, {Boselli}, {Guti{\'e}rrez}, {Hern{\'a}ndez
  Toledo}, {Giovanelli}, \& {Haynes}}]{gavazzi2013_ha3_3}
{Gavazzi}, G., {Savorgnan}, G., {Fossati}, M., {et~al.} 2013{\natexlab{b}},
  \aap, 553, A90

\bibitem[{{Giovanelli} \& {Haynes}(1983)}]{giovanelli1983}
{Giovanelli}, R. \& {Haynes}, M.~P. 1983, \aj, 88, 881

\bibitem[{{Giovanelli} \& {Haynes}(1985)}]{giovanelli1985}
{Giovanelli}, R. \& {Haynes}, M.~P. 1985, \apj, 292, 404

\bibitem[{{Giovanelli} {et~al.}(2005){Giovanelli}, {Haynes}, {Kent},
  {Perillat}, {Saintonge}, {Brosch}, {Catinella}, {Hoffman}, {Stierwalt},
  {Spekkens}, {Lerner}, {Masters}, {Momjian}, {Rosenberg}, {Springob},
  {Boselli}, {Charmandaris}, {Darling}, {Davies}, {Garcia Lambas}, {Gavazzi},
  {Giovanardi}, {Hardy}, {Hunt}, {Iovino}, {Karachentsev}, {Karachentseva},
  {Koopmann}, {Marinoni}, {Minchin}, {Muller}, {Putman}, {Pantoja}, {Salzer},
  {Scodeggio}, {Skillman}, {Solanes}, {Valotto}, {van Driel}, \& {van
  Zee}}]{giovanelli2005}
{Giovanelli}, R., {Haynes}, M.~P., {Kent}, B.~R., {et~al.} 2005, \aj, 130, 2598

\bibitem[{{Girelli} {et~al.}(2020){Girelli}, {Pozzetti}, {Bolzonella},
  {Giocoli}, {Marulli}, \& {Baldi}}]{girelli2020}
{Girelli}, G., {Pozzetti}, L., {Bolzonella}, M., {et~al.} 2020, \aap, 634, A135

\bibitem[{{Goddard} {et~al.}(2017){Goddard}, {Thomas}, {Maraston}, {Westfall},
  {Etherington}, {Riffel}, {Mallmann}, {Zheng}, {Argudo-Fern{\'a}ndez},
  {Bershady}, {Bundy}, {Drory}, {Law}, {Yan}, {Wake}, {Weijmans}, {Bizyaev},
  {Brownstein}, {Lane}, {Maiolino}, {Masters}, {Merrifield}, {Nitschelm},
  {Pan}, {Roman-Lopes}, \& {Storchi-Bergmann}}]{goddard2017}
{Goddard}, D., {Thomas}, D., {Maraston}, C., {et~al.} 2017, \mnras, 465, 688

\bibitem[{{G{\'o}mez} {et~al.}(2003){G{\'o}mez}, {Nichol}, {Miller}, {Balogh},
  {Goto}, {Zabludoff}, {Romer}, {Bernardi}, {Sheth}, {Hopkins}, {Castander},
  {Connolly}, {Schneider}, {Brinkmann}, {Lamb}, {SubbaRao}, \&
  {York}}]{gomez2003}
{G{\'o}mez}, P.~L., {Nichol}, R.~C., {Miller}, C.~J., {et~al.} 2003, \apj, 584,
  210

\bibitem[{{Gonz{\'a}lez Delgado} {et~al.}(2015){Gonz{\'a}lez Delgado},
  {Garc{\'\i}a-Benito}, {P{\'e}rez}, {Cid Fernandes}, {de Amorim},
  {Cortijo-Ferrero}, {Lacerda}, {L{\'o}pez Fern{\'a}ndez}, {Vale-Asari},
  {S{\'a}nchez}, {Moll{\'a}}, {Ruiz-Lara}, {S{\'a}nchez-Bl{\'a}zquez},
  {Walcher}, {Alves}, {Aguerri}, {Bekerait{\'e}}, {Bland-Hawthorn}, {Galbany},
  {Gallazzi}, {Husemann}, {Iglesias-P{\'a}ramo}, {Kalinova},
  {L{\'o}pez-S{\'a}nchez}, {Marino}, {M{\'a}rquez}, {Masegosa}, {Mast},
  {M{\'e}ndez-Abreu}, {Mendoza}, {del Olmo}, {P{\'e}rez}, {Quirrenbach}, \&
  {Zibetti}}]{gonzalez-delgado2015}
{Gonz{\'a}lez Delgado}, R.~M., {Garc{\'\i}a-Benito}, R., {P{\'e}rez}, E.,
  {et~al.} 2015, \aap, 581, A103

\bibitem[{{Goto} {et~al.}(2003){Goto}, {Yamauchi}, {Fujita}, {Okamura},
  {Sekiguchi}, {Smail}, {Bernardi}, \& {Gomez}}]{goto2003}
{Goto}, T., {Yamauchi}, C., {Fujita}, Y., {et~al.} 2003, \mnras, 346, 601

\bibitem[{{Greener} {et~al.}(2020){Greener}, {Arag{\'o}n-Salamanca},
  {Merrifield}, {Peterken}, {Fraser-McKelvie}, {Masters}, {Krawczyk},
  {Boardman}, {Boquien}, {Andrews}, {Brinkmann}, \& {Drory}}]{greener2020}
{Greener}, M.~J., {Arag{\'o}n-Salamanca}, A., {Merrifield}, M.~R., {et~al.}
  2020, \mnras, 495, 2305

\bibitem[{{Gunn} \& {Gott}(1972)}]{gunn1972}
{Gunn}, J.~E. \& {Gott}, J.~Richard, I. 1972, \apj, 176, 1

\bibitem[{{Gwyn}(2008)}]{gwyn2008}
{Gwyn}, S. D.~J. 2008, \pasp, 120, 212

\bibitem[{{Haynes} {et~al.}(2018){Haynes}, {Giovanelli}, {Kent}, {Adams},
  {Balonek}, {Craig}, {Fertig}, {Finn}, {Giovanardi}, {Hallenbeck}, {Hess},
  {Hoffman}, {Huang}, {Jones}, {Koopmann}, {Kornreich}, {Leisman}, {Miller},
  {Moorman}, {O'Connor}, {O'Donoghue}, {Papastergis}, {Troischt}, {Stark}, \&
  {Xiao}}]{haynes2018}
{Haynes}, M.~P., {Giovanelli}, R., {Kent}, B.~R., {et~al.} 2018, \apj, 861, 49

\bibitem[{{Hernquist}(1990)}]{hernquist1990}
{Hernquist}, L. 1990, \apj, 356, 359

\bibitem[{{Herrmann} {et~al.}(2013){Herrmann}, {Hunter}, \&
  {Elmegreen}}]{herrmann2013}
{Herrmann}, K.~A., {Hunter}, D.~A., \& {Elmegreen}, B.~G. 2013, \aj, 146, 104

\bibitem[{{Herrmann} {et~al.}(2016){Herrmann}, {Hunter}, \&
  {Elmegreen}}]{herrmann2016}
{Herrmann}, K.~A., {Hunter}, D.~A., \& {Elmegreen}, B.~G. 2016, \aj, 151, 145

\bibitem[{{Huang} {et~al.}(2012){Huang}, {Haynes}, {Giovanelli}, \&
  {Brinchmann}}]{huang2012}
{Huang}, S., {Haynes}, M.~P., {Giovanelli}, R., \& {Brinchmann}, J. 2012, \apj,
  756, 113

\bibitem[{{Huchra} {et~al.}(1983){Huchra}, {Davis}, {Latham}, \&
  {Tonry}}]{huchra1983}
{Huchra}, J., {Davis}, M., {Latham}, D., \& {Tonry}, J. 1983, \apjs, 52, 89

\bibitem[{{Hunter} \& {Elmegreen}(2006)}]{hunter2006}
{Hunter}, D.~A. \& {Elmegreen}, B.~G. 2006, \apjs, 162, 49

\bibitem[{{Hunter}(2007)}]{matplotlib}
{Hunter}, J.~D. 2007, Computing in Science and Engineering, 9, 90

\bibitem[{{Jedrzejewski}(1987)}]{jedrzejewski1987}
{Jedrzejewski}, R.~I. 1987, \mnras, 226, 747

\bibitem[{{Johnson} {et~al.}(2021){Johnson}, {Leja}, {Conroy}, \&
  {Speagle}}]{johnson2021}
{Johnson}, B.~D., {Leja}, J., {Conroy}, C., \& {Speagle}, J.~S. 2021, \apjs,
  254, 22

\bibitem[{{Junais} {et~al.}(2022){Junais}, {Boissier}, {Boselli}, {Ferrarese},
  {C{\^o}t{\'e}}, {Gwyn}, {Roediger}, {Lim}, {Peng}, {Cuillandre},
  {Longobardi}, {Fossati}, {Hensler}, {Koda}, {Bautista}, {Boquien},
  {Ma{\l}ek}, {Amram}, \& {Roehlly}}]{junais2022}
{Junais}, {Boissier}, S., {Boselli}, A., {et~al.} 2022, \aap, 667, A76

\bibitem[{{Kantharia} {et~al.}(2008){Kantharia}, {Rao}, \&
  {Sirothia}}]{kantharia2008}
{Kantharia}, N.~G., {Rao}, A.~P., \& {Sirothia}, S.~K. 2008, \mnras, 383, 173

\bibitem[{{Kauffmann} {et~al.}(2004){Kauffmann}, {White}, {Heckman},
  {M{\'e}nard}, {Brinchmann}, {Charlot}, {Tremonti}, \&
  {Brinkmann}}]{kauffmann2004}
{Kauffmann}, G., {White}, S. D.~M., {Heckman}, T.~M., {et~al.} 2004, \mnras,
  353, 713

\bibitem[{{Kawata} \& {Mulchaey}(2008)}]{kawata2008}
{Kawata}, D. \& {Mulchaey}, J.~S. 2008, \apjl, 672, L103

\bibitem[{{Kawinwanichakij} {et~al.}(2017){Kawinwanichakij}, {Papovich},
  {Quadri}, {Glazebrook}, {Kacprzak}, {Allen}, {Bell}, {Croton}, {Dekel},
  {Ferguson}, {Forrest}, {Grogin}, {Guo}, {Kocevski}, {Koekemoer}, {Labb{\'e}},
  {Lucas}, {Nanayakkara}, {Spitler}, {Straatman}, {Tran}, {Tomczak}, \& {van
  Dokkum}}]{kawinwanichakij2017}
{Kawinwanichakij}, L., {Papovich}, C., {Quadri}, R.~F., {et~al.} 2017, \apj,
  847, 134

\bibitem[{{Kelkar} {et~al.}(2020){Kelkar}, {Dwarakanath}, {Poggianti},
  {Moretti}, {Monteiro-Oliveira}, {Machado}, {Lima-Neto}, {Fritz}, {Vulcani},
  {Gullieuszik}, \& {Bettoni}}]{kelkar2020}
{Kelkar}, K., {Dwarakanath}, K.~S., {Poggianti}, B.~M., {et~al.} 2020, \mnras,
  496, 442

\bibitem[{{Kenney} {et~al.}(2004){Kenney}, {van Gorkom}, \&
  {Vollmer}}]{kenney2004}
{Kenney}, J. D.~P., {van Gorkom}, J.~H., \& {Vollmer}, B. 2004, \aj, 127, 3361

\bibitem[{{Kennicutt Jr.,}(1989)}]{kennicutt1989}
{Kennicutt Jr.,}, R.~C. 1989, \apj, 344, 685

\bibitem[{{Kennicutt Jr.,}(1998)}]{kennicutt1998_sf}
{Kennicutt Jr.,}, R.~C. 1998, \araa, 36, 189

\bibitem[{{Kim} {et~al.}(2014){Kim}, {Rey}, {Jerjen}, {Lisker}, {Sung}, {Lee},
  {Chung}, {Pak}, {Yi}, \& {Lee}}]{kim2014}
{Kim}, S., {Rey}, S.-C., {Jerjen}, H., {et~al.} 2014, \apjs, 215, 22

\bibitem[{{Koopmann} {et~al.}(2006){Koopmann}, {Haynes}, \&
  {Catinella}}]{koopmann2006}
{Koopmann}, R.~A., {Haynes}, M.~P., \& {Catinella}, B. 2006, \aj, 131, 716

\bibitem[{{Koopmann} \& {Kenney}(2004)}]{koopmann2004_ha_morph}
{Koopmann}, R.~A. \& {Kenney}, J. D.~P. 2004, \apj, 613, 866

\bibitem[{{Larson} {et~al.}(1980){Larson}, {Tinsley}, \&
  {Caldwell}}]{larson1980}
{Larson}, R.~B., {Tinsley}, B.~M., \& {Caldwell}, C.~N. 1980, \apj, 237, 692

\bibitem[{{Lewis} {et~al.}(2002){Lewis}, {Balogh}, {De Propris}, {Couch},
  {Bower}, {Offer}, {Bland-Hawthorn}, {Baldry}, {Baugh}, {Bridges}, {Cannon},
  {Cole}, {Colless}, {Collins}, {Cross}, {Dalton}, {Driver}, {Efstathiou},
  {Ellis}, {Frenk}, {Glazebrook}, {Hawkins}, {Jackson}, {Lahav}, {Lumsden},
  {Maddox}, {Madgwick}, {Norberg}, {Peacock}, {Percival}, {Peterson},
  {Sutherland}, \& {Taylor}}]{lewis2002}
{Lewis}, I., {Balogh}, M., {De Propris}, R., {et~al.} 2002, \mnras, 334, 673

\bibitem[{{Lim} {et~al.}(2020){Lim}, {C{\^o}t{\'e}}, {Peng}, {Ferrarese},
  {Roediger}, {Durrell}, {Mihos}, {Wang}, {Gwyn}, {Cuillandre}, {Liu},
  {S{\'a}nchez-Janssen}, {Toloba}, {Sales}, {Guhathakurta}, {Lan{\c{c}}on}, \&
  {Puzia}}]{lim2020}
{Lim}, S., {C{\^o}t{\'e}}, P., {Peng}, E.~W., {et~al.} 2020, \apj, 899, 69

\bibitem[{{Lin} {et~al.}(2014){Lin}, {Jian}, {Foucaud}, {Norberg}, {Bower},
  {Cole}, {Arnalte-Mur}, {Chen}, {Coupon}, {Hsieh}, {Heinis}, {Phleps}, {Chen},
  {Lee}, {Burgett}, {Chambers}, {Denneau}, {Draper}, {Flewelling}, {Hodapp},
  {Huber}, {Kaiser}, {Kudritzki}, {Magnier}, {Metcalfe}, {Price}, {Tonry},
  {Wainscoat}, \& {Waters}}]{lin2014}
{Lin}, L., {Jian}, H.-Y., {Foucaud}, S., {et~al.} 2014, \apj, 782, 33

\bibitem[{{Lotz} {et~al.}(2021){Lotz}, {Dolag}, {Remus}, \&
  {Burkert}}]{lotz2021}
{Lotz}, M., {Dolag}, K., {Remus}, R.-S., \& {Burkert}, A. 2021, \mnras, 506,
  4516

\bibitem[{{McCarthy} {et~al.}(2008){McCarthy}, {Frenk}, {Font}, {Lacey},
  {Bower}, {Mitchell}, {Balogh}, \& {Theuns}}]{mccarthy2008}
{McCarthy}, I.~G., {Frenk}, C.~S., {Font}, A.~S., {et~al.} 2008, \mnras, 383,
  593

\bibitem[{McKinney {et~al.}(2010)}]{pandas}
McKinney, W. {et~al.} 2010, in Proceedings of the 9th Python in Science
  Conference, Vol. 445, Austin, TX, 51--56

\bibitem[{{McLaughlin}(1999)}]{mclaughlin1999}
{McLaughlin}, D.~E. 1999, \apjl, 512, L9

\bibitem[{{Mei} {et~al.}(2007){Mei}, {Blakeslee}, {C{\^o}t{\'e}}, {Tonry},
  {West}, {Ferrarese}, {Jord{\'a}n}, {Peng}, {Anthony}, \& {Merritt}}]{mei2007}
{Mei}, S., {Blakeslee}, J.~P., {C{\^o}t{\'e}}, P., {et~al.} 2007, \apj, 655,
  144

\bibitem[{{Minchin} {et~al.}(2019){Minchin}, {Taylor}, {K{\"o}ppen}, {Davies},
  {van Driel}, \& {Keenan}}]{minchin2019}
{Minchin}, R.~F., {Taylor}, R., {K{\"o}ppen}, J., {et~al.} 2019, \aj, 158, 121

\bibitem[{{Moore} {et~al.}(1996){Moore}, {Katz}, {Lake}, {Dressler}, \&
  {Oemler}}]{moore1996}
{Moore}, B., {Katz}, N., {Lake}, G., {Dressler}, A., \& {Oemler}, A. 1996,
  \nat, 379, 613

\bibitem[{{Moore} {et~al.}(1998){Moore}, {Lake}, \& {Katz}}]{moore1998}
{Moore}, B., {Lake}, G., \& {Katz}, N. 1998, \apj, 495, 139

\bibitem[{{Moss} \& {Whittle}(1993)}]{moss1993}
{Moss}, C. \& {Whittle}, M. 1993, \apjl, 407, L17

\bibitem[{{Moss} \& {Whittle}(2000)}]{moss2000}
{Moss}, C. \& {Whittle}, M. 2000, \mnras, 317, 667

\bibitem[{{Muzzin} {et~al.}(2013){Muzzin}, {Marchesini}, {Stefanon}, {Franx},
  {McCracken}, {Milvang-Jensen}, {Dunlop}, {Fynbo}, {Brammer}, {Labb{\'e}}, \&
  {van Dokkum}}]{muzzin2013}
{Muzzin}, A., {Marchesini}, D., {Stefanon}, M., {et~al.} 2013, \apj, 777, 18

\bibitem[{{Navarro} {et~al.}(1996){Navarro}, {Frenk}, \& {White}}]{navarro1996}
{Navarro}, J.~F., {Frenk}, C.~S., \& {White}, S. D.~M. 1996, \apj, 462, 563

\bibitem[{{Oemler}(1974)}]{oemler1974}
{Oemler}, Augustus, J. 1974, \apj, 194, 1

\bibitem[{{Peng} {et~al.}(2010){Peng}, {Lilly}, {Kova{\v{c}}}, {Bolzonella},
  {Pozzetti}, {Renzini}, {Zamorani}, {Ilbert}, {Knobel}, {Iovino}, {Maier},
  {Cucciati}, {Tasca}, {Carollo}, {Silverman}, {Kampczyk}, {de Ravel},
  {Sanders}, {Scoville}, {Contini}, {Mainieri}, {Scodeggio}, {Kneib}, {Le
  F{\`e}vre}, {Bardelli}, {Bongiorno}, {Caputi}, {Coppa}, {de la Torre},
  {Franzetti}, {Garilli}, {Lamareille}, {Le Borgne}, {Le Brun}, {Mignoli},
  {Perez Montero}, {Pello}, {Ricciardelli}, {Tanaka}, {Tresse}, {Vergani},
  {Welikala}, {Zucca}, {Oesch}, {Abbas}, {Barnes}, {Bordoloi}, {Bottini},
  {Cappi}, {Cassata}, {Cimatti}, {Fumana}, {Hasinger}, {Koekemoer},
  {Leauthaud}, {Maccagni}, {Marinoni}, {McCracken}, {Memeo}, {Meneux}, {Nair},
  {Porciani}, {Presotto}, \& {Scaramella}}]{peng2010}
{Peng}, Y.-j., {Lilly}, S.~J., {Kova{\v{c}}}, K., {et~al.} 2010, \apj, 721, 193

\bibitem[{P\'erez \& Granger(2007)}]{ipython}
P\'erez, F. \& Granger, B.~E. 2007, Computing in Science and Engineering, 9, 21

\bibitem[{{Poggianti} {et~al.}(2016){Poggianti}, {Fasano}, {Omizzolo},
  {Gullieuszik}, {Bettoni}, {Moretti}, {Paccagnella}, {Jaff{\'e}}, {Vulcani},
  {Fritz}, {Couch}, \& {D'Onofrio}}]{poggianti2016}
{Poggianti}, B.~M., {Fasano}, G., {Omizzolo}, A., {et~al.} 2016, \aj, 151, 78

\bibitem[{{Poggianti} {et~al.}(2017){Poggianti}, {Moretti}, {Gullieuszik},
  {Fritz}, {Jaff{\'e}}, {Bettoni}, {Fasano}, {Bellhouse}, {Hau}, {Vulcani},
  {Biviano}, {Omizzolo}, {Paccagnella}, {D'Onofrio}, {Cava}, {Sheen}, {Couch},
  \& {Owers}}]{poggianti2017}
{Poggianti}, B.~M., {Moretti}, A., {Gullieuszik}, M., {et~al.} 2017, \apj, 844,
  48

\bibitem[{{Pohlen} \& {Trujillo}(2006)}]{pohlen2006}
{Pohlen}, M. \& {Trujillo}, I. 2006, \aap, 454, 759

\bibitem[{{Quadri} {et~al.}(2012){Quadri}, {Williams}, {Franx}, \&
  {Hildebrandt}}]{quadri2012}
{Quadri}, R.~F., {Williams}, R.~J., {Franx}, M., \& {Hildebrandt}, H. 2012,
  \apj, 744, 88

\bibitem[{Reback {et~al.}(2020)Reback, McKinney, jbrockmendel, den Bossche,
  Augspurger, Cloud, gfyoung, Sinhrks, Hawkins, Klein, Roeschke, Tratner, She,
  Petersen, Ayd, MomIsBestFriend, Garcia, Schendel, Hayden, Jancauskas, Saxton,
  Battiston, McMaster, Seabold, chris b1, h~vetinari, Hoyer, Dong, Overmeire,
  \& Winkel}]{pandasv1.0.1}
Reback, J., McKinney, W., jbrockmendel, {et~al.} 2020, pandas-dev/pandas:
  Pandas 1.1.1, Zenodo, doi:10.5281/zenodo.3993412

\bibitem[{{Rhee} {et~al.}(2017){Rhee}, {Smith}, {Choi}, {Yi}, {Jaff{\'e}},
  {Candlish}, \& {S{\'a}nchez-J{\'a}nssen}}]{rhee2017}
{Rhee}, J., {Smith}, R., {Choi}, H., {et~al.} 2017, \apj, 843, 128

\bibitem[{{Rich} {et~al.}(2011){Rich}, {Kewley}, \& {Dopita}}]{rich2011}
{Rich}, J.~A., {Kewley}, L.~J., \& {Dopita}, M.~A. 2011, \apj, 734, 87

\bibitem[{{Roberts} {et~al.}(2019){Roberts}, {Parker}, {Brown}, {Joshi},
  {Hlavacek-Larrondo}, \& {Wadsley}}]{roberts2019}
{Roberts}, I.~D., {Parker}, L.~C., {Brown}, T., {et~al.} 2019, \apj, 873, 42

\bibitem[{{Roberts} {et~al.}(2021){Roberts}, {van Weeren}, {McGee}, {Botteon},
  {Ignesti}, \& {Rottgering}}]{roberts2021_group}
{Roberts}, I.~D., {van Weeren}, R.~J., {McGee}, S.~L., {et~al.} 2021, \aap,
  652, A153

\bibitem[{{Roediger} \& {Br{\"u}ggen}(2006)}]{roediger2006}
{Roediger}, E. \& {Br{\"u}ggen}, M. 2006, \mnras, 369, 567

\bibitem[{{Roediger} \& {Courteau}(2015)}]{roediger2015}
{Roediger}, J.~C. \& {Courteau}, S. 2015, \mnras, 452, 3209

\bibitem[{{Saitoh} {et~al.}(2009){Saitoh}, {Daisaka}, {Kokubo}, {Makino},
  {Okamoto}, {Tomisaka}, {Wada}, \& {Yoshida}}]{saitoh2009}
{Saitoh}, T.~R., {Daisaka}, H., {Kokubo}, E., {et~al.} 2009, \pasj, 61, 481

\bibitem[{{S{\'a}nchez} {et~al.}(2012){S{\'a}nchez}, {Kennicutt}, {Gil de Paz},
  {van de Ven}, {V{\'\i}lchez}, {Wisotzki}, {Walcher}, {Mast}, {Aguerri},
  {Albiol-P{\'e}rez}, {Alonso-Herrero}, {Alves}, {Bakos}, {Bart{\'a}kov{\'a}},
  {Bland-Hawthorn}, {Boselli}, {Bomans}, {Castillo-Morales}, {Cortijo-Ferrero},
  {de Lorenzo-C{\'a}ceres}, {Del Olmo}, {Dettmar}, {D{\'\i}az}, {Ellis},
  {Falc{\'o}n-Barroso}, {Flores}, {Gallazzi}, {Garc{\'\i}a-Lorenzo},
  {Gonz{\'a}lez Delgado}, {Gruel}, {Haines}, {Hao}, {Husemann},
  {Igl{\'e}sias-P{\'a}ramo}, {Jahnke}, {Johnson}, {Jungwiert}, {Kalinova},
  {Kehrig}, {Kupko}, {L{\'o}pez-S{\'a}nchez}, {Lyubenova}, {Marino},
  {M{\'a}rmol-Queralt{\'o}}, {M{\'a}rquez}, {Masegosa}, {Meidt},
  {Mendez-Abreu}, {Monreal-Ibero}, {Montijo}, {Mour{\~a}o}, {Palacios-Navarro},
  {Papaderos}, {Pasquali}, {Peletier}, {P{\'e}rez}, {P{\'e}rez}, {Quirrenbach},
  {Rela{\~n}o}, {Rosales-Ortega}, {Roth}, {Ruiz-Lara},
  {S{\'a}nchez-Bl{\'a}zquez}, {Sengupta}, {Singh}, {Stanishev}, {Trager},
  {Vazdekis}, {Viironen}, {Wild}, {Zibetti}, \& {Ziegler}}]{sanchez2012}
{S{\'a}nchez}, S.~F., {Kennicutt}, R.~C., {Gil de Paz}, A., {et~al.} 2012,
  \aap, 538, A8

\bibitem[{{Sazonova} {et~al.}(2020){Sazonova}, {Alatalo}, {Lotz}, {Rowlands},
  {Snyder}, {Boone}, {Brodwin}, {Hayden}, {Lanz}, {Perlmutter}, \&
  {Rodriguez-Gomez}}]{sazonova2020}
{Sazonova}, E., {Alatalo}, K., {Lotz}, J., {et~al.} 2020, \apj, 899, 85

\bibitem[{{Sazonova} {et~al.}(2021){Sazonova}, {Alatalo}, {Rowlands},
  {Deustua}, {French}, {Heckman}, {Lanz}, {Lisenfeld}, {Luo}, {Medling},
  {Nyland}, {Otter}, {Petric}, {Snyder}, \& {Urry}}]{sazonova2021}
{Sazonova}, E., {Alatalo}, K., {Rowlands}, K., {et~al.} 2021, \apj, 919, 134

\bibitem[{{Schindler} {et~al.}(1999){Schindler}, {Binggeli}, \&
  {B{\"o}hringer}}]{schindler1999}
{Schindler}, S., {Binggeli}, B., \& {B{\"o}hringer}, H. 1999, \aap, 343, 420

\bibitem[{{Serra} {et~al.}(2023){Serra}, {Maccagni}, {Kleiner}, {Moln{\'a}r},
  {Ramatsoku}, {Loni}, {Loi}, {de Blok}, {Bryan}, {Dettmar}, {Frank}, {van
  Gorkom}, {Govoni}, {Iodice}, {J{\'o}zsa}, {Kamphuis}, {Kraan-Korteweg},
  {Loubser}, {Murgia}, {Oosterloo}, {Peletier}, {Pisano}, {Smith}, {Trager}, \&
  {Verheijen}}]{serra2023}
{Serra}, P., {Maccagni}, F.~M., {Kleiner}, D., {et~al.} 2023, \aap, 673, A146

\bibitem[{{S{\'e}rsic}(1963)}]{sersic1963}
{S{\'e}rsic}, J.~L. 1963, Boletin de la Asociacion Argentina de Astronomia La
  Plata Argentina, 6, 41

\bibitem[{{Simionescu} {et~al.}(2017){Simionescu}, {Werner}, {Mantz}, {Allen},
  \& {Urban}}]{simionescu2017}
{Simionescu}, A., {Werner}, N., {Mantz}, A., {Allen}, S.~W., \& {Urban}, O.
  2017, \mnras, 469, 1476

\bibitem[{{Smith} {et~al.}(2010){Smith}, {Lucey}, {Hammer}, {Hornschemeier},
  {Carter}, {Hudson}, {Marzke}, {Mouhcine}, {Eftekharzadeh}, {James},
  {Khosroshahi}, {Kourkchi}, \& {Karick}}]{smith2010}
{Smith}, R.~J., {Lucey}, J.~R., {Hammer}, D., {et~al.} 2010, \mnras, 408, 1417

\bibitem[{{Solanes} {et~al.}(2002){Solanes}, {Sanchis}, {Salvador-Sol{\'e}},
  {Giovanelli}, \& {Haynes}}]{solanes2002}
{Solanes}, J.~M., {Sanchis}, T., {Salvador-Sol{\'e}}, E., {Giovanelli}, R., \&
  {Haynes}, M.~P. 2002, \aj, 124, 2440

\bibitem[{{Sorgho} {et~al.}(2017){Sorgho}, {Hess}, {Carignan}, \&
  {Oosterloo}}]{sorgho2017}
{Sorgho}, A., {Hess}, K., {Carignan}, C., \& {Oosterloo}, T.~A. 2017, \mnras,
  464, 530

\bibitem[{{Steinhauser} {et~al.}(2016){Steinhauser}, {Schindler}, \&
  {Springel}}]{steinhauser2016}
{Steinhauser}, D., {Schindler}, S., \& {Springel}, V. 2016, \aap, 591, A51

\bibitem[{{Stone} {et~al.}(2021){Stone}, {Arora}, {Courteau}, \&
  {Cuillandre}}]{stone2021}
{Stone}, C.~J., {Arora}, N., {Courteau}, S., \& {Cuillandre}, J.-C. 2021,
  \mnras, 508, 1870

\bibitem[{{Strazzullo} {et~al.}(2019){Strazzullo}, {Pannella}, {Mohr}, {Saro},
  {Ashby}, {Bayliss}, {Bocquet}, {Bulbul}, {Khullar}, {Mantz}, {Stanford},
  {Benson}, {Bleem}, {Brodwin}, {Canning}, {Capasso}, {Chiu}, {Gonzalez},
  {Gupta}, {Hlavacek-Larrondo}, {Klein}, {McDonald}, {Noordeh}, {Rapetti},
  {Reichardt}, {Schrabback}, {Sharon}, \& {Stalder}}]{strazzullo2019}
{Strazzullo}, V., {Pannella}, M., {Mohr}, J.~J., {et~al.} 2019, \aap, 622, A117

\bibitem[{{Taylor}(2005)}]{topcat}
{Taylor}, M.~B. 2005, in Astronomical Society of the Pacific Conference Series,
  Vol. 347, Astronomical Data Analysis Software and Systems XIV, ed.
  P.~{Shopbell}, M.~{Britton}, \& R.~{Ebert}, 29

\bibitem[{{Tonnesen} \& {Bryan}(2009)}]{tonnesen2009}
{Tonnesen}, S. \& {Bryan}, G.~L. 2009, \apj, 694, 789

\bibitem[{{Tonnesen} {et~al.}(2007){Tonnesen}, {Bryan}, \& {van
  Gorkom}}]{tonnesen2007}
{Tonnesen}, S., {Bryan}, G.~L., \& {van Gorkom}, J.~H. 2007, \apj, 671, 1434

\bibitem[{{Trujillo} {et~al.}(2020){Trujillo}, {Chamba}, \&
  {Knapen}}]{trujillo2020}
{Trujillo}, I., {Chamba}, N., \& {Knapen}, J.~H. 2020, \mnras, 493, 87

\bibitem[{{van der Burg} {et~al.}(2020){van der Burg}, {Rudnick}, {Balogh},
  {Muzzin}, {Lidman}, {Old}, {Shipley}, {Gilbank}, {McGee}, {Biviano},
  {Cerulo}, {Chan}, {Cooper}, {De Lucia}, {Demarco}, {Forrest}, {Gwyn},
  {Jablonka}, {Kukstas}, {Marchesini}, {Nantais}, {Noble}, {Pintos-Castro},
  {Poggianti}, {Reeves}, {Stefanon}, {Vulcani}, {Webb}, {Wilson}, {Yee}, \&
  {Zaritsky}}]{vanderburg2020}
{van der Burg}, R. F.~J., {Rudnick}, G., {Balogh}, M.~L., {et~al.} 2020, \aap,
  638, A112

\bibitem[{{van der Walt} {et~al.}(2011){van der Walt}, {Colbert}, \&
  {Varoquaux}}]{numpy}
{van der Walt}, S., {Colbert}, S.~C., \& {Varoquaux}, G. 2011, Computing in
  Science and Engineering, 13, 22

\bibitem[{{Villanueva} {et~al.}(2022){Villanueva}, {Bolatto}, {Vogel}, {Brown},
  {Wilson}, {Zabel}, {Ellison}, {Stevens}, {Jim{\'e}nez Donaire}, {Spekkens},
  {Tharp}, {Davis}, {Parker}, {Roberts}, {Basra}, {Boselli}, {Catinella},
  {Chung}, {Cortese}, {Lee}, \& {Watts}}]{villanueva2022}
{Villanueva}, V., {Bolatto}, A.~D., {Vogel}, S., {et~al.} 2022, \apj, 940, 176

\bibitem[{Virtanen {et~al.}(2020)Virtanen, Gommers, Oliphant, Haberland, Reddy,
  Cournapeau, Burovski, Peterson, Weckesser, Bright, {van der Walt}, Brett,
  Wilson, Millman, Mayorov, Nelson, Jones, Kern, Larson, Carey, Polat, Feng,
  Moore, {VanderPlas}, Laxalde, Perktold, Cimrman, Henriksen, Quintero, Harris,
  Archibald, Ribeiro, Pedregosa, {van Mulbregt}, \& {SciPy 1.0
  Contributors}}]{scipy}
Virtanen, P., Gommers, R., Oliphant, T.~E., {et~al.} 2020, Nature Methods, 17,
  261

\bibitem[{{Vollmer} {et~al.}(2004){Vollmer}, {Beck}, {Kenney}, \& {van
  Gorkom}}]{vollmer2003}
{Vollmer}, B., {Beck}, R., {Kenney}, J. D.~P., \& {van Gorkom}, J.~H. 2004,
  \aj, 127, 3375

\bibitem[{{Vollmer} {et~al.}(2001){Vollmer}, {Cayatte}, {Balkowski}, \&
  {Duschl}}]{vollmer2001}
{Vollmer}, B., {Cayatte}, V., {Balkowski}, C., \& {Duschl}, W.~J. 2001, \apj,
  561, 708

\bibitem[{{Vollmer} \& {Huchtmeier}(2007)}]{vollmer2007}
{Vollmer}, B. \& {Huchtmeier}, W. 2007, \aap, 462, 93

\bibitem[{{Vollmer} {et~al.}(2013){Vollmer}, {Soida}, {Beck}, {Chung},
  {Urbanik}, {Chy{\.z}y}, {Otmianowska-Mazur}, \& {Kenney}}]{vollmer2013}
{Vollmer}, B., {Soida}, M., {Beck}, R., {et~al.} 2013, \aap, 553, A116

\bibitem[{{Vollmer} {et~al.}(2010){Vollmer}, {Soida}, {Chung}, {Beck},
  {Urbanik}, {Chy{\.z}y}, {Otmianowska-Mazur}, \& {van Gorkom}}]{vollmer2010}
{Vollmer}, B., {Soida}, M., {Chung}, A., {et~al.} 2010, \aap, 512, A36

\bibitem[{{Watkins} {et~al.}(2019){Watkins}, {Laine}, {Comer{\'o}n}, {Janz}, \&
  {Salo}}]{watkins2019}
{Watkins}, A.~E., {Laine}, J., {Comer{\'o}n}, S., {Janz}, J., \& {Salo}, H.
  2019, \aap, 625, A36

\bibitem[{{Watts} {et~al.}(2023){Watts}, {Cortese}, {Catinella}, {Brown},
  {Wilson}, {Zabel}, {Roberts}, {Davis}, {Thorp}, {Chung}, {Stevens},
  {Ellison}, {Spekkens}, {Parker}, {Bah{\'e}}, {Villanueva},
  {Jim{\'e}nez-Donaire}, {Bisaria}, {Boselli}, {Bolatto}, \& {Lee}}]{watts2023}
{Watts}, A.~B., {Cortese}, L., {Catinella}, B., {et~al.} 2023, \pasa, 40, e017

\bibitem[{{Weinmann} {et~al.}(2006){Weinmann}, {van den Bosch}, {Yang}, \&
  {Mo}}]{weinmann2006}
{Weinmann}, S.~M., {van den Bosch}, F.~C., {Yang}, X., \& {Mo}, H.~J. 2006,
  \mnras, 366, 2

\bibitem[{{Wetzel} {et~al.}(2012){Wetzel}, {Tinker}, \& {Conroy}}]{wetzel2012}
{Wetzel}, A.~R., {Tinker}, J.~L., \& {Conroy}, C. 2012, \mnras, 424, 232

\bibitem[{{Willmer}(2018)}]{willmer2018}
{Willmer}, C. N.~A. 2018, \apjs, 236, 47

\bibitem[{{Yoon} {et~al.}(2017){Yoon}, {Chung}, {Smith}, \&
  {Jaff{\'e}}}]{yoon2017}
{Yoon}, H., {Chung}, A., {Smith}, R., \& {Jaff{\'e}}, Y.~L. 2017, \apj, 838, 81

\bibitem[{{Yoshida} {et~al.}(2002){Yoshida}, {Yagi}, {Okamura}, {Aoki},
  {Ohyama}, {Komiyama}, {Yasuda}, {Iye}, {Kashikawa}, {Doi}, {Furusawa},
  {Hamabe}, {Kimura}, {Miyazaki}, {Miyazaki}, {Nakata}, {Ouchi}, {Sekiguchi},
  {Shimasaku}, \& {Ohtani}}]{yoshida2002}
{Yoshida}, M., {Yagi}, M., {Okamura}, S., {et~al.} 2002, \apj, 567, 118

\bibitem[{{Zabel} {et~al.}(2022){Zabel}, {Brown}, {Wilson}, {Davis}, {Cortese},
  {Parker}, {Boselli}, {Catinella}, {Chown}, {Chung}, {Deb}, {Ellison},
  {Jim{\'e}nez-Donaire}, {Lee}, {Roberts}, {Spekkens}, {Stevens}, {Thorp},
  {Tonnesen}, \& {Villanueva}}]{zabel2022}
{Zabel}, N., {Brown}, T., {Wilson}, C.~D., {et~al.} 2022, \apj, 933, 10

\end{thebibliography}

\begin{appendix}
\section{Masking of contaminating sources}
\label{app:mask}
\FloatBarrier

We show in Table \ref{tab:mask} the default \textsc{photutils} parameters used in each step of the masking run. Note that in certain cases, parameters were changed from these defaults after visual inspection and identification of over- or under-deblending.

\begin{table}[hbt!]

\caption{\textsc{photutils} parameters used in object masking routine}
\label{tab:mask}
\centering
\begin{tabular}{cccc}
\hline
\hline
Parameter & `Hot' mode &  `Cool' mode  &  `Cold' mode  \\ 
\hline

\textsc{nsigma}  &  97th percentile  &  2$\sigma$  & 2$\sigma$   \\
\textsc{npixels}  &  1 pixel  &  20 arcsec$^2$  & 20 pixels   \\
\textsc{contrast}  &  $4\times 10^{-1}$  &  $10^{-4}$  & ...  \\
\textsc{nlevels}  &  32  &  32  & ... \\
\textsc{mask}  &  Original  &  Original  & Object mask \\

\hline\hline
    \end{tabular}

\end{table}

\begin{figure}[hbt!]
\centering
\includegraphics[width=\columnwidth]{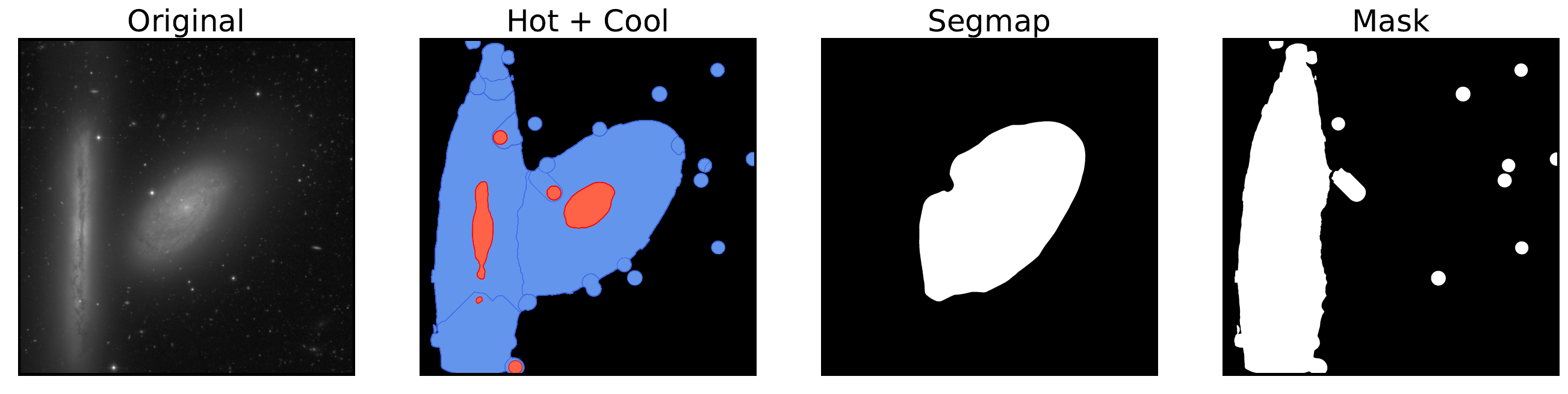}
\caption{Masking routine for VCC 483. From left to right: original \rband{} postage stamp, hot (red) and cool (blue) runs, final segmentation map and final mask. The red (hot) regions are bright peaks in the images, whereas the blue (cool) regions are extended surfaces with a low surface brightness threshold.}
\label{fig:mask}
\end{figure}

To help visualise the masking process, we show in Fig.~\ref{fig:mask} an example of the masking process on VCC~483 which has a close neighbour, VCC~497. The first panel shows the \rband{} postage stamp centred around VCC~483. In the second panel, we show regions detected in the `hot' run in red and the `cool' run in blue. These two runs create an object mask, with any blue regions beyond $4r_{\rm e,i}$ from the centre being masked, as well as any blue regions containing a red region (except for the central red region pertaining to the target galaxy). Using this mask, the segmentation map and mask are created in the final `cold' run and are shown in the third and fourth panel.

\vfill
\section{Correcting profiles for inclination effects}
\label{app:inc}

The majority of galaxies in the VESTIGE sample have moderate axis ratios between 0.5 and 0.8 - however there are a number of galaxies that are more substantially inclined. Radially averaged profiles of disk galaxies will vary depending on the inclination of the galaxy. The simplest model to correct for inclination effect is to consider the change in projected area for a galaxy with $i=0^{\circ}$ to $i>0 ^{\circ}$. Since the projected area of a disk galaxy with $i=0 ^{\circ}$ is a circle with area $A=\pi a^2$, where $a$ is the semi-major axis then the annulus used to measure surface brightness would have area of $A=\pi \left( a_{out}^2 - a_{in}^2 \right)$, where $a_{out}$ and $a_{in}$ represent the semi major axis (radius) of the outer and inner rings of the annulus, respectively. A galaxy with non-zero inclination will have a projected area of $A=\pi ab$ where $b$ is the semi-minor axis and $b=a \sec{\! \left(i\right)}$. Thus, the annuli used to measure the surface brightness will have area $A=\pi \left( a_{out}b_{out} - a_{in}b_{in} \right) = \pi \left( a_{out}^2 - a_{in}^2 \right)\sec{\left(i\right)}$. The line-of-sight flux measured within the annulus is constant (neglecting inclination effects on extinction), and the surface brightness is measured as $\mu = -2.5\log{\left( F/A \right)} + ZP = m + 2.5\log{\left( A \right)} + ZP$. Therefore, the difference in surface brightness $\mu$ as a function of inclination angle is simply:
\begin{equation}
\begin{split}
    \mu_{i=0 ^{\circ}} - \mu_{i=i'} ^{\circ} &= 2.5 \log{\left( \frac{A_{i=0}^{\circ}}{A_{i=i'} ^{\circ}} \right)} \\ 
    &= 2.5 \log{\left( \frac{\pi \left( a_{out}^2 - a_{in}^2 \right)}{\pi \left( a_{out}^2 - a_{in}^2 \right)\sec{\left( i \right) }} \right)} \\ 
    &= 2.5 \log{\left( \frac{1}{\sec{\left( i \right) }} \right)} \\ 
    &= 2.5 \log{\left( \frac{a}{b} \right)}.
\end{split}
\end{equation}

\noindent This simple correction assumes a disk with negligible height and also assumes that internal dust extinction is not a function of inclination. For the purposes of this analysis, we expect that both of these assumptions suffice for $b/a > 0.3$. After removing these highly inclined objects from our analysis, we find negligible differences in the overall results, so we choose to leave them in our sample.

\section{Examples of peculiar galaxies}
\label{app:peculiar}

\subsection{\texorpdfstring{Starburst galaxies with \ha{} filaments}{Starburst galaxies with H-alpha filaments}}

\begin{figure*}
\centering
\includegraphics[width=2\columnwidth]{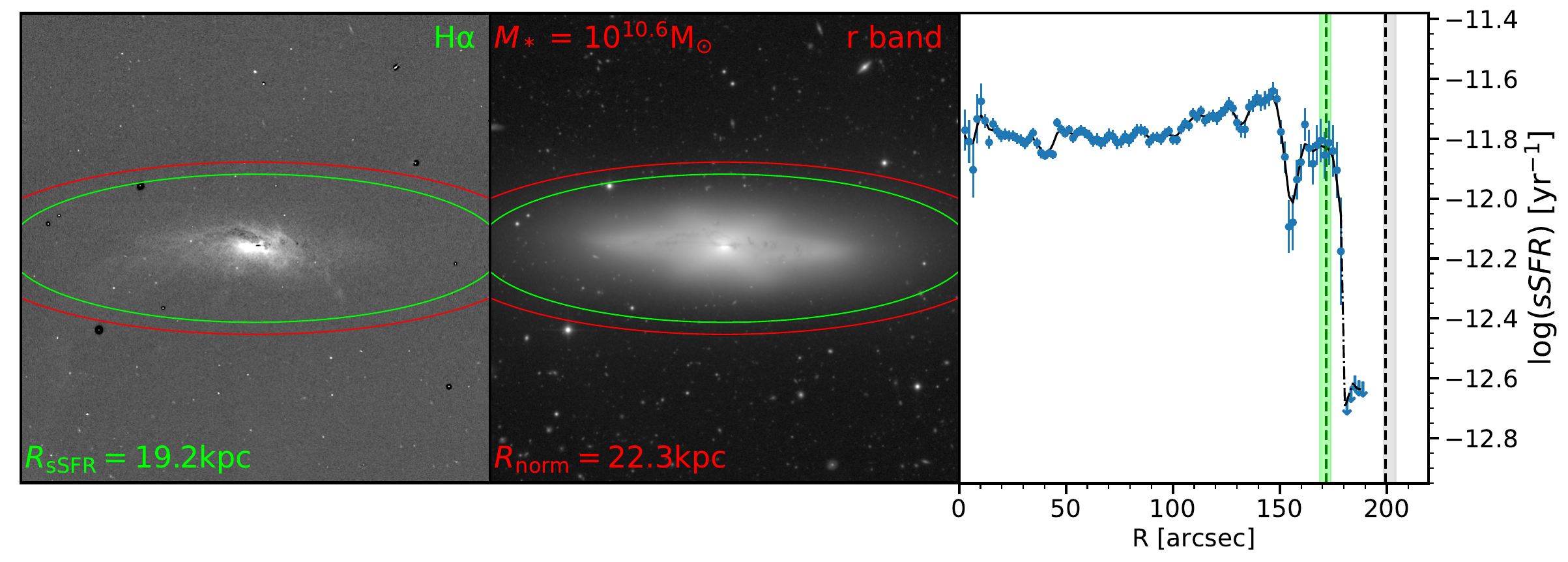}
\caption{\ha{} NB emission (\textit{left}), \rband{} optical emission (\textit{centre}), and sSFR profile (\textit{right}) for VCC~1190. In the left and centre postage stamps, the green ellipse outlines \rssfr{} while the red ellipse outlines \rnorm{}. On the sSFR plots, we show \rssfr{} and its associated uncertainty range as a green vertical line and band, and \rnorm{} and its uncertainty range as a grey vertical line and band.}
\label{fig:vcc1190}
\end{figure*}

Starburst galaxies with \ha{} filaments were not excluded from our sample. The detection of \ha{}-emitting filaments extending beyond the central starburst give these galaxies \rssfr{}/\rnorm{} values close to unity; however, these galaxies do not have an intact SF disk. Fig.~\ref{fig:vcc1190} shows VCC~1190 as an example, in the \ha{} (left) and \rband{} (right). This galaxy is massive, extremely gas-deficient and yet has \rssfr{}/\rnorm{}~$\sim 0.9$.

\FloatBarrier

\begin{figure*}
\centering
\includegraphics[width=2\columnwidth]{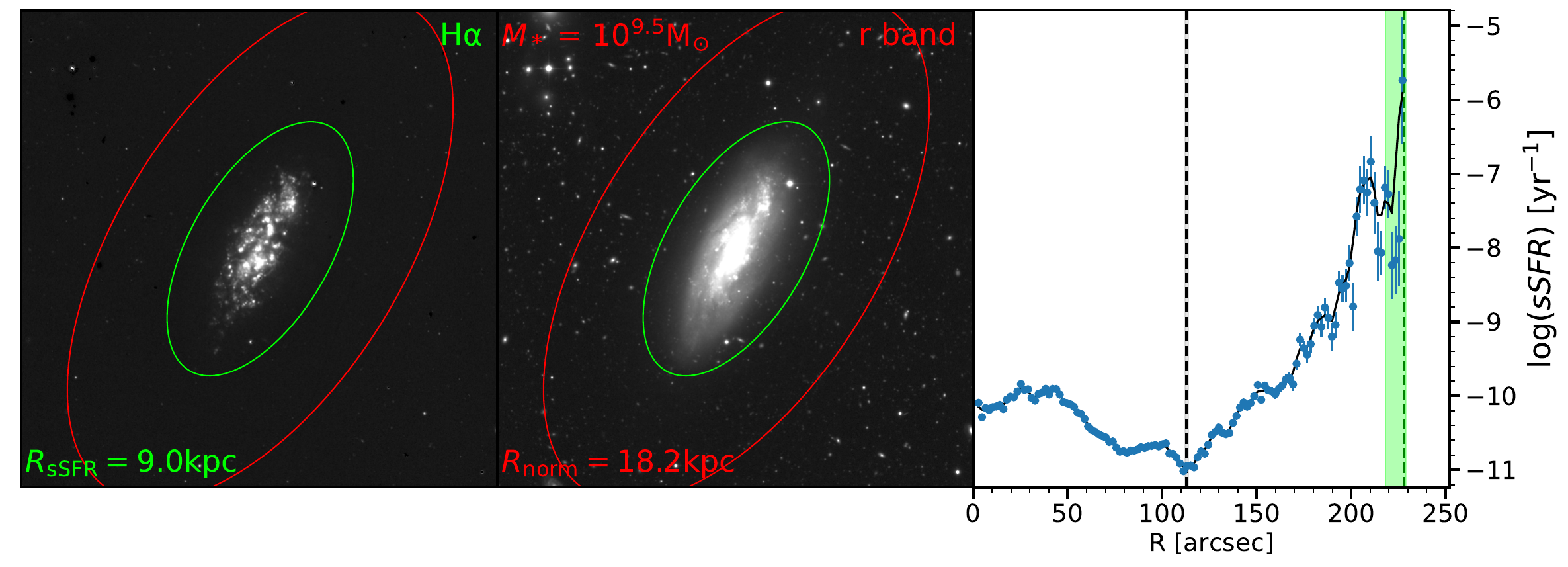}
\caption{Same as Fig.~\ref{fig:vcc1190} but for VCC~465. The value shown for \rssfr{} is visually a poor representation of the last valid local maximum on the sSFR plot, and drastically overestimates the extent of the \ha{} emission.}
\label{fig:vcc465}
\end{figure*}
\FloatBarrier

\subsection{\texorpdfstring{Galaxies with unphysical \rssfr{}}{Galaxies with unphysical Rssfr}}

In a small number of cases (18), visual inspection of the sSFR profiles showed features that caused the measured \rssfr{} value to be unphysical. Typically this was due to a spike in the sSFR values at large galactocentric radii. This could be due to local background variations or artefacts such as bright star halos which were not entirely masked. Additionally, local variation in the continuum subtraction could bias the \ha{} flux in low-SB regions to be higher than the true value. All of these galaxies were removed from our sample. We show an example of VCC~497 in Fig.~\ref{fig:vcc465}, with the \ha{} image on the left with an ellipse showing the unphysical \rssfr{} and the sSFR profile on the right. Clearly the green ellipse outlining \rssfr{} radius extends well beyond the \ha{} emission.

\section{Galaxy and quenching models}
\label{app:model}

To define the radial profiles of the gas and SFR, we use the sub-kpc Kennicutt-Schmidt law defined by \citet{bigiel2008} along with the gas density threshold concept used throughout this work. Choosing an example galaxy to have $M_*=10^{9.5}~\rm{M_{\odot}}$, our fits to the $\Sigma_*(R_{\rm edge})-M_*$ and $R_{\rm edge}-M_*$ relations from \citetalias{chamba2022} give a threshold $\Sigma_*(R_{\rm edge})=0.6~\rm{M_{\odot}~pc^{-2}}$ and $R_{\rm edge}=8.1~\text{kpc}$. Calculating the scale radius of the exponential profile from the critical \smsd{} value and the total stellar mass, we land on a scale radius $r_s=1.3~\text{kpc}$.

From \citet{huang2012}, we have scaling relations for H\textsc{i} and SFR as a function of stellar mass:

\begin{equation}
    \log{(M_{\rm H\textsc{i}})} = \begin{cases}
    0.712\log{(M_*)} + 3.117, & \log{(M_*)} \leq 9~\rm{M_{\odot}} \\
    0.276\log{(M_*)} + 7.042, & \log{(M_*)} > 9~\rm{M_{\odot}} 
    \end{cases},
\label{eq:mass_HI}
\end{equation}

\begin{equation}
    \log{(sSFR)} = \begin{cases}
    -0.149\log{(M_*)} - 8.207, & \log{(M_*)} \leq 9.5~\rm{M_{\odot}} \\
    -0.759\log{(M_*)} - 2.402, & \log{(M_*)} > 9.5~\rm{M_{\odot}} 
    \end{cases}.
\label{eq:mass_sfr}
\end{equation}

\begin{figure}[hbt!]
\includegraphics[scale=0.25]{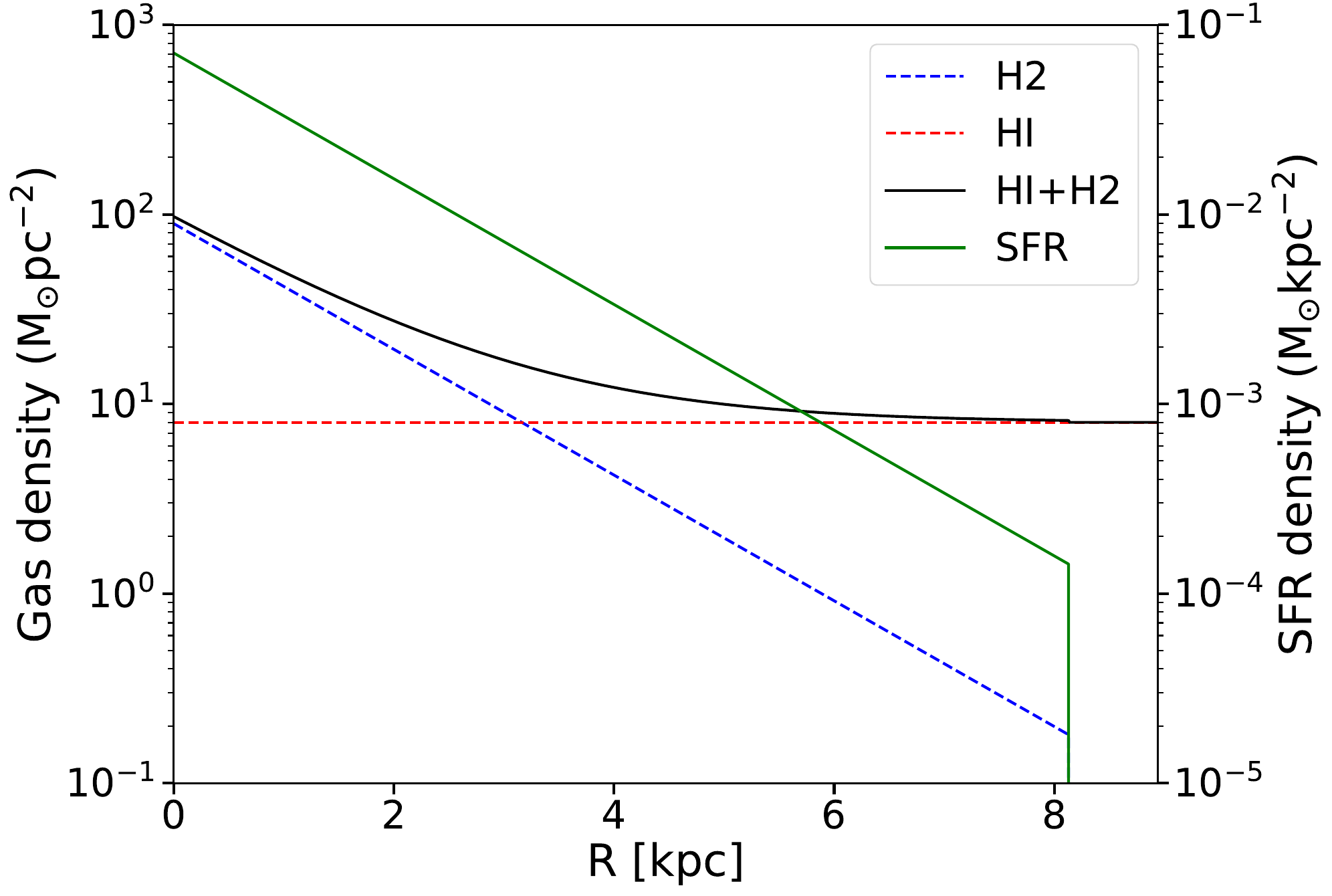}
\caption{Initial SFR (green), H\textsc{i} (red), H\textsubscript{2} (blue) and total gas (black) density profiles for a $10^{9.5} \rm{M_{\odot}}$ galaxy.}
\label{fig:initial_profiles}
\end{figure}

\noindent For the example galaxy, $M_*=10^{9.5}~\rm{M_{\odot}}$ gives $\log{(M_{\rm H\textsc{i}})}=9.7~\rm{M_{\odot}}$ and $~\text{SFR}=0.75~\text{M}_{\odot}~\text{yr}^{-1}$. We can then build a \sfrd{} profile with the same shape as the \smsd{} profile if we assume a flat sSFR profile. Following \citet{bigiel2008}, $\Sigma_{\rm SFR} = 10^{-3.1} \Sigma_{\rm H\textsubscript{2}}$ (with \sfrd{} in $\rm{M_{\odot}~kpc^{-2}}$ and $\Sigma_{\rm H\textsubscript{2}}$ in $\rm{M_{\odot}~pc^{-2}}$) and galaxies have relatively flat H\textsc{i} profiles with a value around $8~\rm{M_{\odot}~pc^{-2}}$. This allows us to determine an H\textsubscript{2} profile which also has an edge at 8.1~kpc, and an H\textsc{i} profile that extends beyond the SF disk to 13.5~kpc (assuming the total H\textsc{i} content follows the scaling relations in \citet{huang2012}, our galaxy has $M_{\rm H\textsc{i}} = 10^{9.7}~\rm{M_{\odot}}$). The initial gas and SFR density profiles for the galaxy are shown in Fig.~\ref{fig:initial_profiles}. Since the edge of the galaxy occurs at 8.1~kpc and beyond this point there is only H\textsc{i} gas, the total gas density threshold for star formation is determined by this value of H\textsc{i} which is $8~\rm{M_{\odot}~pc^{-2}}$.

\FloatBarrier
\FloatBarrier

\begin{figure*}
\centering
\includegraphics[scale=0.3]{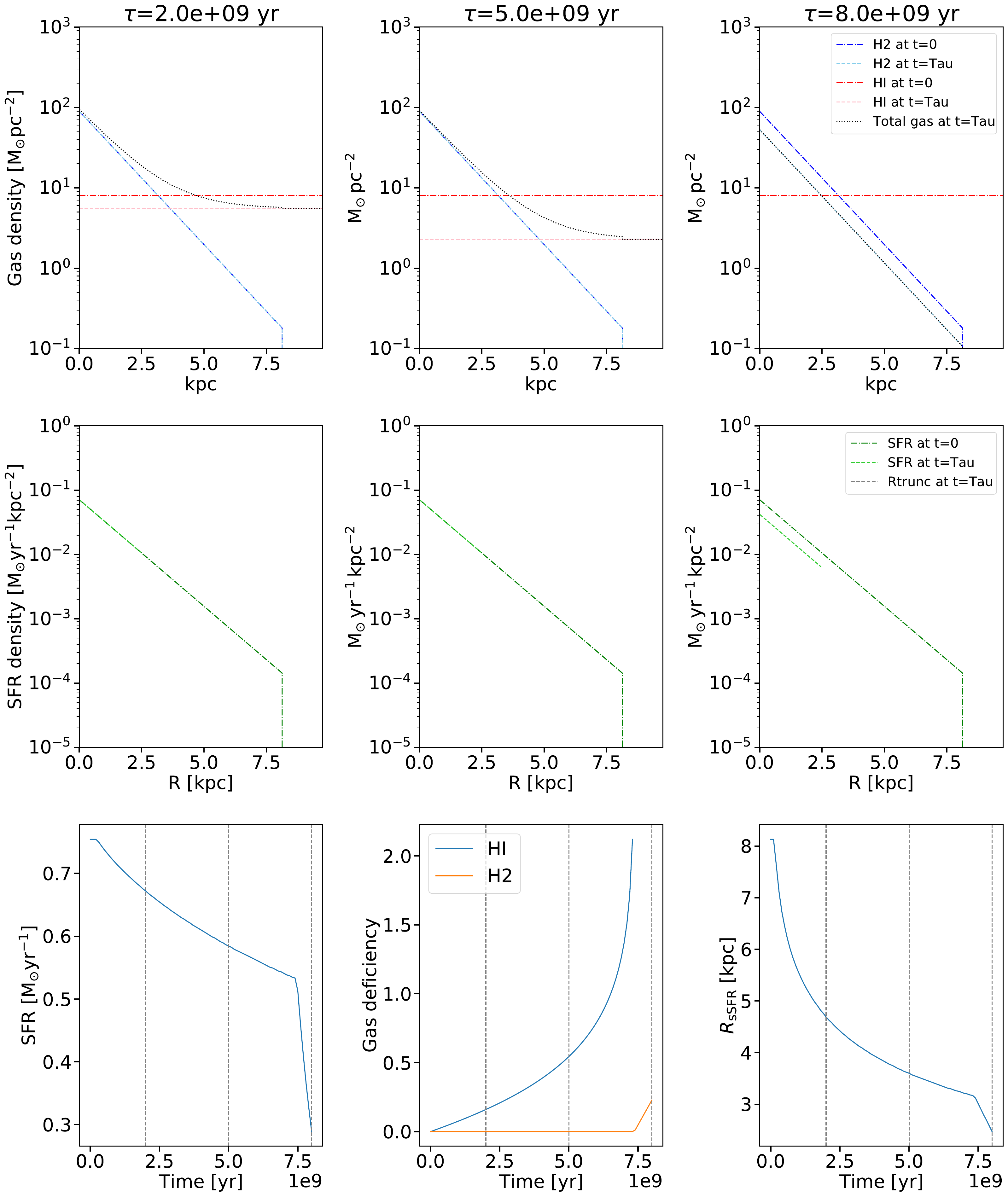}
\caption{\textit{Top}: Gas density profiles at three timesteps for starvation model. Initial profiles for H\textsc{i} and H\textsubscript{2} are shown in red and blue, respectively while evolved profiles are shown in pink and light blue, respectively. Evolved total gas density profile is shown in black. \textit{Middle}: SFR density profiles at three timesteps, where the initial profile is shown in green and the evolved profile in light green. The grey vertical line shows \rssfr{}. \textit{Bottom}: Time evolution of global SFR, gas deficiency and \rssfr{}.}
\label{fig:model2}
\end{figure*}

We assume that there is no radial dependence of the H\textsc{i} to H\textsubscript{2} conversion, and that the global H\textsc{i} content just drops by whatever amount necessary to replenish the H\textsubscript{2} at each timestep. As such, the H\textsc{i} profile drops but remains constant at all radii. 
In Fig.~\ref{fig:model2}, we show the gas and SFR density profiles in the first and second rows, respectively. Each column shows a different timestep (2, 5 and 8 Gyr after infall). On the SFR density plots, we show the \rssfr{} determined by where the total gas density falls below the threshold value of $8~\rm{M_{\odot}~pc^{-2}}$. In the bottom row, we show the evolution of the global SFR, gas deficiency and \rssfr{}. We see that in this model, truncations occur early as the H\textsc{i} begins to deplete, however it takes several Gyr for the global SFR to drop significantly; by this time, the galaxy is completely anemic in terms of its H\textsc{i} gas content.

Our RPS model assumes a galaxy falling into the cluster radially and face-on, such that the direction of the ram-pressure force is orthogonal to the plane of the disk. The efficiency of RPS can vary moderately as a function of inclination angle \citep[e.g.][]{roediger2006, steinhauser2016}. The restoring force holding the gas to the galactic disk is given by

\begin{equation}
    f = \left[g_*(\rm r) + g_{\rm H\textsc{i}}(r) + g_{\rm H\textsubscript{2}}(r) + g_{\rm DM}(r)\right]\,\Sigma_{\rm H\textsc{i}}(r),
\end{equation}

\noindent where $g_*(\rm r)$, $g_{\rm H\textsc{i}}(r)$, $g_{\rm H\textsubscript{2}}(r)$ and $g_{\rm DM}(r)$ are the contributions to the gravitational acceleration from the different galactic components (stars, H\textsc{i}, H\textsubscript{2} and dark matter), and $\Sigma_{\rm H\textsc{i}}(r)$ is the surface density of H\textsc{i} gas. Since our stellar component is simply an exponential disk, it follows that
\begin{equation}
g_*(r)=2\pi G \Sigma_*(r) = 2\pi G \Sigma_{0,*} e^{\rm -r/r_s},
\label{eq:g_star}
\end{equation}

\noindent where $\Sigma_{0,*}$ is the central stellar mass surface density and $r_s$ is the scale radius. Our H\textsubscript{2} profile is also an exponential so it will take the same form. Our H\textsc{i} profile is flat and as such $\Sigma_{\rm H\textsc{i}}(r) = \Sigma_{\rm 
0,HI}$. For the dark matter halo, we deviate from \citet{roberts2019} only in that we use a Navarro-Frenk-White (NFW; \citealt{navarro1996}) profile instead of a Hernquist profile \citep{hernquist1990}, with
\begin{equation}
    \Phi(r, \, z) = \frac{4\pi G \rho_0 a^3}{\sqrt{r^2 +  z^2}} \ln{\left(1 + \frac{\sqrt{r^2 + z^2}}{a}\right)},
\label{eq:nfw_pot}
\end{equation}

\noindent where we determine the halo mass, $M_{\rm h}$, from the stellar-to-halo mass relation of \citet{girelli2020}, which allows the determination of $R_{\rm 200}$. The scale radius, $a=R_{\rm 200}/c$, where $c$ is the concentration determined from the concentration-halo mass relation from \citet{diemer2015}. $\rho_0$ is the central density of the NFW profile which can be be determined from $M_{\rm h}$, $R_{\rm 200}$ and $a$. We calculate the gravitational acceleration of the dark matter profile by taking the maximum of the derivative along the z axis, where z is the direction perpendicular to the disk:

\begin{equation}
\begin{split}
    g_{\rm DM}(r)
    &= \max_{\rm z} \frac{\partial \Phi(r,~z)}{\partial z} \\ 
    &= \max_{\rm z} 4 \pi G \rho_{0} a^{3} z \left[ \frac{\ln{\left(a + \sqrt{r^2 + z^2} \right)} - \ln(a)}{\left(r^2 + z^2 \right)^{3/2}} \right. \\
    &\hspace{2.65cm} \left. -\frac{1  }{\left(  r^2 + z^2 \right) \left( a + \sqrt{r^2 + z^2}\right)} \right].
\end{split}
\end{equation}

To determine the ram-pressure at a given point in phase space, we use the beta profile fit from \citet{schindler1999}, with $r_c=12.9~\text{kpc}$ and $\beta=0.47$, and determine the central ICM density from \citet{simionescu2017} to be $n_0 = 2.33 \times 10^{-2}~\text{cm}^{-3}$. We then can determine the ICM mass density based on this relation from \citet{boselli2022}: 
\begin{equation}
    \rho_{\rm ICM}(r) = 1.92 \, \mu \, n_{\rm e}(r) \, m_{\rm p} = 1.15 \, n_{\rm e}(r) \, m_{\rm p},
\label{eq:icm}
\end{equation}

\noindent where $\mu=0.6$.

\FloatBarrier
\FloatBarrier

\clearpage
\onecolumn
\section{Additional radial profile plots}

\label{app:moreplots}
\FloatBarrier
 
\begin{figure*}[hbt!]
\centering
\includegraphics[width=1.\columnwidth]{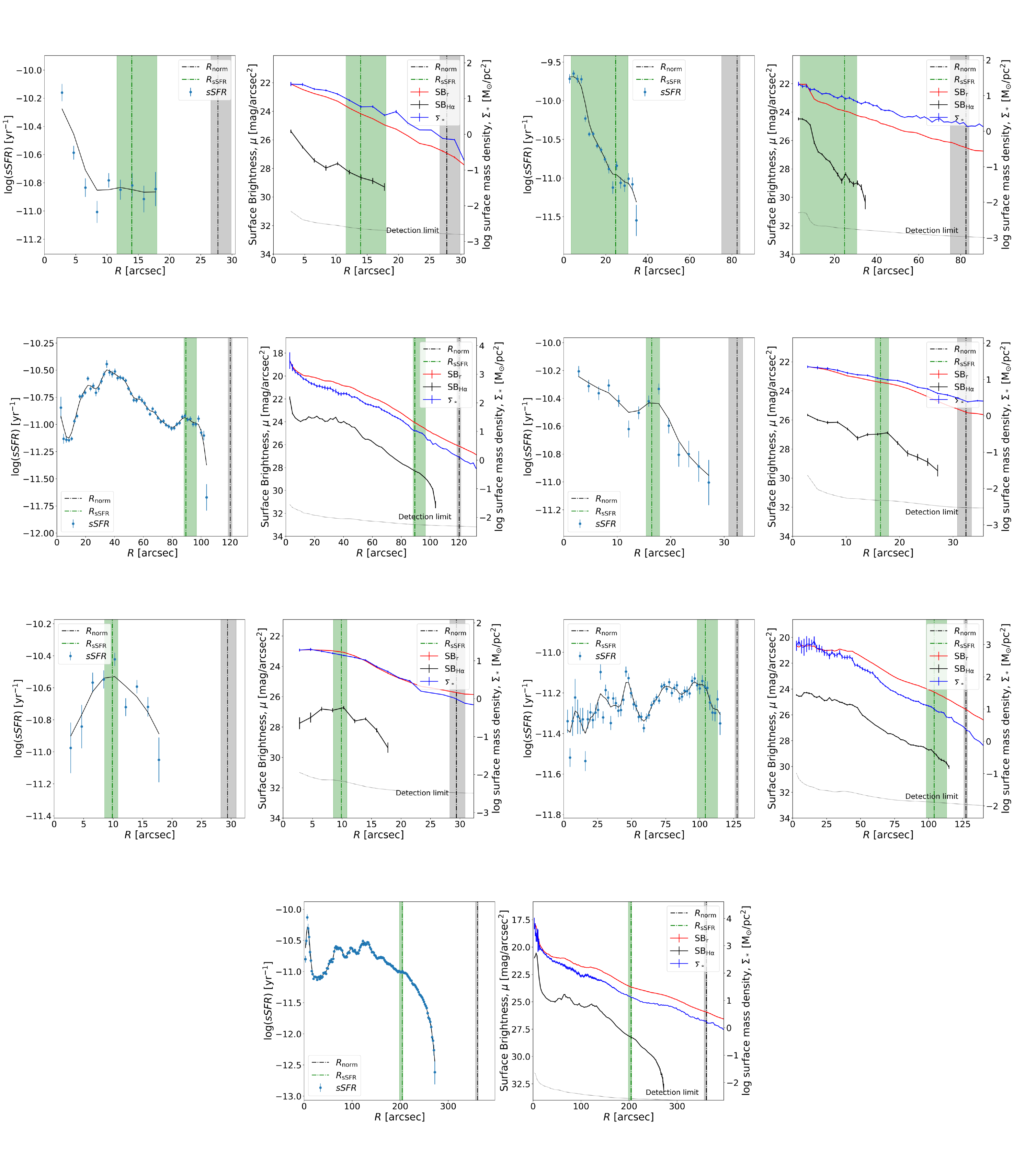}
\caption{Same as Fig. 1 but for the galaxies shown in Fig. \ref{fig:stamps}. From left to right and top to bottom: VCC 304, VCC 2037, VCC 157 (NGC 4208), VCC 1605, VCC 565, VCC 1868 (NGC 4607) and VCC 596 (NGC 4321). VCC 865 (NGC 4396) is not shown as it is featured in Fig. \ref{fig:VCC865_prof}.}
\label{fig:radialplots}
\end{figure*}

We show in Fig. \ref{fig:radialplots} the equivalent of Fig. \ref{fig:VCC865_prof} for the 7 additional galaxies whose postage stamps are shown in Fig. \ref{fig:stamps}.

\FloatBarrier

\end{appendix}
\end{document}